\documentclass[11pt]{article}
\usepackage{jheppub}
\usepackage{amssymb}
\usepackage{amsfonts}
\usepackage{amsmath}
\usepackage{graphicx}
\usepackage{caption}
\usepackage{subcaption}
\usepackage[usenames,dvipsnames,svgnames,table]{xcolor}
\usepackage{tikz}
\usetikzlibrary{matrix}

\def\FOmega{\overline{\underline{\Omega}}}
\def\IA{\mathbb{A}}

\def\IC{\mathbb{C}}

\def\IZ{{\mathbb{Z}}}
\def\IR{{\mathbb{R}}}
\def\IS{{\mathbb{S}}}
\def\IT{{\mathbb{T}}}
\def\IP{\mathbb{P}}

\def\MS {{\rm MS}}

\def\CQ {{\cal Q}}

\def\CN{{\cal N}}
\def\CK {{\cal K}}
\def\CN {{\cal N}}

\def\CF {{\cal F}}

\def\CP {{\cal P }}

\def\CW {{\cal W}}

\def\CH {{\cal H}}
\def\CB {{\cal B}}
\def\CS {{\cal S}}

\def\CK{{\cal K}}
\def\CU{{\cal U}}

\def\tSigma{\tilde{\Sigma}}
\def\tGamma{\tilde{\Gamma}}
\def\plift{\upsilon}

\def\one{{\hbox{ 1\kern-.8mm l}}}

\def\be{\bar{e}}

\def\hX{\hat{X}}
\def\hY{\hat{Y}}

\def\wr{\mathfrak{wr}}

\def\Tr{{\rm Tr}}

\def\be{ \begin{equation} }
\def\ee{ \end{equation}}

\def\fg{\mathfrak{g}}

\title{Wall-Crossing Invariants from Spectral Networks}

\author[1]{Pietro Longhi}
\affiliation[1]{Department of Physics and Astronomy,
Uppsala University,\\
Uppsala, Sweden}

\emailAdd{pietro.longhi@physics.uu.se}

\abstract{
A new construction of BPS monodromies for 4d ${\mathcal N}=2$ theories of class ${\mathcal S}$ is introduced.
A novel feature of this construction is its manifest invariance under Kontsevich-Soibelman wall crossing, in the sense that no information on the 4d BPS spectrum is employed.
The BPS monodromy is encoded by topological data of a finite graph, embedded into the UV curve $C$ of the theory.
The graph arises from a degenerate limit of spectral networks, constructed at maximal intersections of walls of marginal stability in the Coulomb branch of the gauge theory.
The topology of the graph, together with a notion of framing, encode equations that determine the monodromy. 
We develop an algorithmic technique for solving the equations, and compute the monodromy in several examples.
The graph manifestly encodes the symmetries of the monodromy, providing some support for conjectural relations to specializations of the superconformal index.
For $A_1$-type theories, the graphs encoding the monodromy are ``dessins d'enfants'' on $C$, the corresponding Strebel differentials coincide with the quadratic differentials that characterize the Seiberg-Witten curve.
}

\begin{document}

\maketitle

\section{Introduction}

Advances on wall crossing have been seminal for a wealth of developments that reshaped our understanding of BPS spectra of supersymmetric gauge theories \cite{Cecotti:1992rm, Kontsevich:2008fj, Gaiotto:2008cd, Gaiotto:2009hg, Alim:2011kw}.
There are several incarnations of the wall crossing phenomenon, taking place in theories in diverse dimensions, possibly in presence of supersymmetric defects of various types.
A universal feature of wall crossing across all these contexts, is the existence of an invariant quantity, known as the \emph{BPS monodromy}, which controls how BPS spectra change across moduli spaces.
Wall crossing formulae are an extremely powerful tool for computing BPS spectra, reducing the task of computing several (often, infinitely many) spectra to a single one, for a given theory.
At the same time, they have shifted the focus away from BPS states, highlighting the more fundamental role of BPS monodromies.
More recently, importance of the monodromy as a characterizing attribute of a quantum field theory has also emerged through conjectural relations to supersymmetric indices \cite{Cecotti:2010qn, Cecotti:2010fi, Iqbal:2012xm, Cordova:2015nma, Cordova:2016uwk, Cecotti:2015lab}.
Nevertheless there are certain aspects of BPS monodromies that remain deeply puzzling.
On the one hand, a direct field theoretic interpretation has been elusive so far, despite the close connection to BPS states.
On the other hand, the precise relation to Schur indices of 4d $\CN=2$ gauge theories is also rather mysterious, in particular outside the context of superconformal theories.
Somewhat puzzling is even the definition of the BPS monodromy, schematically it is the product
\be\label{eq:schematic-def}
	\mathbb{U} = \prod_{\gamma,m}^{\nwarrow}\Phi((-y)^m \hY_{\gamma})^{a_m(\gamma)}\,,
\ee
ordered by decreasing phase of the $\CN=2$ central charge $\arg Z_\gamma$, where $|a_m(\gamma)|$ counts BPS states with charge $\gamma$ and spin $m$. 
The manifest dependence on the BPS spectrum obscures both invariance under wall crossing, and transformation properties of $\mathbb{U}$ under global symmetries of a theory.

In this paper we introduce a new construction of the BPS monodromy for theories of class $\CS$, which differs from (\ref{eq:schematic-def}) in two key aspects. First, the construction is manifestly invariant under wall crossing, in the sense that it does not involve any data about BPS states. In fact our construction develops at loci of Coulomb branches where the BPS spectrum is \emph{ill-defined}, on walls of marginal stability.
Second, the construction shares a key property of Schur indices of class $\CS$ theories: it depends on topological data related to the UV curve that defines the theory. 
As we show with some examples, a consequence of this topological nature is that the monodromy is manifestly symmetric under permutations of identical punctures, reflecting symmetry of the index under the exchange of the corresponding fugacities.

\begin{figure}[h!]
\begin{subfigure}{0.3\textwidth}
        \includegraphics[width=\textwidth]{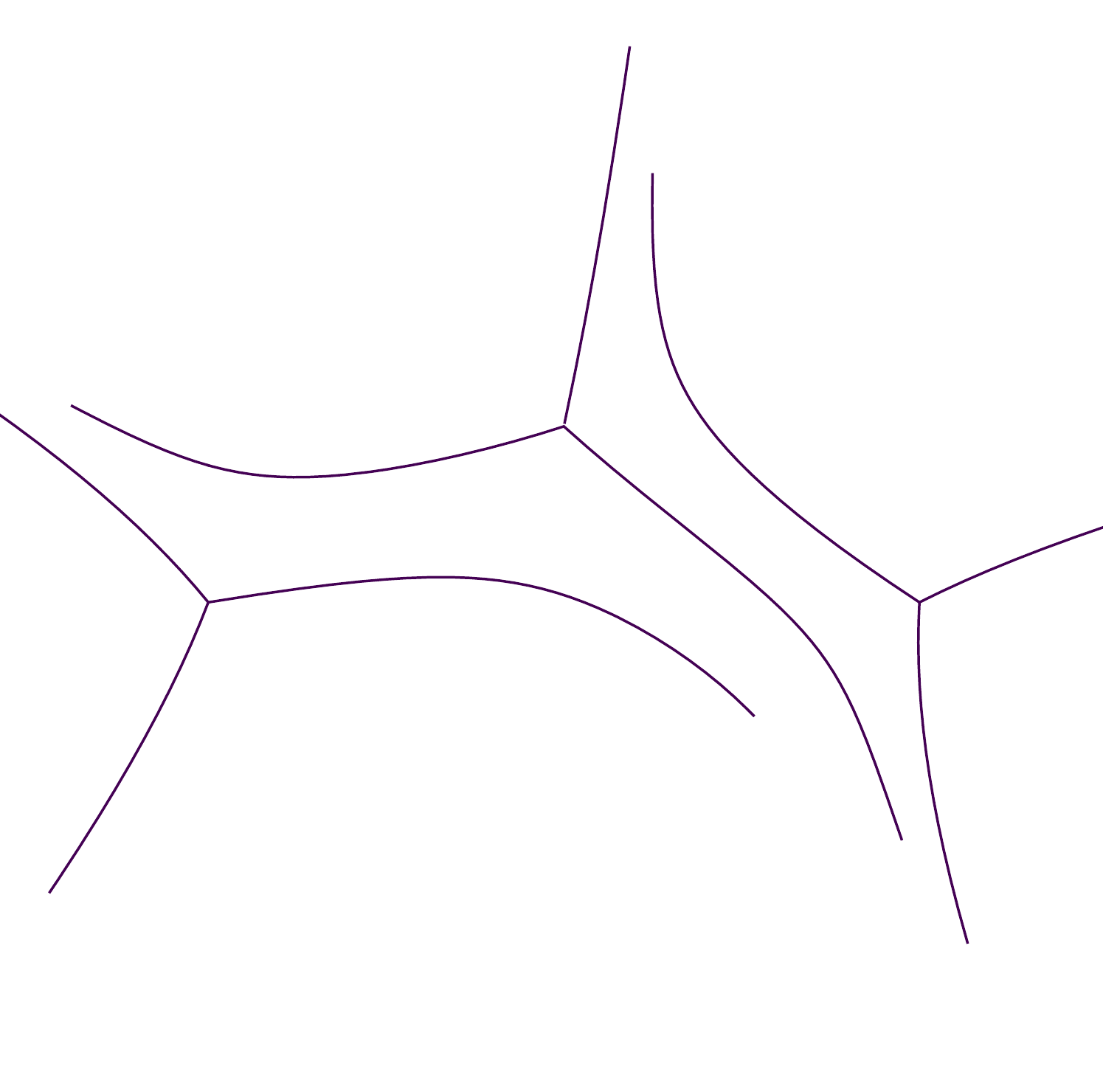}
\end{subfigure}
\hfill
\begin{subfigure}{0.3\textwidth}
        \includegraphics[width=\textwidth]{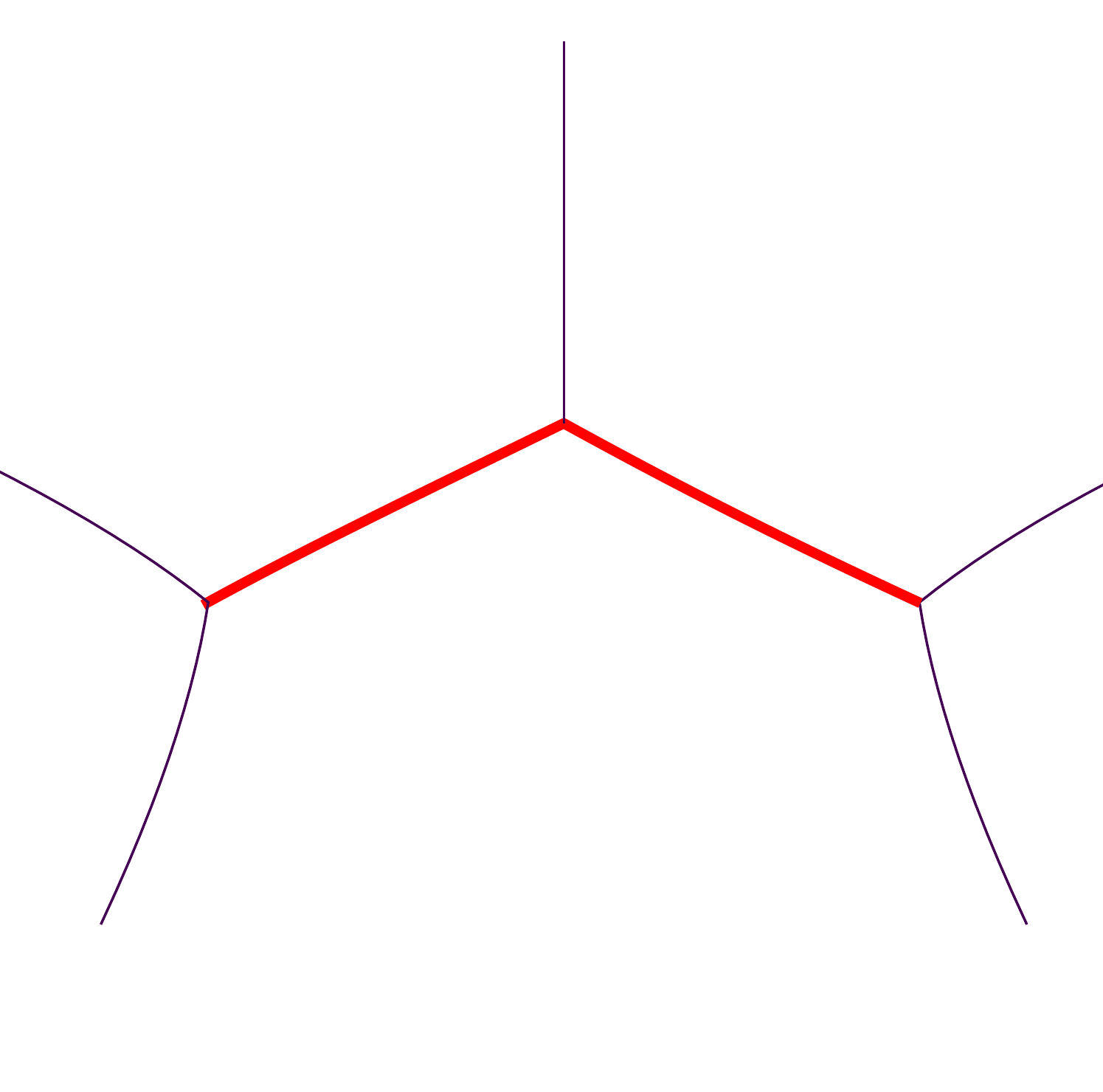}
\end{subfigure}
\hfill
\begin{subfigure}{0.3\textwidth}
        \includegraphics[width=\textwidth]{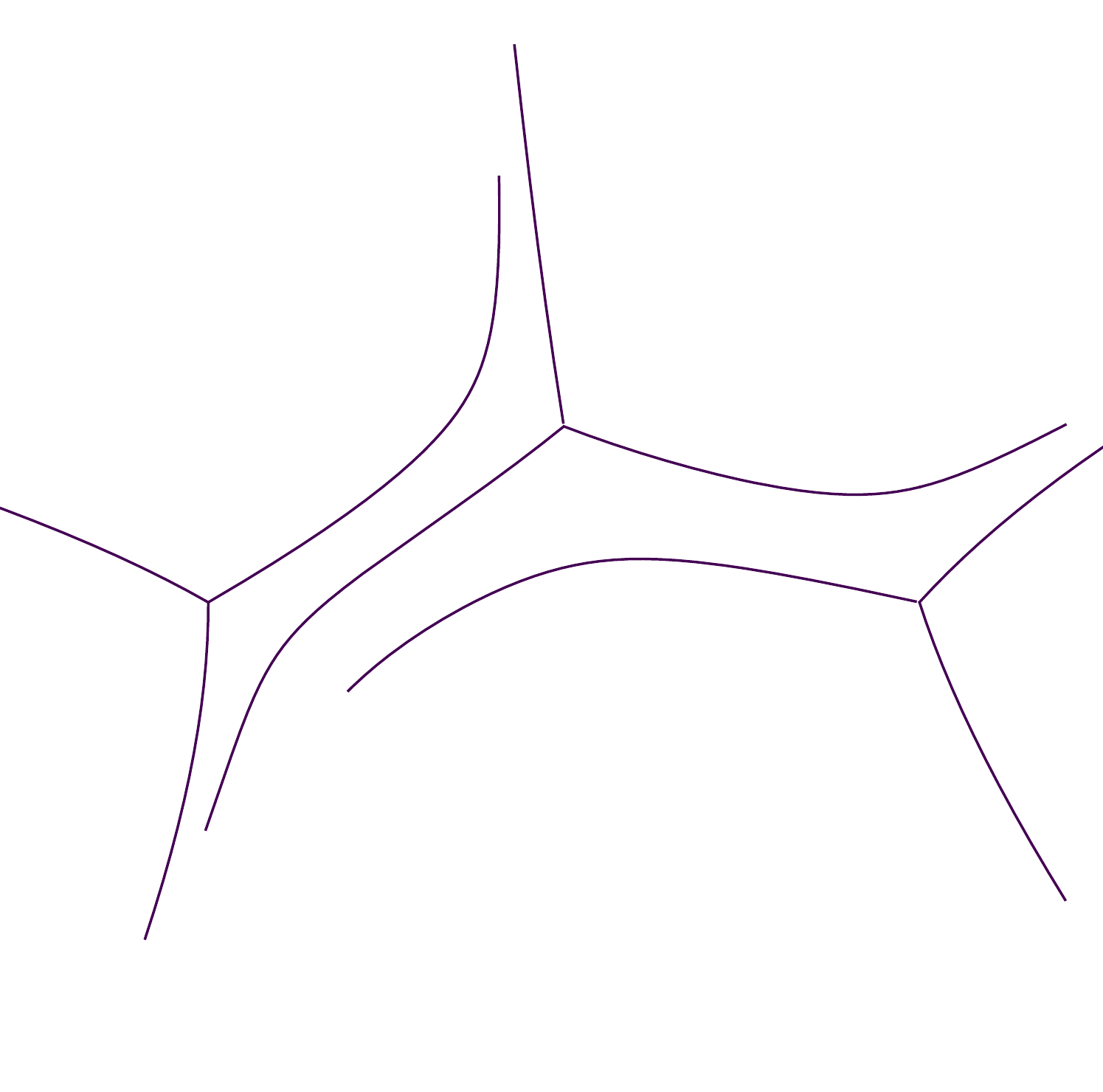}
\end{subfigure}
\caption{ 
A spectral network $\CW(\vartheta, u_c)$ for $u_c\in\CB_c$, at phases $\vartheta_c-\epsilon, \vartheta_c,\vartheta_c+\epsilon$. 
At the critical phase $\vartheta_c$ the topology of the network degenerates, as shown in the central frame. 
Degenerate edges, marked in red, make up the critical graph $\CW_c$.
}
\label{fig:example-graph}
\end{figure}

An important part of our construction is a certain graph $\CW_c$ embedded in the Riemann surface $C$ defining a class $\CS$ theory.
The relevance of such graphs in the context of BPS spectra emerged in the upcoming joint work \cite{network-quiver}, where they are used to obtain BPS quivers. These graphs also played a key role in \cite{Hollands:2013qza} in a construction of Fenchel-Nielsen coordinates on moduli spaces of flat connections on $C$.
Physically, $\CW_c$ emerges from the spectral network of the gauge theory at a special locus $\CB_c$ in the Coulomb branch $\CB$.
Recall that a spectral network is a collection of real curves on $C$, 
endowed with certain combinatorial data for each curve \cite{Gaiotto:2012rg, Galakhov:2014xba, Longhi:2016rjt}. 
The shape of a network $\CW(\vartheta,u)$  is controlled by a phase $\vartheta$ and a Coulomb vacuum $u\in\CB$, an important property is that its topology degenerates for special values of $\vartheta$: at phases of central charges of BPS states.
At the special locus $\CB_c$ the phases of central charges are all degenerate and coincide with either $\vartheta_c$ or $\vartheta_c+\pi$, corresponding to central charges of all BPS particles and their CPT conjugates. 
In other words $\CB_c$ is a maximal intersection of walls of marginal stability, where the whole BPS spectrum becomes ill-defined (a precise definition is given in Section \ref{sec:special-locus}).
The topology of the spectral network therefore degenerates only at the critical phases $\vartheta_c,\vartheta_c+\pi$. 
The \emph{critical graph} $\CW_c$ is the degenerate sub-network inside $\CW(\vartheta_c,u_c)$, an example is shown in Figure \ref{fig:example-graph}.
The BPS monodromy is entirely determined by the topology of $\CW_c$, together with a notion of framing, consisting of a cyclic ordering of the edges at each node. 
In the case of $A_1$ theories this data defines a \emph{ribbon graph}, and this  corresponds to a well-known object: at the special locus $\CB_c$ the Seiberg-Witten curve is described by a {Strebel differential} $\phi_2$, the critical graph $\CW_c$ is the corresponding \emph{dessin d'enfants}  \cite{1998math.ph..11024M, Ashok:2006du, Ashok:2006br, He:2015vua}.\footnote{More precisely, this holds when $C$ is a Riemann surfaces with regular punctures. The situation is slightly generalized in presence of irregular ones.}
In higher rank $ADE$ theories the special locus $\CB_c$ is instead specified by a collection of multi-differentials, and the graphs have a richer structure, involving new types of nodes.
Connections between 4d $\CN=2$ gauge theories and dessins d'enfants appeared previously in \cite{Ashok:2006du}, and more recently in \cite{He:2015vua} in a context closely related to spectral networks.

The fact that the locus $\CB_c$ contains any information about the BPS monodromy may seem surprising at first, since the BPS spectrum of the 4d gauge theory is not well-defined there.
In fact, to make sense of our construction it is essential to probe the theory with BPS surface defects, and to study the enlarged Hilbert space of 2d-4d BPS states \cite{Gaiotto:2011tf}.
These are supersymmetric field configurations which interpolate between vacua of the 2d defect theory at its spatial boundaries (located at $\{x^1=x^2=0,x^3 = \pm\infty\}$), classified by the net flux of the 4d IR abelian gauge fields through the $S^2$ at spatial infinity.
The spectrum of 2d-4d states does not suffer from ambiguities at $\CB_c$, since their central charges are generically different from those of 4d BPS states, due to the extra boundary conditions for field configurations imposed by the defect vacua.\footnote{We are slightly abusing terminology: the notion of ``surface defects'' adopted in this context includes coupled 2d-4d systems with $x^3$-dependent couplings for the 2d theory, which would be more appropriately termed ``supersymmetric interfaces'' \cite{Gaiotto:2011tf}. Accordingly, what we are currently referring to as ``2d-4d states'' would correspond to ``framed 2d-4d states''.
For the sake of simplicity here we deliberately abuse terminology, noting that the abuse is only mild, since in the limit of interest to us the two notions essentially coincide. Nevertheless, in the main body of the paper we shall be more precise, and the two concepts will be fully distinguished from each other.}
For this reason the 2d-4d spectrum is well-defined at $u_c$, in turn this confers a new physical meaning to the BPS monodromy of the 4d theory: it is the ``2d-4d wall crossing'' jump of the  spectrum of 2d-4d states at $\vartheta_c$.
The insight that wall crossing of 2d-4d BPS states encodes the BPS spectrum of the 4d theory is due to \cite{Gaiotto:2011tf}, here we limit ourselves to a mild extrapolation: even if 4d BPS states are not defined for $u\in\CB_c$, the jumps of 2d-4d states still capture the 4d BPS monodromy. In fact, invariance of $\mathbb{U}$ under wall crossing suggests that it should admit a definition even on walls of marginal stability, including the locus $\CB_c$, we claim that 2d-4d wall crossing provides such an interpretation.

Going back to the graph $\CW_c$, we can explain its relevance to computing $\mathbb{U}$ by recalling the physical interpretation of spectral networks. 
The combinatorial data attached to a spectral network encodes the spectrum of 2d-4d BPS states for a particular type of surface defect, termed \emph{canonical defect} \cite{Gaiotto:2011tf, Gaiotto:2012rg, Longhi-Park-2d}.
Moreover this data is determined entirely by the \emph{topology} of the network $\CW(\vartheta, u)$, and exhibits wall crossing behavior simultaneously with the topological degeneration of the network, at the critical phase $\vartheta_c$ in our setup.
A nontrivial but important fact is that $\mathbb{U}$ can be constructed by retaining only a specific piece of the spectral network, the \emph{degenerate sub-network} $\CW_c$. 
Our main result is the derivation of a set of equations associated to edges of the graph, which characterize the monodromy:
\be\label{eq:main-result}
	\mathbb{U} \, Q^{(-)}(p,y) \,   = \,Q^{(+)}(p,y) \, \mathbb{U} \,,\qquad \qquad \forall p \in \{\text{edges of }\CW_c\}\,,
\ee
where $Q^{(\pm)}(p,y)$ are formal generating series of 2d-4d states, entirely determined by the topology and framing of $\CW_c$.
Here $y$ is a fugacity for the ${\mathfrak{so}(2)}$ rotational symmetry preserved by the surface defect, it is a deformation parameter introduced in the context of spectral networks ``with spin'' in \cite{Galakhov:2014xba}, whose framework underlies the definition of $Q^{(\pm)}(p,y)$.
We choose to characterize the monodromy through equations (\ref{eq:main-result}), as opposed to giving a usual factorization into quantum dilogarithms, since the latter would inevitably obscure fundamental properties of $\mathbb{U}$. 
Nevertheless, we also provide an algorithmic method for solving these equations, which allows to compute refined BPS spectra efficiently from $Q^{(\pm)}(p,y)$, everywhere on the Coulomb branch.

\begin{figure}[h!]
\begin{center}
\includegraphics[width=0.180\textwidth]{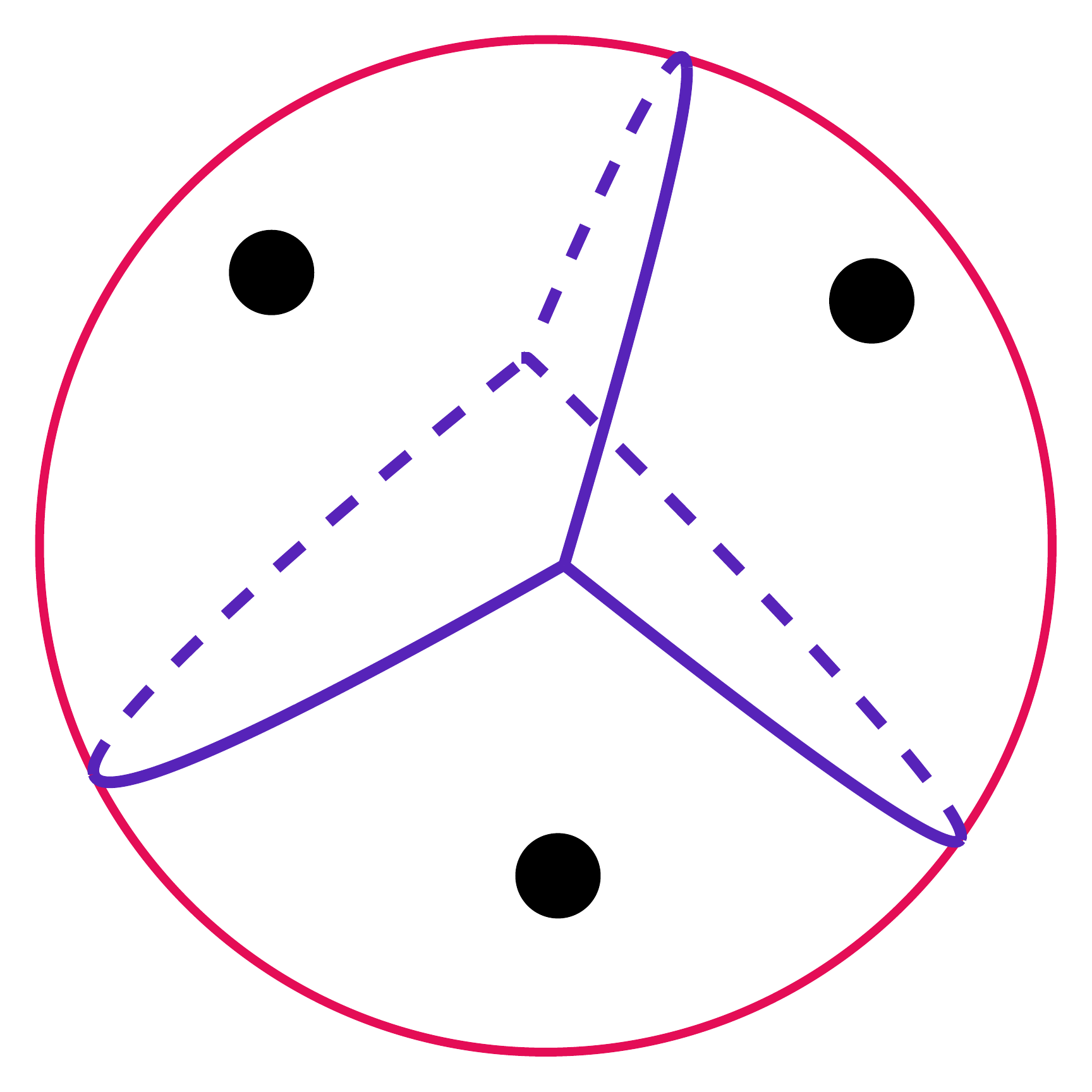}
\caption{Critical graph of the $T_2$ theory. The $S_3$ symmetry is generated by a cyclic permutation of the three edges (corresponding to a rotation of the sphere by $2\pi/3$ about the axis through the graph nodes), and by the transposition of two edges (a rotation of the sphere by $\pi$ about the axis going through the puncture between them).}
\label{fig:T2-intro}
\end{center}
\end{figure}

Besides the natural 2d-4d wall crossing interpretation, there is a simple way of thinking about (\ref{eq:main-result}), which becomes evident if we rewrite them as $Q^{(+)} = \mathbb{U} Q^{(-)} \mathbb{U}^{-1}$. These equations characterize a transformation $\mathbb{U}$ acting on certain formal variables $\hY_\gamma$ (in terms of which $Q^{(\pm)}(p,y)$  are formulated), by giving its action on a ``canonical basis'', namely the set of $Q^{(-)}(p,y)$ for each edge $p\in\CW_c$.
This language it suitable for stating an important caveat: the equations (\ref{eq:main-result}) truly characterize $\mathbb{U}$ only if it acts \emph{freely} on the formal variables.
In many interesting examples this condition is violated, the most extreme case being theories containing only gauge-neutral BPS states (charged under flavor symmetries), in which $\mathbb{U}$ simply commutes with $Q^{(\pm)}$.
We show how even in such cases $\mathbb{U}$ is still determined by $Q^{(\pm)}$, although not just through equations (\ref{eq:main-result}), but a by mild generalization whose details are given in the main body of the paper. 
An important point is that $Q^{(\pm)}$ are entirely determined by topological data of $\CW_c$, this means that symmetries of the graph must be manifest in the generating functions $Q^{(\pm)}$, and therefore appear manifestly as symmetries of $\mathbb{U}$.
While these symmetries are often hidden by the usual definition (\ref{eq:schematic-def}), the graph $\CW_c$ instead manifestly elucidates their origin.
A neat example is the critical graph of the $T_2$ theory shown in Figure \ref{fig:T2-intro}. 
In this case the graph has a manifest $S_3$ permutation symmetry, which nicely fits with the conjectural relation between $\mathbb{U}$ and specializations of the superconformal index: for example the Schur index is known to be a symmetric function of the three fugacities associated to the punctures \cite{Gaiotto:2012xa}.

Along with the construction of the BPS monodromy $\mathbb{U}$, we develop a parallel story for its ``{classical}'' counterpart $\mathbb{S}$ which controls wall crossing of 4d BPS indices. 
The latter is in fact a specialization of $\mathbb{U}$ to $y=1$, however there is one difference worth of mention. 
On the one hand, the generating functions $Q^{(\pm)}(p)$ for the classical monodromy are rather easy to compute by hand from $\CW_c$, thanks to a simplified formalism of spectral networks rules, which we develop and illustrate through several examples.
On the other hand, the $Q^{(\pm)}(p,y)$ are often better obtained with the aid of a computer, for this purpose we developed a software that computes them for arbitrary graphs, this is provided in \cite{python-code}.
On a practical level, having a method for constructing $\mathbb{S}$ exactly, significantly facilitates the computation of the non-commutative BPS monodromy $\mathbb{U}$.\\
A construction for $\mathbb{S}$ was previously developed in \cite{Gaiotto:2009hg}, based on ideal triangulations of $C$ for class $\CS$ theories of type $A_1$. %
For this particular sub-class of theories, our construction is certainly related to the one of \cite{Gaiotto:2009hg}, in fact the critical graph $\CW_c$ appears to be dual to ideal triangulations of $C$ in all the examples we considered. 
On the other hand, a refined version of $\mathbb{S}$ was constructed in \cite{Longhi:2012mj}, which captures the spectrum of 2d-4d BPS states for canonical surface defects of $A_1$ theories. 
In this case it is less clear if there is a connection to the present work, but it is natural to ask whether the construction of \cite{Longhi:2012mj} admits a generalization to higher rank theories which relies on critical graphs.

\medskip

The paper is organized as follows.
In Section \ref{sec:background} we collect some background on wall crossing, with an emphasis on the approach to studying BPS spectra by probing a theory with surface defects, pioneered by Gaiotto, Moore and Neitzke. A self-contained introduction to spectral networks is included.\\
Section \ref{sec:monodromy-construction} contains the main results.
We review the definitions of the critical locus $\CB_c$ and the critical graph $\CW_c$ following \cite{network-quiver}.
We then introduce the main physical idea behind our construction, how 2d-4d BPS states capture the 4d monodromy even on this locus of marginal stability.
After this we explain the construction of both classical and quantum monodromies from the graph. An algorithmic technique for solving the monodromy equations in factorized form (\ref{eq:schematic-def}) is also given.\\
In Section \ref{sec:examples} we illustrate our constructions with several examples, obtaining classical and quantum monodromies by deriving the monodromy equations and solving them explicitly.\\
We conclude in Section \ref{sec:coda} by highlighting open questions and potential applications raised by this paper, both of physical and mathematical nature.
In the Appendix we collect some conventions and several technical results, worth of mention is the formalism developed in Appendix \ref{sec:soliton-computations}, which may be of separate interest for handling complicated computations in generic spectral networks.

\section{Some background on wall crossing}\label{sec:background}
In this section we collect useful background on wall crossing in 4d $\CN=2$ gauge theories. 
We provide a self-contained review of the framework of spectral networks, with a particular emphasis on their physical interpretation in terms of framed 2d-4d wall crossing.
For the sake of brevity we shall deliberately avoid any technical details, clear expositions of these can be found both in the original paper \cite{Gaiotto:2012rg} and in \cite[\S2]{Galakhov:2013oja}.

\subsection{Four dimensional wall crossing}\label{sec:review-4d-wc}

Wall crossing in 4d $\CN=2$ gauge theories is the phenomenon by which the spectrum of BPS states is piecewise smooth across the Coulomb branch $\CB$. 
The spectrum changes discontinuously across real-codimension one loci known as \emph{wall of marginal stability}, along which the phases of central charges of two BPS states are equal
\be
	\MS(\gamma_1,\gamma_2) := \{u\in\CB \,,\ \arg Z_{\gamma_1}(u)=\arg Z_{\gamma_2}(u)\}\,.
\ee
Physically these jumps correspond to the appearance (or disappearance) of boundstates with charges $n_1 \gamma_1+n_2\gamma_2$, for certain positive integers $n_1,n_2$.
Changes in the spectrum are determined by the integer-valued electromagnetic pairing $\langle\gamma_1,\gamma_2\rangle$. 
Wall crossing formulae provide a precise description of the jumps in terms of protected indices for BPS states.
One such index is the \emph{Protected Spin Character} (PSC) introduced in \cite{Gaiotto:2010be}, which is defined as
\be\label{eq:PSC-def}
	\Omega(\gamma,u;y) := \Tr_{\mathfrak h_\gamma} y^{2J_3}(-y)^{2I_3} = \sum_{m\in\IZ} a_m(\gamma,u)\, (-y)^m\,.
\ee
$J_3, I_3$ are Cartan generators of $\mathfrak{so}(3)$ and $\mathfrak{su}(2)_R$, while $\mathfrak h_\gamma$ denotes the Clifford vacuum\footnote{More details about this notation can be found  in \cite{Galakhov:2013oja}.} of $\CH^{\rm BPS}_\gamma$, the sector of BPS states with charge $\gamma$. The second equality defines the integers $a_m(\gamma,u)$.
To express the wall crossing formula we introduce formal variables $\hY_\gamma$ with relations 
\be
	\hY_{\gamma}\hY_{\gamma'}=y^{\langle\gamma,\gamma'\rangle}\hY_{\gamma+\gamma'}\,,
\ee
together with the following ordered product of quantum dilogarithms (conventions for these are spelled out in Appendix \ref{sec:active-passive})
\be\label{eq:factorized-U}
	\mathbb{U} := \prod^{\nwarrow}_{\gamma\in\Gamma_+}\prod_{m\in\IZ} \Phi((-y)^m \, \hY_\gamma)^{a_m(\gamma,u)}\,.
\ee
$\mathbb{U}$ is the \emph{BPS monodromy}, also known as the ``motivic (or quantum) spectrum generator'' \cite{Dimofte:2009bv, Dimofte:2009tm, Gaiotto:2010be}, its definition manifestly depends on $u$ both through the ordering of factors, and through the PSC coefficients $a_m(\gamma,u)$.
The arrow $\nwarrow$ indicates that  factors are ordered with $\arg Z_\gamma(u)$ increasing to the left. $\Gamma_+$ denotes a choice of half-lattice within the lattice of gauge and flavor charges $\Gamma$, reflecting the distinction between particles and anti-particles corresponding to a choice of half plane on the complex plane of central charges.
\footnote{Two different choices are related by conjugation of $\mathbb{U}$ by quantum dilogarithms;  the wall crossing formula holds for any choice of half plane.}
The statement of  Kontsevich and Soibelman's  wall crossing formula is  that $\mathbb{U}$ is actually independent of $u$, implying that across MS walls both the phase ordering and the spectrum jump in ways that balance each other \cite{Kontsevich:2008fj}.

In the following we will actually work with a $\IZ$-extension of the lattice $\Gamma$, denoted by $\tGamma$. 
$\tGamma$ is endowed with a natural projection map $\tilde\pi_*:\tGamma\to\Gamma$, whose kernel has rank one, ${\rm{ker}} \,\tilde\pi_*\simeq\IZ$. 
There is a canonical choice of generator for the kernel (i.e. among the two choices differing by a sign), which we denote by $H$. The DSZ pairing $\langle\,,\,\rangle$ extends to $\tGamma$ by taking $H$ to be in the annihilator, i.e. $\langle\tilde\gamma,\tilde\gamma'\rangle_{\tGamma} \equiv \langle\gamma,\gamma'\rangle_{\Gamma}$ where $\gamma,\gamma'$ are the canonical projections.
We then introduce formal variables associated to $\tGamma$, subject to the following relations
\be\label{eq:extended-formal-variables-motivic}
	\hY_{\tilde\gamma}\hY_{\tilde\gamma'} = y^{\langle\tilde\gamma,\tilde\gamma'\rangle}\hY_{\tilde\gamma+\tilde\gamma'} \,, \qquad \hY_{\tilde\gamma + n H} = (-y)^{n} \hY_{\tilde\gamma}\,.
\ee
Thanks to the second relation, and to the existence of a canonical section $\plift:\Gamma\to\tGamma$, it is possible to trade back the $\hY_{\tilde\gamma}$ for the $\hY_\gamma$,
for details see Appendix \ref{sec:homology-conventions}.
Therefore a reader who is unfamiliar with the extended lattice $\tGamma$ may safely ignore all  $\ \tilde{}\ $: much of what follows could be rewritten using $\gamma$ instead of $\tilde\gamma$, at the cost of using explicitly the section $\plift$.
We will work with $\tGamma$ because it is more natural in our context.

There are two specializations of these formal variables that arise naturally, they fit in the following commutative diagram:
\be
\label{graph:variables-relations}
\begin{tikzpicture}
  \matrix (m) [matrix of math nodes,row sep=3em,column sep=4em,minimum width=2em]
  {
      & \hY_{\tilde\gamma} & \\
      X_{\tilde\gamma} &  & \hX_{\gamma} \\};
  \path[-stealth]
    (m-1-2) edge node [left] {$y\to 1$} (m-2-1)
    (m-1-2) edge node [right] {$y\to -1$} (m-2-3)
     (m-2-1) edge node [above] {$\rho(\tilde\gamma)$} (m-2-3)
     (m-2-3) edge node [above] {} (m-2-1);
\end{tikzpicture}
\ee
The $X_{\tilde\gamma}$ are precisely the variables employed in \cite{Gaiotto:2012rg}, they arise naturally in the context of spectral networks.
The $\hX_{\gamma}$ on the other hand are independent of $H$ (as can be seen by taking $y\to -1$ in (\ref{eq:extended-formal-variables-motivic})) therefore they really depend on $\gamma:=\tilde\pi_*(\tilde\gamma)\in \Gamma$.
Note that, while commutative, the $\hX$ are subject to the twisted product rule $\hX_\gamma\hX_{\gamma'} = (-1)^{\langle\gamma,\gamma'\rangle}\hX_{\gamma+\gamma'}$. These in fact coincide with variables employed in \cite{Gaiotto:2009hg}.
The two specializations are related to each other by a quadratic refinement (a sign $\rho(\tilde\gamma)$ which is canonically fixed by the section $\plift$), this will be sometimes employed to switch between $X$ and $\hX$ below. Further details can be found in Appendix \ref{sec:homology-conventions}.
The specialization to $y=-1$ is natural, since the PSC reduces to the integer-valued second helicity supertrace, or \emph{BPS index}
\be
	\Omega(\gamma,u;y) \mathop{\longrightarrow}^{y\to-1}_{} \Omega(\gamma,u) = \Tr_{\mathfrak{h}_{\gamma}}(-1)^{2J_3}\,.
\ee
In this limit the BPS monodromy has a counterpart known as the \emph{spectrum generator} $\mathbb{S}$ \cite{Gaiotto:2009hg, Longhi:2012mj}. The spectrum generator is an ordered product of transformations weighted by BPS indices
\be
	\IS  = \prod^{\searrow}_{\gamma\in\Gamma_+} \CK_\gamma^{\Omega(\gamma,u)}\,,
\ee
with
\be
	\CK_\gamma^{\Omega} \,: \hX_{\gamma'} \,\to \hX_{\gamma'} (1-\hX_\gamma)^{\langle\gamma,\gamma'\rangle\,\Omega}\,.
\ee
The transformation $\CK_\gamma$ captures the $y\to-1$ limit of the conjugation of $\hY_{\tilde\gamma'}$ by quantum dilogarithms, as we review in Appendix \ref{sec:active-passive}.

\subsection{Surface defects and 2d-4d wall crossing}\label{sec:defects-review}

In presence of BPS surface defects, the BPS spectrum of a 4d $\CN=2$ theory is enhanced by a new kind of BPS states, known as \emph{2d-4d BPS states} \cite{Gaiotto:2011tf}.
There is a rich interplay between 4d \emph{``vanilla"} BPS states and 2d-4d BPS states, the essence of which is the \emph{2d-4d wall crossing phenomenon}. 
The physics of 2d-4d states underlies the physical interpretation of spectral networks, 
and certain aspects of it will be especially important for explaining our construction of the BPS monodromy.
In this section we give a concise but self-contained review of this story, more details can be found in \cite{Klein, Gaiotto:2011tf, Longhi-Park-2d}.

We consider surface defects defined as 2d-4d systems, consisting of a 2d $\CN=(2,2)$ quantum field theory living on $\IR^{1,1}\subset\IR^{1,3}$ (parameterized by $x^0, x^3$) coupled to the 4d $\CN=2$ theory in the bulk.
Degrees of freedom on the defect include chiral fields transforming under a flavor symmetry $G$ which is gauged by the 4d vector multiplets, whose restriction to the defect couples to the 2d chirals.
Due to $\CN=2$ supersymmetry of the bulk theory, the scalars in the vectormultiplet couple to the 2d chirals in the guise of \emph{twisted masses} \cite{Hanany:1997vm}.
On the 4d Coulomb branch the vevs of 4d vectormultiplets lift the 2d chirals, leaving behind no massless two-dimensional degrees of freedom.
At fixed $u$ the 2d theory has a finite number of massive vacua, controlled by contributions of $u$ to the effective twisted superpotential $\widetilde\CW(u)$. 
A vacuum of the 2d-4d system is characterized both by discrete 2d vacua and continuous 4d moduli, the global structure of this space of vacua is quite rich. 
2d-4d vacua are fibered nontrivially over the 4d Coulomb branch, the analytic structure of $\widetilde\CW(u)$ can be studied using resolvent techniques as in \cite{Gaiotto:2013sma}, where it was shown that the chiral ring equations resemble the Seiberg-Witten curve of the 4d theory.\footnote{More precisely, this reference considered 2d chiral matter coupled both to 2d vector multiplets and 4d vector multiplets, with complexified Fayet-Ilioupolous terms turned on for the former.}

Theories of class $\CS$ admit a \emph{canonical} type of surface defects. Every such theory is characterized by a triplet of data: an $ADE$-type  Lie algebra $\mathfrak{g}$, a punctured Riemann surface $C$, and puncture data $D$. 
The latter fix the boundary conditions for worldvolume fields of the 6d (2,0) theory compactified on $C$ with a partial topological twist \cite{Gaiotto:2009we, Gaiotto:2009hg}.
The UV definition of the canonical surface defect $\mathbf{S}_z$ includes a complex coupling parameterized by $z\in C$ and a choice of representation for $\mathfrak{g}$ \cite{Gaiotto:2009fs, Alday:2009fs, Hori:2013ewa, Longhi-Park-2d, Gadde:2013ftv}, in the rest of this paper we will restrict to the first fundamental representation for simplicity, but will let $\fg$ be of generic $ADE$ type.
The Seiberg-Witten curve of a class $\CS$ theory is identified with the spectral curve of a Hitchin system associated to the data $(\mathfrak{g},C,D)$.
In particular, the spectral curve is presented as a ramified covering  $\pi: \Sigma_u \to C$, and
the discrete set of 2d vacua corresponds to the preimage $\pi^{-1}(z) = \{z_i\}_i \subset\Sigma_u$ \cite{Gaiotto:2009fs, Gaiotto:2011tf, Longhi-Park-2d}. 
The global structure of the space of 2d-4d vacua is therefore captured by a beautifully rich geometric picture, involving both monodromies of the fiber $\pi^{-1}(z)\subset\Sigma_u$ over $C$, together with those of $\Sigma_u$ over $\CB$.

2d-4d BPS states are supersymmetric field configurations of 2d and 4d fields, interpolating between vacuum $(u,z_i)$ at $x^3=-\infty$ and vacuum $(u,z_j)$ at $x^3=+\infty$.
Topological charges of 2d-4d states are classified by relative homology classes $\Gamma_{ij}(z) := H_1^{\rm rel}(\Sigma_u, (z_i,z_j),\IZ)$ of oriented paths on $\Sigma_u$ beginning at $z_i$ and terminating at $z_j$.%
\footnote{
Topological charges of 2d-4d states are classified by a pair $ij$ of vacua, but also include contributions from the 2d flavor symmetry $G$, which on the 4d Coulomb branch is broken to a Cartan torus. The 2d flavor charges are identified with the 4d gauge and flavor charges, classified by the lattice $\IZ^{2r+f} \sim H_1(\Sigma_u,\IZ)$. The overall space of 2d-4d charges is therefore a torsor for $H_1(\Sigma_u,\IZ)$ \cite{Gaiotto:2011tf}.}
The supersymmetry central charge of a 2d-4d state with charge $a\in\Gamma_{ij}(z)$ is 
\be\label{eq:2d-4d-central-charge}
	Z_{a} = \frac{1}{\pi}\int_{a} \lambda_{}
\ee
where $\lambda_{}$ is the Seiberg-Witten differential on $\Sigma_u$.\footnote{This relation between the topological charge $a$ of 2d-4d states and their central charge $Z_a$ is a generalization of the analogous one for 4d BPS states \cite{Witten:1978mh, Seiberg:1994rs}.}

Given two surface defects $\mathbf{S}_{z_-}, \mathbf{S}_{z_+}$ both supported at $x_1=x_2=0$, a UV supersymmetric interface interpolating between them is a domain wall junction along which the coupling varies from $z_-$ for $x^3\to-\infty$ to $z_+$ for $x^3\to+\infty$.
The profile $z(x^3)$ is classified by relative homotopy classes of paths $\wp\subset C$ from $z_-$ to $z_+$. 
A supersymmetric interface is additionally parameterized by an angle $\vartheta \in \IR/2\pi\IZ$, which labels a linear combination of supercharges it preserves \cite{Gaiotto:2011tf}.
In presence of a supersymmetric interface $L_{\wp,\vartheta}$, vacua of the theory are determined by \emph{two} points $z_{-,i},z_{+,j}\in \Sigma_u$ together with the 4d Coulomb moduli $u$.
The spectrum of BPS states gets further enhanced by the presence of \emph{framed 2d-4d states}: these are supersymmetric field configurations interpolating between vacua $z_{-,i}$ and $z_{+,j}$, their charges classified by relative homology classes in $\Gamma_{ij}(z_{-},z_{+}) := H_1^{\rm rel}(\Sigma_u, (z_{-,i},z_{+,j}),\IZ)$. The central charges of framed BPS states are once again given by equation (\ref{eq:2d-4d-central-charge}), with $a\in \Gamma_{ij}(z_{-},z_{+}) $.

Framed BPS states are good ``probes'' of both 2d-4d and vanilla BPS states. 
In particular, the spectrum of framed 2d-4d states depends on $\vartheta$, in addition to $z$ and $u$.
The dependence on $\vartheta$ is piecewise-continuous, with jumps occurring at the phases of central charges of both 2d-4d and vanilla BPS states (this is the 2d-4d framed wall crossing phenomenon of \cite{Gaiotto:2011tf}).
Jumps of the 2d-4d famed spectrum are conveniently described in terms of a generating function 
\be\label{eq:framed-2d-4d-generating-function}
	F(\wp,\vartheta,u) := \sum_{i,j}\sum_{a
	} \FOmega(L_{\wp,\vartheta}, u, a) \, X_a\,,
\ee
where we momentrarily introduced the formal variables $X_a$ labeled by charges of framed 2d-4d states, a more precise definition will appear in the next section. %
The $\FOmega(L_{\wp,\vartheta}, u, a)$ are degeneracies counting framed 2d-4d states, see e.g. \cite[\S4]{Gaiotto:2012rg}.
As already mentioned, $F(\wp,\vartheta,u)$ is discontinuous in $\vartheta$, across phases of central charges of both 2d-4d and vanilla BPS states, the precise form of these jumps will be reviewed in the next section. 
However the key point is that knowledge of $F(\wp,\vartheta,u)$ for different $(\wp,\vartheta)$ allows to reconstruct the entire 2d-4d and vanilla BPS spectra, through the study of its jumps. 
Spectral networks provide both a way to compute $F(\wp,\vartheta,u)$, and to study these jumps efficiently.

\subsection{Spectral networks}

Spectral networks are combinatorial objects associated to ramified coverings of Riemann surfaces  introduced in \cite{Gaiotto:2012rg}.
Given a ramified covering $\pi:\Sigma\to C$, the data of a network $\CW(\vartheta,u)$ consists of two main parts: 
geometric data of real-dimension one trajectories on $C$ known as ``{$\CS$-walls}'', and ``soliton data'' of (relative) homology classes of paths on $\Sigma$ attached to each $\CS$-wall.
An example of spectral network is given below in Figure \ref{fig:AD3-strong-coupling}.

The geometry of {$\CS$-walls} is determined by the projection map $\pi$ and by a phase $\vartheta$.
$\CS$-walls may originate either from branch points of $\pi$ or from intersections of other $\CS$-walls. 
Upon choosing a trivialization for $\pi$, an $\CS$-wall originating from a branch point carries the label $ij$, corresponding to two sheets of $\Sigma$ which meet at the branch point.\footnote{Here it is assumed that branch points are of square-root type, therefore sheets of $\Sigma$ can only meet in pairs in correspondence of each branch point.
Spectral networks can be defined also when higher-degree branch points are present, but for simplicity we assume branch points to be of square-root type.}
More precisely, the shape of each $\CS$-wall is determined by the constraint that the difference $\lambda_j-\lambda_i$, between values of the Seiberg-Witten differential evaluated at the two sheets $i,j$ above $z\in C$, has a fixed phase along the wall, which is determined by $\vartheta$. 
This geometric constraint determines the shapes of $\CS$-walls, and descends naturally from the BPS equations for 2d-4d solitons \cite{Gaiotto:2011tf, Longhi-Park-2d}.
The soliton data carried by each $\CS$-wall is a set of pairs $(a,\mu(a))$. The former are relative homology classes of open paths on $\tSigma$, the circle bundle $\tilde\pi:\tSigma\to\Sigma$
\be
	a\in\tGamma_{ij}(z)\,,\qquad \tGamma_{ij}(z):=H_1^{\text{rel}}(\tSigma,(\tilde z_i, \tilde z_j),\IZ)\,,
\ee
where $\tilde z_i, \tilde z_j$ are lifts of a generic point $z\in C$ on the wall's trajectory to sheets $i,j$ of $\tSigma$.\footnote{Labels $i$ and $j$ correspond to the type of $\CS$-wall, as previously described. The position of $\tilde z_{i}, \tilde z_{j}$ in the $S^1$ fiber of $\tSigma$ is fixed by the tangent direction of the $\CS$-wall: parallel to the $\CS$-wall's orientation for $\tilde z_j$ and opposite for $\tilde z_i$}
The $\mu(a)$ are integers, their values are fixed by the global topology of the network according to a set of rules, see for example \cite{Galakhov:2013oja}. 
Spectral networks admit a precise physical interpretation: the soliton data of an $\CS$-wall of type $ij$ passing through $z$ counts  2d-4d BPS states in presence of a canonical surface defect $\mathbf{S}_z$, whose central charges $Z_a$ have phase $\vartheta$.

Framed 2d-4d BPS states have geometric counterparts too, and their generating function $F(\wp,\vartheta,u)$ for a supersymmetric interface  $L_{\wp,\vartheta}$ can be computed from the network data. 
Given a path $\wp$ on $C$, $F(\wp,\vartheta,u)$ is computed by considering all intersections of $\wp$ with walls of the spectral network $\CW(\vartheta,u)$, and constructing all ``detours'' obtained by composing the lifts of $\wp$ to $\Sigma$ with paths $a$ belonging to soliton data of the intersected walls. 
This construction of $F$ correctly reproduces its behavior under the framed 2d-4d wall-crossing phenomenon, allowing to compute the degeneracies $\FOmega(L_{\wp,\vartheta}, u, a)$ from a combination of soliton data of $\CS$-walls and intersections of $\wp\cap\CW(\vartheta,u)$.
Jumps of $F$ occur in fact when the intersections $|\wp\cap\CW(\vartheta,u)|$ jump, i.e. when $z_-$ or $z_+$ are crossed by some $\CS$-wall at phase $\vartheta$. 
Physically this corresponds to the presence of 2d-4d BPS states supported on $\mathbb{S}_{z_-}$ or $\mathbb{S}_{z_+}$ respectively.
A second, but not less important, manifestation of framed 2d-4d wall crossing is visible in spectral networks: for certain phases $\vartheta$ the topology of the network $\CW(\vartheta,u)$ jumps discontinuously, causing a jump of the soliton data of $\CS$-walls, which in turn determines $F(\wp,\vartheta,u)$. 
These occurrences were called $\CK$-wall jumps  in \cite{Gaiotto:2012rg}, an example is shown in Figure \ref{fig:AD3-strong-coupling}. 

\begin{figure}[h!]
\begin{subfigure}{0.3\textwidth}
        \includegraphics[width=\textwidth]{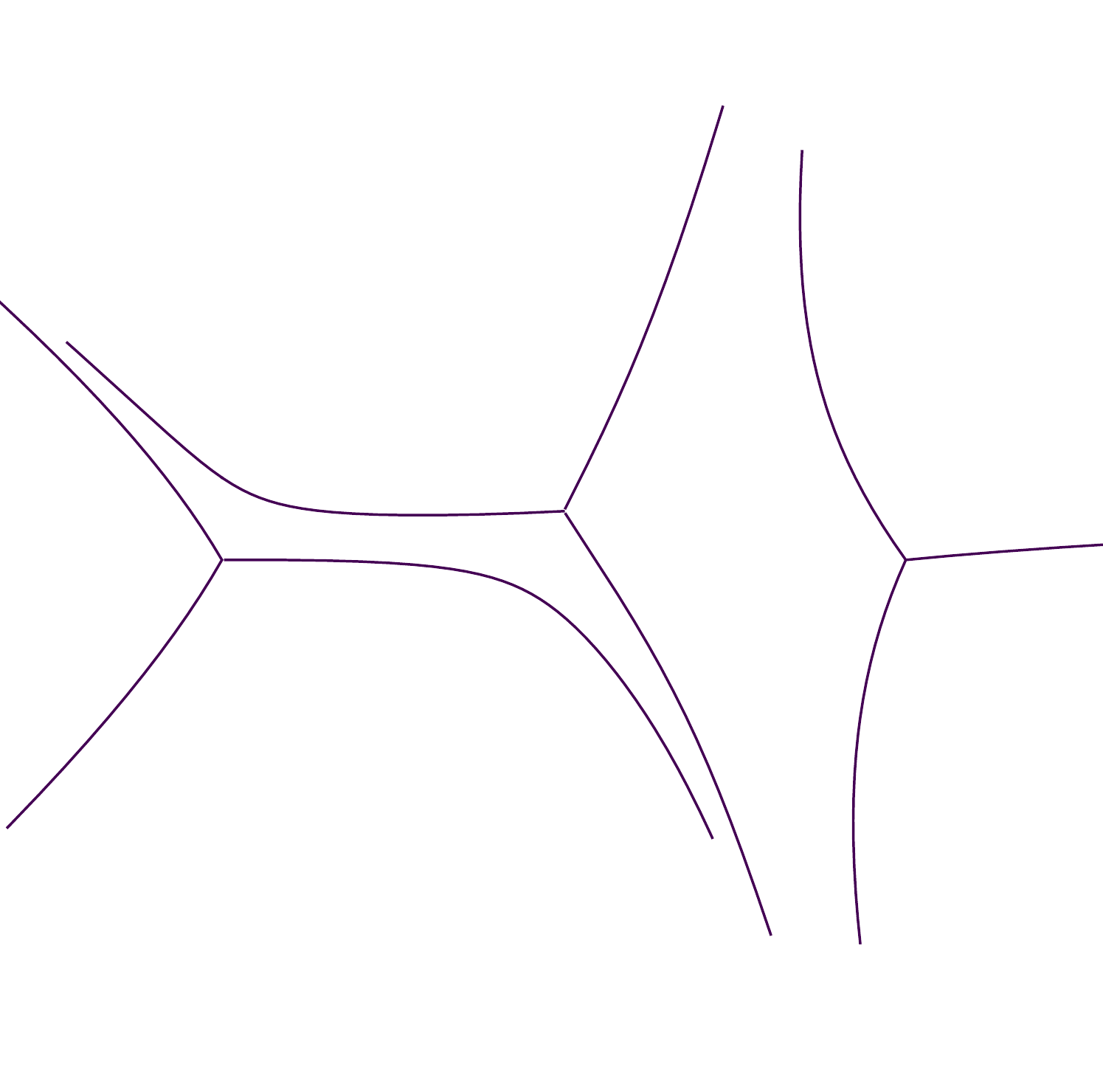}
\end{subfigure}
\hfill
\begin{subfigure}{0.3\textwidth}
        \includegraphics[width=\textwidth]{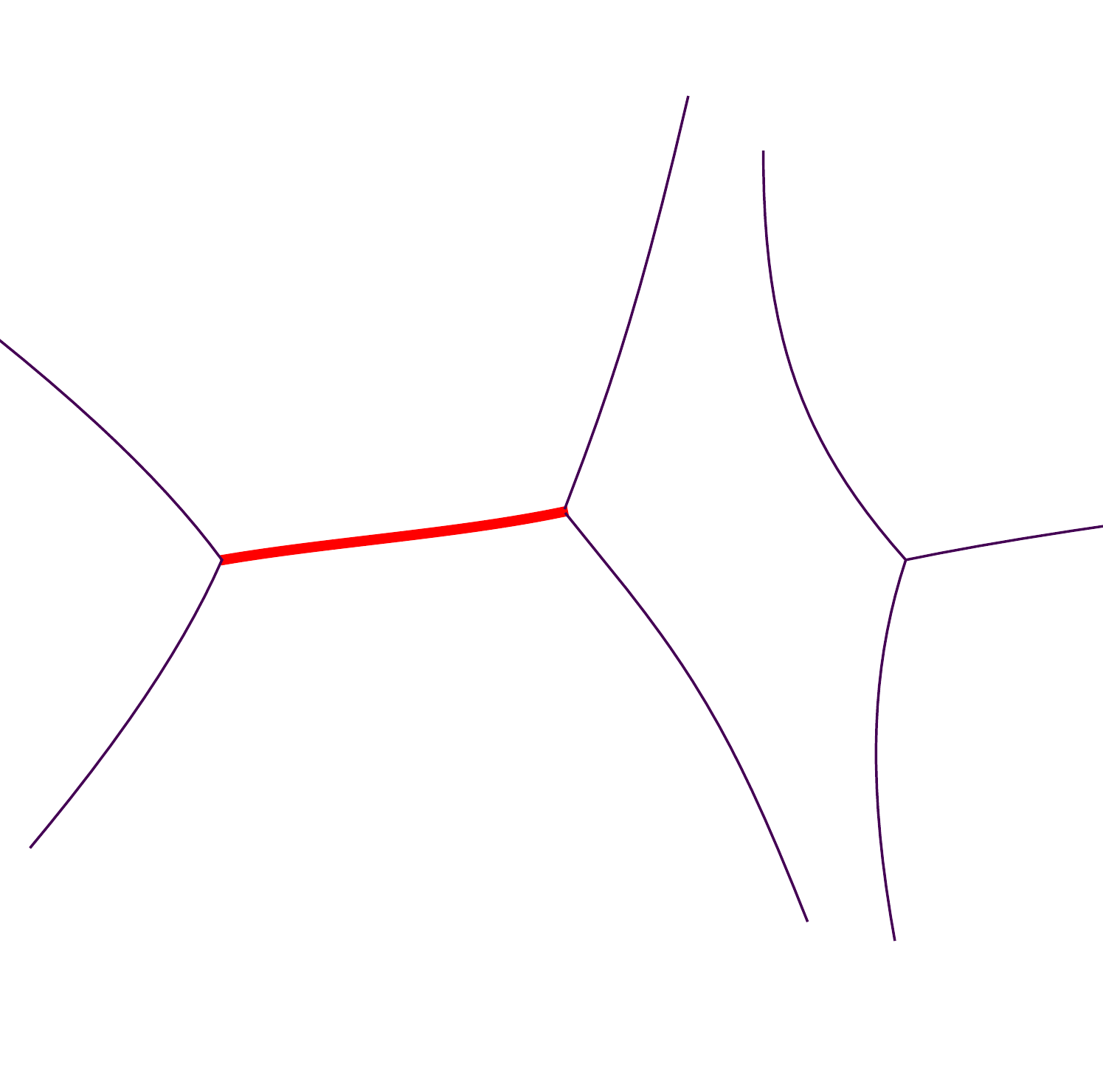}
\end{subfigure}
\hfill
\begin{subfigure}{0.3\textwidth}
        \includegraphics[width=\textwidth]{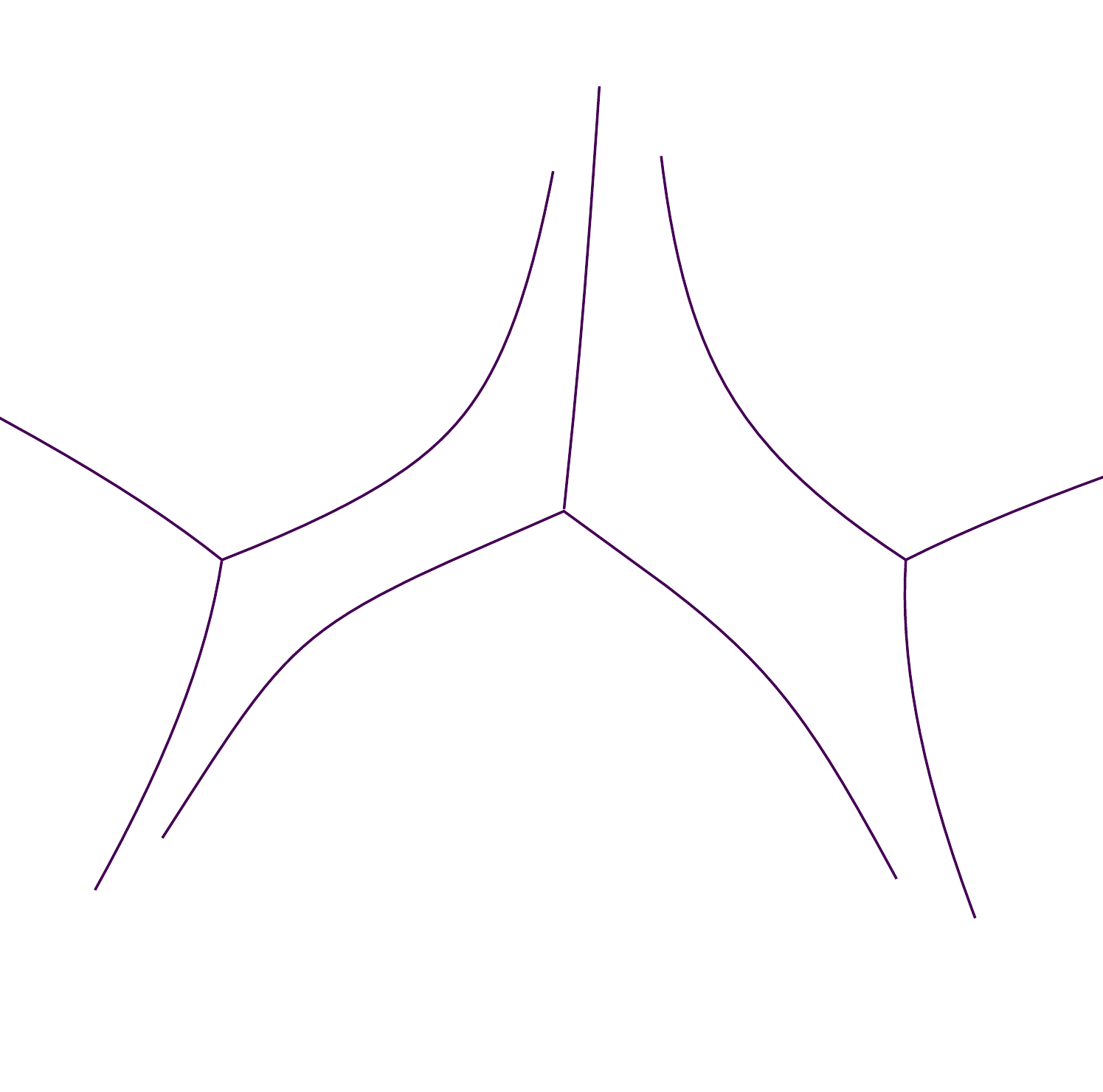}
\end{subfigure}\\
\begin{subfigure}{0.3\textwidth}
        \includegraphics[width=\textwidth]{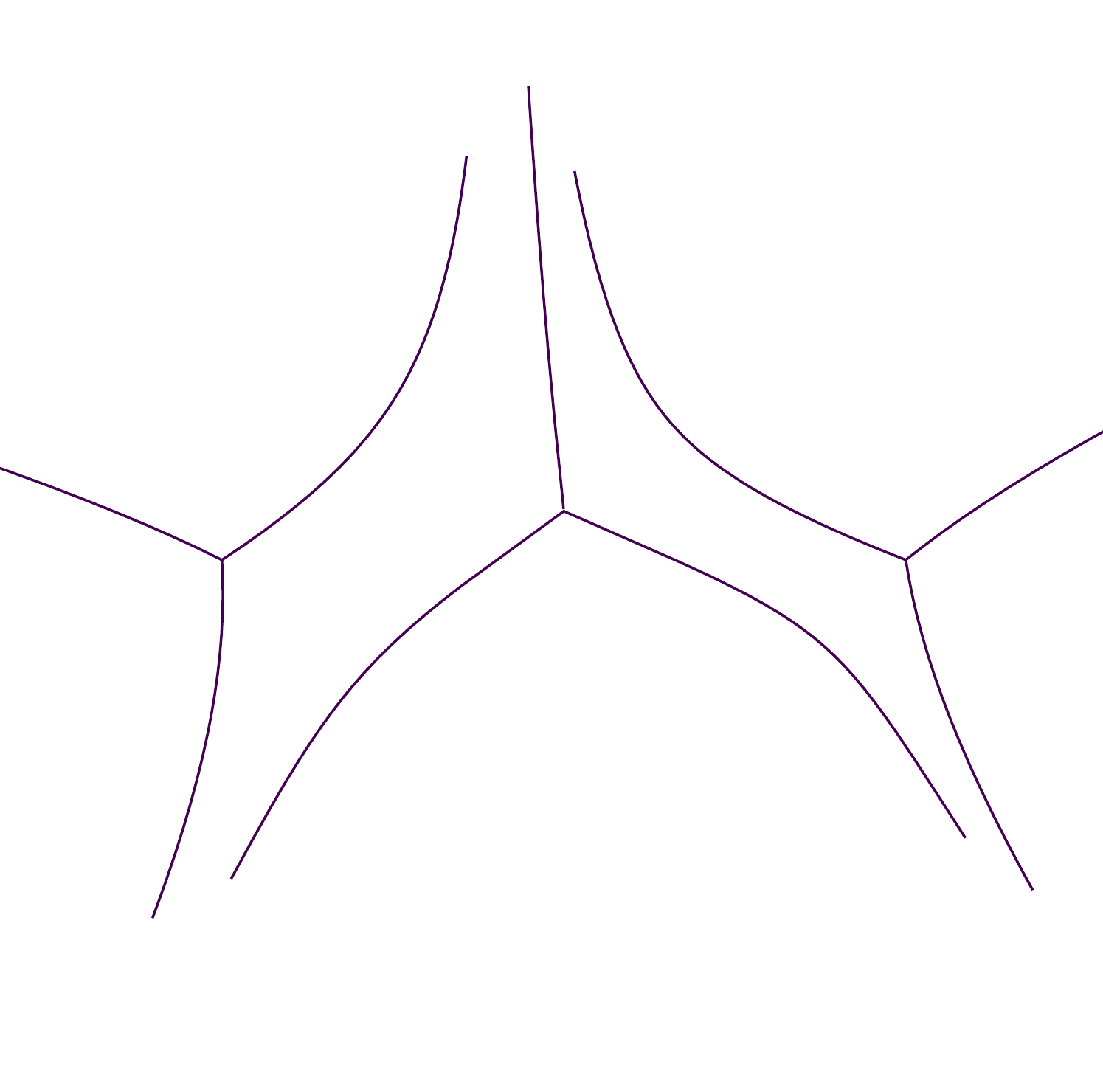}
\end{subfigure}
\hfill
\begin{subfigure}{0.3\textwidth}
        \includegraphics[width=\textwidth]{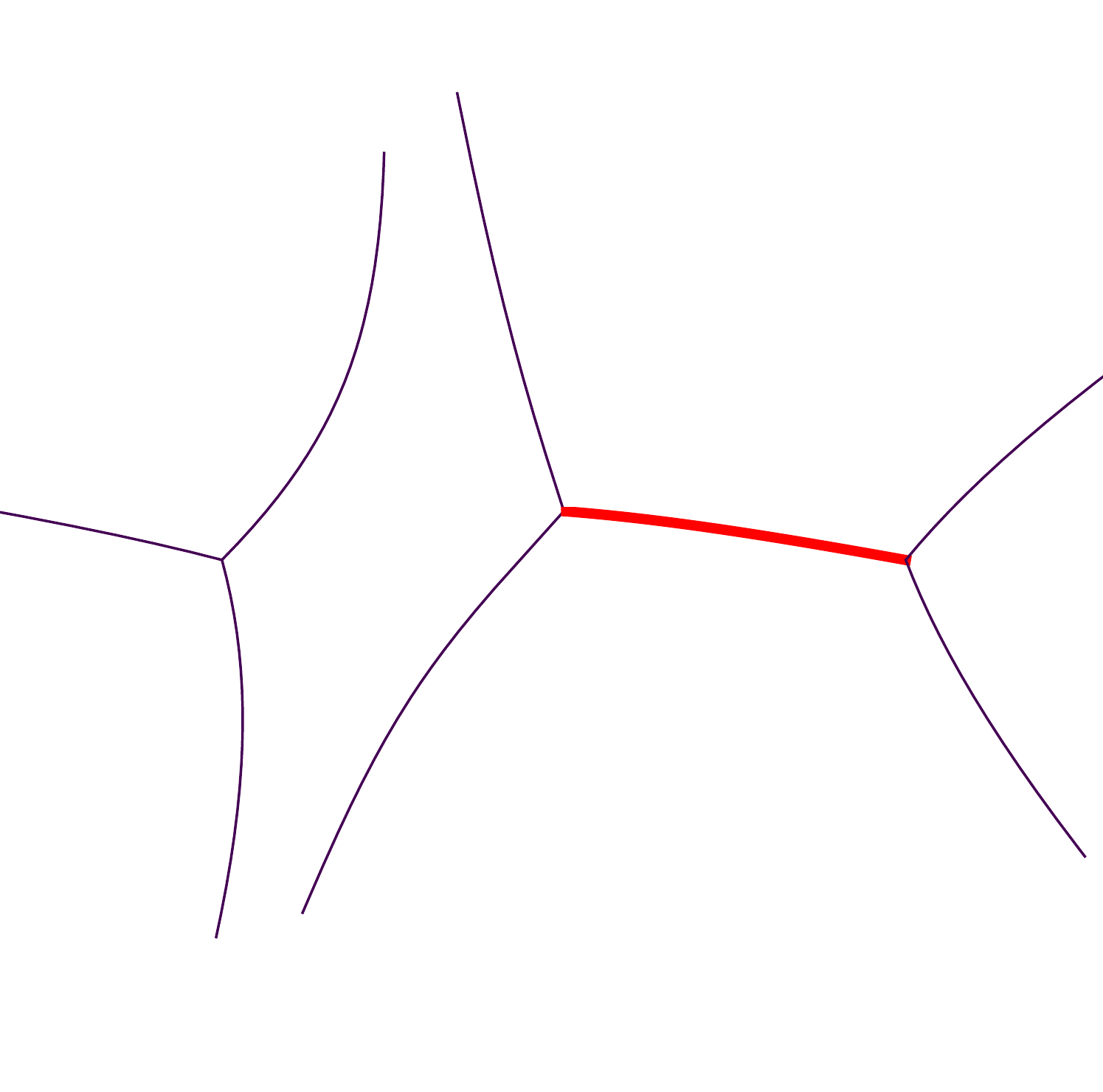}
\end{subfigure}
\hfill
\begin{subfigure}{0.3\textwidth}
        \includegraphics[width=\textwidth]{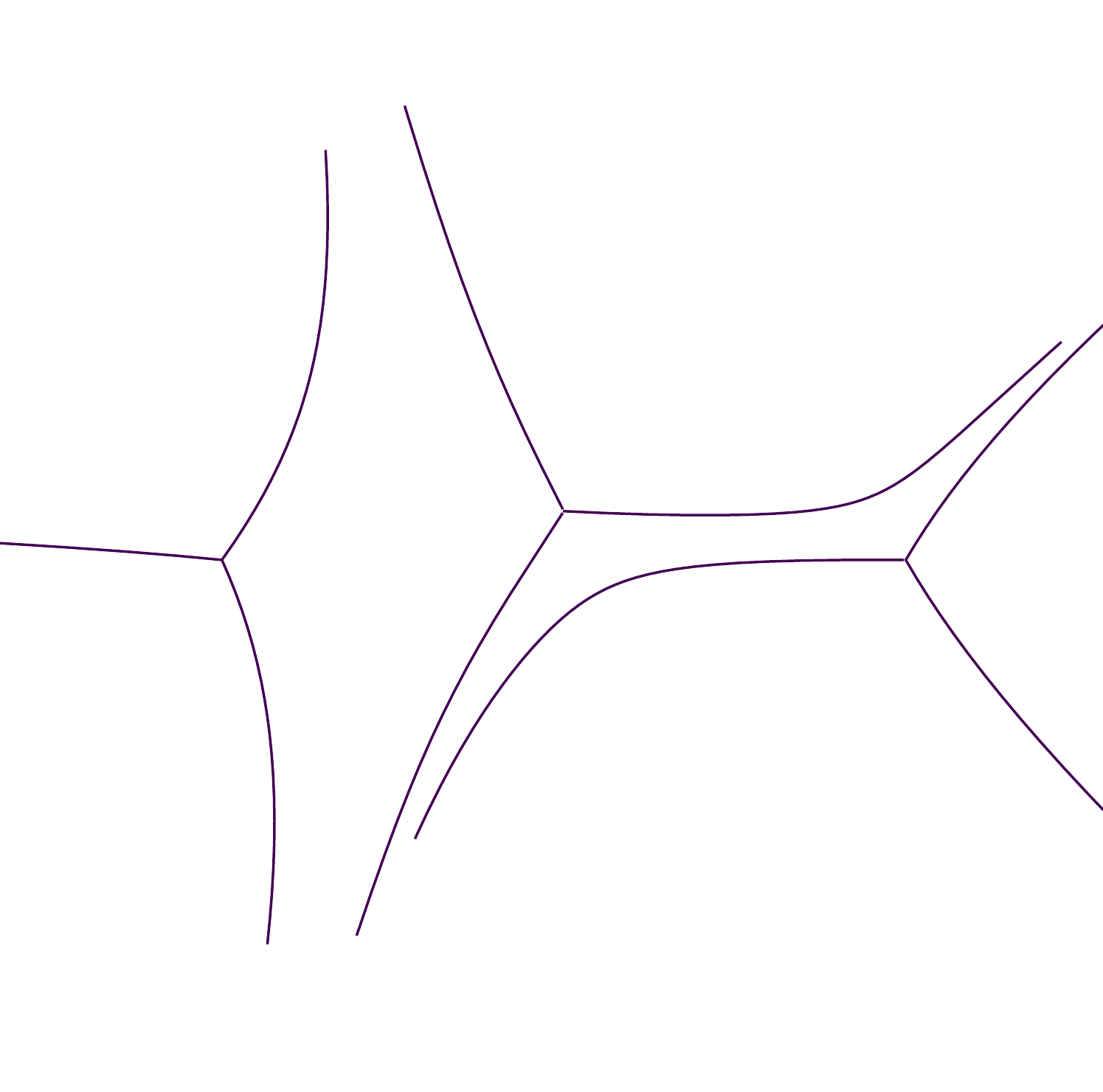}
\end{subfigure}
\caption{Spectral network for the Argyres-Douglas theory $AD_3$, in nomenclature from \cite{Gaiotto:2009hg}, at six distinct phases. $u$ is chosen in the strong coupling chamber, while the phase $\vartheta$ increases from the top left to the bottom right. Two $\CK$-wall jumps can be seen: one in the second frame and another in the fifth frame, they correspond to the two vanilla BPS states at strong coupling. 2-way streets are marked in red. 
}
\label{fig:AD3-strong-coupling}
\end{figure}

As mentioned at the end of the previous section, the jumps of $F(\wp,\vartheta,u)$ encode the BPS degeneracies of both vanilla and 2d-4d BPS states. 
Since we are mainly interested in vanilla degeneracies we will focus on jumps of $\CK$-wall type.
These occur at particular values of $\vartheta$, at which two or more $\CS$-walls become parallel to each other with opposite orientations, forming ``2-way streets". Examples of 2-way streets are  shown in red in Figure \ref{fig:AD3-strong-coupling}.
2-way streets are important because their soliton data alone contains enough information to compute the BPS indices of  vanilla BPS states $\Omega(\gamma,u)$  \cite{Gaiotto:2012rg}.
For each 2-way street $p$ one considers all concatenations of 2d-4d soliton charges supported on the $\CS$-walls that make up $p$. 
Let $ij/ji$ be the types of underlying $\CS$-walls, then one construct the generating series
\be\label{eq:Q-def-2-way}
	Q(p) = 1 + \sum_{a\in\tGamma_{ij}(z)}\sum_{b\in\tGamma_{ji}(z)} \mu(a) \mu(b) X_a X_b\,,
\ee
where we introduced $X_a$, an extension of formal variables $X_{\tilde\gamma}$ to the homology path algebra on $\tSigma$. 
These obey $X_aX_b = X_{a b}$ if $a$ and $b$ can be concatenated, $X_a X_b =0$ otherwise. In addition they have a well-defined product with the formerly introduced $X_{\tilde\gamma}$ as $X_{\tilde\gamma} X_a = X_a X_{\tilde \gamma} = X_{a+\tilde \gamma}$. More details on these variables can be found in \cite{Gaiotto:2012rg, Galakhov:2013oja}.\\
Each pair of $a$ and $b$ appearing in (\ref{eq:Q-def-2-way}) can actually be concatenated into \emph{closed} cycles within a specific homology class $ab \in H_1(\tSigma,\IZ)$.
In fact, it was argued in \cite{Gaiotto:2012rg} that $Q(p)$ admits a unique factorization of the form
\be
	Q(p) = \prod_{n\geq 1} (1-X_{n \tilde\gamma})^{\alpha_{n \gamma}(p)}
\ee
for a suitable\footnote{More precisely, $\tilde\gamma$ was defined as the \emph{preferred lift} of $\gamma=\tilde\pi_*(\tilde \gamma)$ in \cite[\S 6.3]{Gaiotto:2012rg}. In the language of appendix \ref{sec:homology-conventions} this would correspond to taking $\tilde\gamma = \plift(\gamma)$.} $\tilde\gamma\in \tGamma$, whereas $\gamma$ denotes its projection by the pushforward $\tilde\pi_*$. 
The charge $\tilde \gamma$ generates a one-dimensional sub-lattice which contains all concatenations of the form $ab$, simultaneously for \emph{all} two-way streets $p$ appearing at the same phase $\vartheta$ (in fact this phase coincides with $\arg Z_{\gamma}$, where $\gamma=\tilde\pi_*(\tilde\gamma)$).
The coefficients $\alpha_{\gamma}(p)$ determine a closed path on $\Sigma$, defined as the formal sum over all two-way streets 
\be\label{eq:GMN-L}
	L(n\gamma) = \sum_p \alpha_{n\gamma}(p)\, p_\Sigma
\ee 
of canonical lifts $p_\Sigma$ to sheets $i$ and $j$ of $\Sigma$ ($ij/ji$ being the types of a two-way street $p$, which also fix the orientation of the lifted segments \cite{Gaiotto:2012rg}). 
The homology class of $L(n\gamma)$ is furthermore an integer multiple of $n\gamma$, and the BPS index is the multiplicity constant 
\be\label{eq:BPS-index-networks}
	\Omega(n\gamma,u) = [L(n\gamma)] \, /\, n\gamma\,.
\ee
In view of the constructions that will be developed later in this paper, it is important to stress that the existence of a one-dimensional sub-lattice of $\tGamma$ which contains all concatenations $ab$ is guaranteed in this case by assuming that $u$ is \emph{generic}, i.e. away from walls of marginal stability. 
Of course, this is also a necessary requirement for making sense of $\Omega(\gamma,u)$, which would be ill-defined on MS walls.

To conclude our brief review on spectral networks, let us mention an extension of this framework that captures the spin of BPS states.  
The main ideas of this extension will play an important role in this paper.
In \cite{Galakhov:2014xba} it was observed that spectral networks contain much more information about 2d-4d soliton spectra than formerly utilized. 
In particular, the geometry of $\CS$-walls contains information about the \emph{actual} soliton paths, as opposed to the somewhat coarse classification by homology,
therefore a refined classification of 2d-4d charges based on \emph{restricted regular homotopy} was  introduced. 
Thanks to the refined classification, a new invariant known as \emph{writhe} was associated to these equivalence classes, counting self-intersections of a path with signs.
It was then proposed that the writhe of detour paths, which appear in the generating function $F(\wp,\vartheta,u)$ (see (\ref{eq:framed-2d-4d-generating-function})) should be identified with (twice) the $J_3$-eigenvalue of the corresponding framed 2d-4d states.
This identification allowed to compute \emph{refined} framed 2d-4d degeneracies from spectral networks, distinguishing states with different spin.
Finally, Protected Spin Characters of vanilla BPS states can be computed by studying the jumps of these refined framed 2d-4d degeneracies.
In the following we shall follow this line of reasoning, by computing the whole BPS monodromy from jumps of refined degeneracies, which in turn will be obtained from spectral networks through the framework of \cite{Galakhov:2014xba}.

\section{BPS monodromy from marginal stability}\label{sec:monodromy-construction}

\subsection{A special locus on the Coulomb branch}\label{sec:special-locus}

Bare masses, and other UV parameters of a theory, are useful handles on the geometry of Coulomb branches. 
In particular, sensitivity of the geometry to masses is reflected in the relative positions of singularities, which usually signal the presence of charged massless degrees of freedom.
An effective application of this observation is the practice of engineering 4d $\CN=2$ SCFTs by tuning masses in such a way to induce collision of mutually non-local singularities \cite{Seiberg:1994aj, Argyres:1995xn, Minahan:1996fg, Minahan:1996cj, Argyres:2015ffa}.
Following \cite{network-quiver}, in this paper we shall tune these parameters to achieve a different goal. 
We wish to restrict our attention to Coulomb branches with a specific property: that there exists a locus $\CB_c$ where all central charges have degenerate phases
\be\label{eq:B-c-def}
	\CB_c := \{ u \in\CB \, | \ \arg Z_\gamma \in\{ \vartheta_c, \vartheta_c+\pi\}   \,,\ \forall \gamma\}\,
\ee
for some $u$-dependent phase $\vartheta_c$.
Closely related loci played a key role in the construction of Fenchel-Nielsen coordinates through spectral networks in \cite{Hollands:2013qza}.

Leaving $\vartheta_c$ arbitrary, the condition in (\ref{eq:B-c-def}) amounts to imposing $2r+f-1$ real constraints ($2r$ is the rank of the IR gauge charge lattice, $f$ is the rank of the flavor charge lattice). 
These equations have generally no solution on $\CB$, which has real dimension $2r$.
However including masses as tunable parameters, we can formulate the constraints on the enlarged parameter space $\widetilde \CB$ of real dimension $2r+2f$. 
(We could enlarge the parameter space by including more UV moduli, such as gauge couplings, but we will leave them out for simplicity).
We shall denote by $\CB_m\subset\widetilde\CB$ the slice corresponding to a Coulomb branch for fixed values of masses, the critical locus within $\CB_m$ will be denoted by $\CB_{c,m}$.
In $\widetilde \CB$ our condition identifies an $f+1$-dimensional submanifold $\widetilde \CB_c \subset \widetilde \CB$.
Then $\CB_{c,m} \equiv \CB_m \cap \widetilde \CB_c$ is intersection of $\widetilde\CB_c$ with the generic Coulomb slice at fixed masses $\CB_m$. 
This intersection is not generic, because half of mass parameters (the phases) have been fixed both in defining $\CB_m$ and in defining $\widetilde\CB_c$. 
Taking this into account the generic dimension of the intersection between $\widetilde\CB_c \cap \CB_m$ is one, when it exists. 
In fact, if all masses are set to zero (in asymptotically free theories), we just require that $2r$ gauge central charges all have the same phase (which is left as a free parameter), this generically gives a one dimensional locus $\CB_{m,c}\subset\CB_{m=0}$.
Note that $\CB_{m,c}$ lies at the intersection of (at least) $2r+f-1$  walls of marginal stability, this locus has dimension one on the massless slice $\CB_{m=0}$. 
In rank 1 theories (i.e. theories with a Coulomb branch of complex-dimension 1) MS walls have real dimension one, therefore they overlap entirely on the locus $\CB_{m,c}$; moreover since MS walls generally pass through singularities on $\CB_m$, so does $\CB_{m,c}$.
If some masses are nonzero, but all have the same phase $\vartheta_c$, we must instead require that $2r$ central charges have the phase $\vartheta_c$, which generically gives a zero-dimensional locus on $\CB_m$. When masses have different phases we can expect no solution in general.

\begin{figure}[h!]
\begin{center}
\includegraphics[width=.4\textwidth]{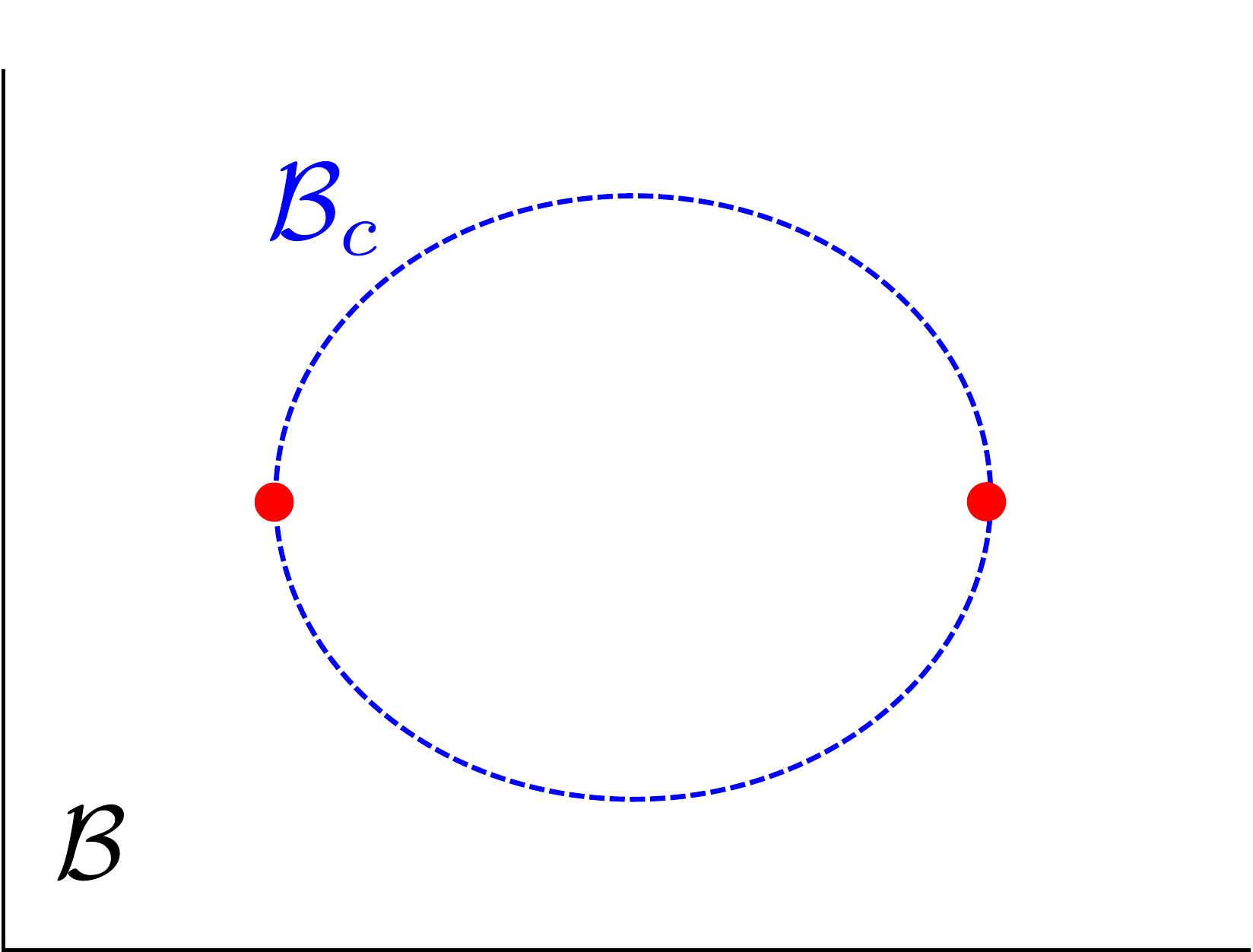}
\caption{A schematic picture of the Coulomb branch of SU(2) SYM theory, or of the original Argyres-Douglas $AD_3$ theory. The two singularities are marked in red, the wall of marginal stability is the  dashed blue line.}
\label{fig:flavorless}
\end{center}
\end{figure}

The simplest examples of such a critical locus can be found in rank one gauge theories with trivial flavor symmetry, such as SU(2) super Yang-Mills theory or the Argyres-Douglas theory \cite{Seiberg:1994rs,Argyres:1995jj}.\footnote{More properly, a certain embedding of the latter into a larger theory, as engineered in \cite{Gaiotto:2009hg}.}
In both cases the Coulomb branch has a single wall of marginal stability dividing a strong coupling chamber from a weak coupling chamber, and the critical locus $\CB_c$ corresponds to the MS wall itself as shown in Figure \ref{fig:flavorless}.

Another example of such a critical locus can be found in SU(2) SQCD with $N_f=1,2,3$. 
It was shown in \cite{Bilal:1996sk} that when all masses are zero, all singularities arrange on a circular curve on $\CB$, which is the only wall of marginal stability. In this case too, the MS wall is $\CB_c$.
In section \ref{sec:examples} we will examine a few more examples of such special loci. 
It would be very interesting to gain a systematic understanding of their existence and structure,
we leave these questions to future work.

\subsection{BPS states at $\CB_c$}\label{sec:2d-4d-physics}
%

The critical locus $\CB_c$ lies at the intersection of walls of marginal stability, therefore the 4d ``vanilla'' BPS spectrum is ill-defined for $u\in\CB_c$.
However, this is not so for the spectrum of 2d-4d BPS states in presence of a generic surface defect $\mathbf{S}_z$, since their central charges (\ref{eq:2d-4d-central-charge}) are not necessarily at the same phases as the 4d central charges. Schematically, the central charge of a 2d-4d BPS state with charge $a\in\tGamma_{ij}(z)$ gets contributions both from the 2d effective superpotential $\widetilde\CW(u)$ and from the 2d twisted masses (which are identified with 4d central charges)
\be
	Z_a(u) \sim \widetilde\CW_j(u)- \widetilde\CW_i(u) + Z_\gamma(u)\,.
\ee
As reviewed in Section \ref{sec:defects-review}, the values of $\widetilde\CW_{i},\widetilde\CW_{j}$ depend both on $u$ and on $z\in C$. Therefore choosing $z$ generic will resolve the phase-degeneracy of the $Z_\gamma(u)$, thanks to the contributions from $\widetilde\CW$, even for $u\in\CB_c$.
More precisely, we will choose $z$ such that the following condition is satisfied 
\be\label{eq:2d-4d-MS-condition}
	\arg Z_a \neq \arg Z_b\quad\forall \ a\in\Gamma_{ij}(z)\,,\ b\in\Gamma_{jk}\,.
\ee
This condition ensures that we are not on a wall of marginal stability for the 2d-4d BPS states, and therefore that the 2d-4d spectrum is well defined.\footnote{This condition \emph{also} implies that phases of central charges of all $ij$ and $ji$ solitons must be different.}
In section \S\ref{sec:monodromy-graphs} we will argue that this condition is generally satisfied by most $z\in C$.

\begin{figure}[h!]
\begin{center}
\includegraphics[width=0.55\textwidth]{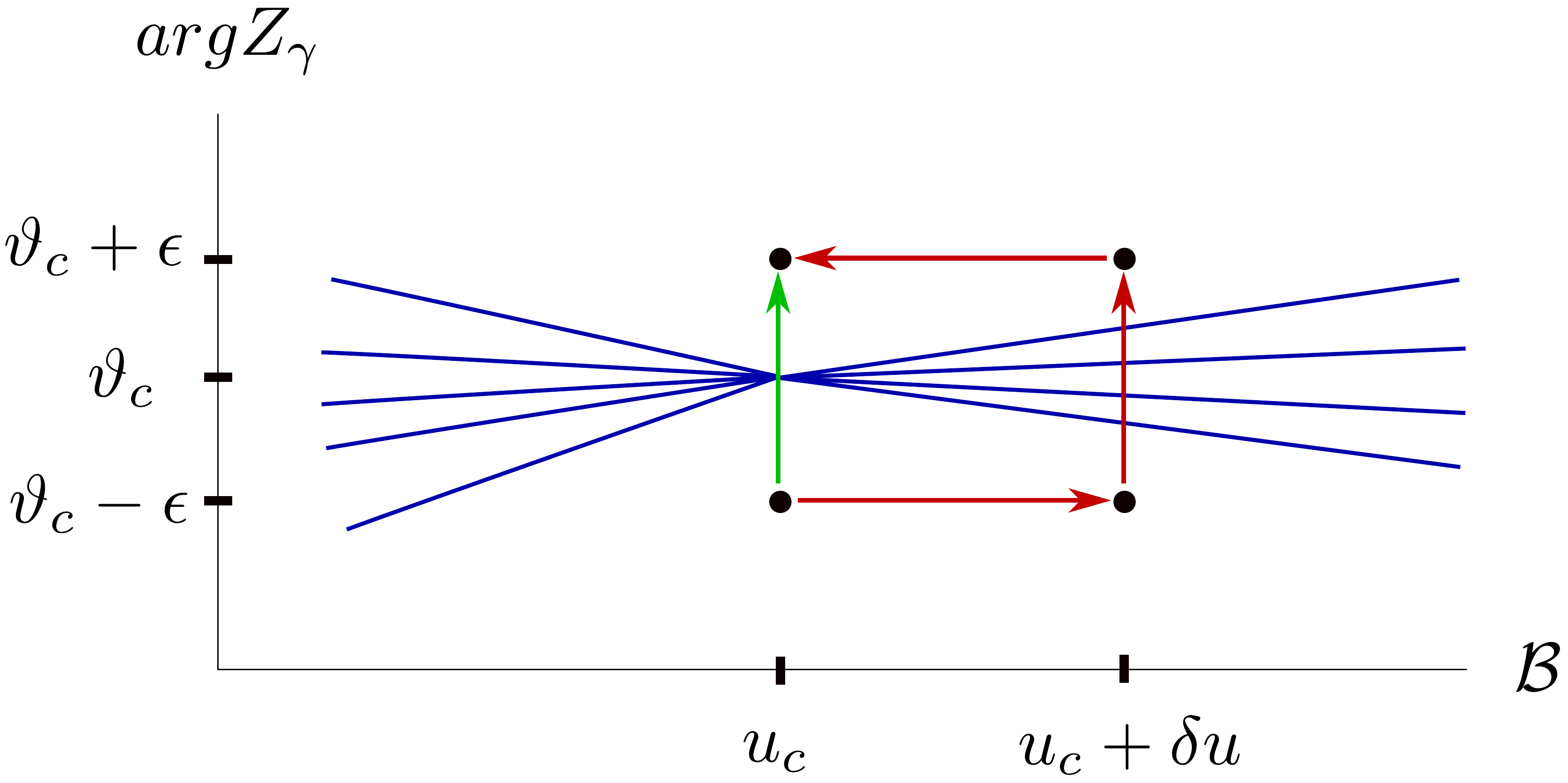}
\caption{Two paths from $(u_c,\vartheta_c-\epsilon)$ to $(u_c,\vartheta_c+\epsilon)$. The space of 2d-4d vacua is smooth within the region enclosed by the two paths, therefore the transitions of 2d-4d framed BPS spectra encountered along each path must be equivalent.}
\label{fig:path-around}
\end{center}
\end{figure}

Next we consider two half-defects, located at $z_\pm$, and joined by a supersymmetric interface $L_{\wp,\vartheta}$ parameterized by a path $\wp\subset C$ interpolating between $z_\pm$.
Let us fix $u_c\in\CB_c$ and study the discontinuities of the framed 2d-4d spectrum as a function of the phase $\vartheta$. 
There will be several distinct jumps in correspondence of phases of 2d-4d states, moreover there will also be a jump at the phase $\vartheta_c$, corresponding to both 2d-4d and vanilla 4d BPS states.\footnote{Although 4d BPS states are ill-defined at $u_c$, the occurrence of the jump of  framed 2d-4d spectra persists at this point, as will be argued momentarily.} 
We shall focus on the infinitesimal phase interval between $\vartheta_c-\epsilon$ and $\vartheta_c+\epsilon$ and study the jump at $\vartheta_c$ alone.
To clarify its physical interpretation, we turn on a small generic perturbation away from $u_c$ without crossing any marginal stability wall for framed degeneracies.\footnote{There are examples in which this is actually not possible, however it is possible in most cases we are aware of.} At $u_c+\delta u$ framed 2d-4d degeneracies undergo several jumps of the standard ``$\CK$-wall'' type, as one takes $\vartheta$ from $\vartheta_c-\epsilon\to\vartheta_c+\epsilon$. 
After these $\CK$-wall jumps we may then bring back $u$ to $u_c$, the spectrum of framed degeneracies must be the same as the one we obtained with the single jump considered previously, because the moduli space of 2d-4d vacua is (by assumption) smooth in the small region around $(u_c,\pi^{-1}(z))\subset \CB\times\Sigma$. A schematic diagram illustrating this reasoning is depicted in Figure \ref{fig:path-around}.

Since the single jump at $(u_c,\vartheta_c)$ must be equivalent to the overall sequence of $\CK$-wall jumps that would be observed at generic $u$, it captures the whole 4d vanilla spectrum at $u_c+\delta u$, hence the whole BPS monodromy.
Our next task will be to build a framework for efficiently computing the BPS monodromy from the jump of framed 2d-4d degeneracies at the critical locus, without any reference to other points like $u_c+\delta u$ nor to any BPS spectrum.
As we will show below, this jump is qualitatively different from the standard $\CK$-wall jumps of \cite{Gaiotto:2012rg} in several regards. For example there is no simple interpretation of the jump in terms of halo formation (see \cite{Gaiotto:2010be, Denef:2007vg, Andriyash:2010qv, Manschot:2010qz}) because several mutually non-local 4d states contribute simultaneously to the jump.

\subsection{Critical graphs}
\label{sec:monodromy-graphs}

\begin{figure}[h!]
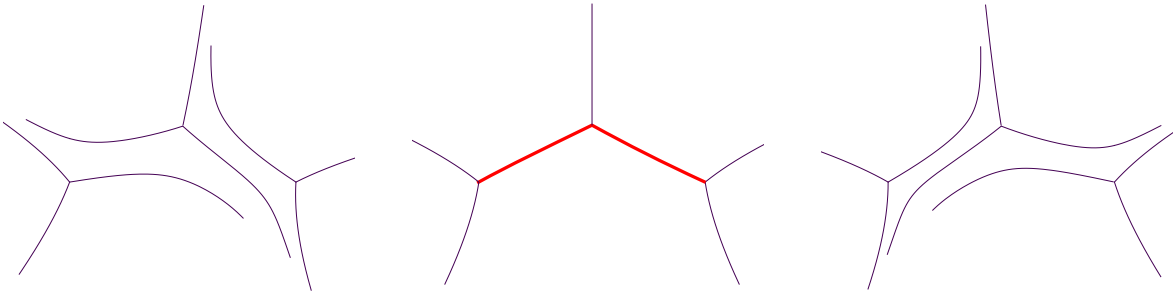

\begin{subfigure}{0.3\textwidth}
        \includegraphics[width=\textwidth]{figures/AD3_lvl2_26_mod.pdf}
\end{subfigure}
\hfill
\begin{subfigure}{0.3\textwidth}
        \includegraphics[width=\textwidth]{figures/AD3_lvl2_30_mod.pdf}
\end{subfigure}
\hfill
\begin{subfigure}{0.3\textwidth}
        \includegraphics[width=\textwidth]{figures/AD3_lvl2_33_mod.pdf}
\end{subfigure}
\caption{Spectral network for the Argyres-Douglas theory $AD_3$, at three distinct phases. $u$ is chosen on the critical locus $\CB_c$, shown in figure \ref{fig:flavorless}. 
A single $\CK$-wall jumps can be seen at the phase $\vartheta_c$, two 2-way streets appear, they are marked in red in the second frame. 
Note that both 2-way streets appearing in the jump had previously appeared in the strong coupling chamber at separate phases, see figure \ref{fig:AD3-strong-coupling}.
}
\label{fig:AD3-MS}
\end{figure}

Here we review the construction of critical graphs from degenerate spectral networks proposed in \cite{network-quiver}.
Let $u_c\in\CB_c$ be a point on the critical locus, and let $\vartheta_c$ be the corresponding critical phase.\footnote{Actually there are two critical phases: $\vartheta_c$ and $\vartheta_c+\pi$, everything that follows applies to either. The two settings are related by an involution on the lattice of 4d charges, taking $\gamma\to -\gamma$.}
The family of spectral networks at the critical point $\CW(u_c,\vartheta)$ exhibits a single jump, in correspondence of the critical phase $\vartheta_c$. 
At this jump several 2-way streets appear simultaneously, as shown in Figure \ref{fig:AD3-MS}:
we shall denote by $\CW_c$ the sub-network made of these.
This definition of critical graphs is a natural generalization of the ``$\CK$-wall jumps'' originally studied in \cite{Gaiotto:2012rg}. It is also very close to that of the ``Fenchel-Nielsen networks" studied in \cite{Hollands:2013qza}.\footnote{One difference is that in that reference no leaves of the Strebel foliation are allowed to end on punctures, see \cite[\S 3.3]{Hollands:2013qza}.}

\begin{figure}
\begin{center}
\includegraphics[width=.7\textwidth]{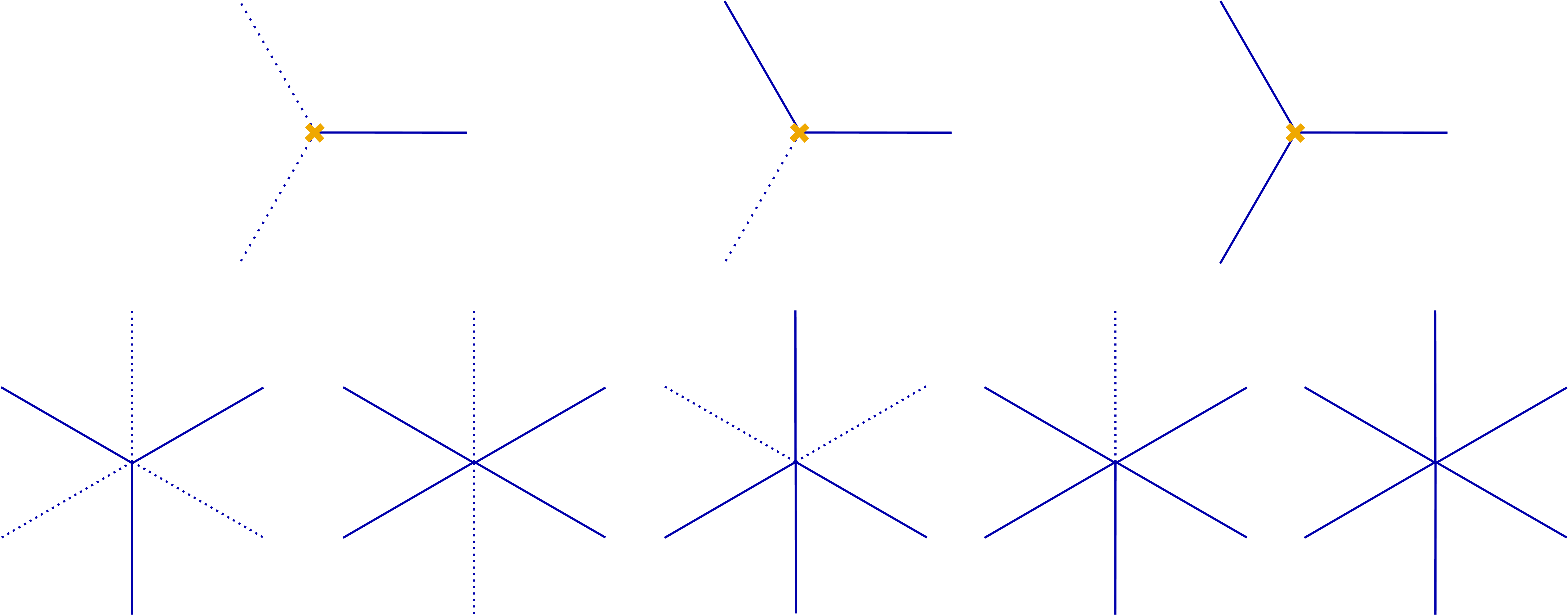}
\caption{The allowed node types for a critical graph. 
At the top are three types of branch point nodes, while joint-type nodes are at the bottom.
Branch points have 3 slots, joints have 6.
Solid lines denote 2-way streets, while dotted lines denote regular $
\CS$-walls. Only 2-way streets are included in the critical graph.
However, the cyclic positions of empty slots are part of the graph data, for example this distinguishes between two types of joints with four 2-way streets.
}
\label{fig:all-nodes}
\end{center}
\end{figure}

Our construction of BPS monodromies will rely entirely on \emph{topological} data of $\CW_c$,
which can therefore be regarded as a graph: edges are the 2-way streets, nodes can be either \emph{branch points} or \emph{joints}. A classification of allowed nodes is given in Figure \ref{fig:all-nodes}. 
There is also additional information that plays an important role in this paper: a certain notion of ``framing''.
More precisely, the topological data that will enter the construction of BPS monodromies includes
\begin{enumerate}
\item A set of nodes, of types shown in Figure \ref{fig:all-nodes}
\item Edges connecting pairs of nodes
\item The cyclic ordering of edges at each node, including empty slots.
\end{enumerate}
In the case of $A_1$ theories no joints are allowed, therefore nodes consist only of branch points. In this case $\CW_c$ is just a graph with bi-valent or tri-valent vertices, with cyclic ordering of edges at each vertex: therefore our critical graphs reduce in this case to  \emph{ribbon graphs}.
In fact, at the critical loci $\CB_c$, the Seiberg-Witten curve is described by a \emph{Strebel quadratic differential} \cite{Hollands:2013qza}. In turn such a differential is known to correspond to a specific ribbon graph, namely a \emph{dessin d'enfants} for the UV curve $C$ \cite{1998math.ph..11024M}.

Having described what the spectral network looks like at the critical locus, we are now in a position to explain how to choose supersymmetric interfaces properly, such that condition (\ref{eq:2d-4d-MS-condition}) is satisfied. 
This amounts to choosing $z_\pm$ away from intersections of $\CS$-walls of the spectral network for any $\vartheta$.
In addition, since this condition also implies that all $ij$ and $ji$ solitons must have central charges of different phases, it means that $z_\pm$ must also be away from all 2-way streets that appear on the network, i.e. $z$ must lie off $\CW_c$.

\subsection{Classical monodromy from critical graphs}\label{sec:classical-monodromy-equations}
The degenerate network $\CW(u_c,\vartheta_c)$ admits two resolutions: the American one consists of perturbing the phase to $\vartheta_c^- = \vartheta_c-\epsilon $, the British one corresponds to the opposite perturbation $\vartheta_c^+ = \vartheta_c+\epsilon$.
For example, in Figure \ref{fig:AD3-MS} the left frame corresponds to the American resolution, while the right frame corresponds to the British one.
In either resolution the spectral network is generic, and its topology determines the soliton content $(a,\mu(a))$ on all $\CS$-walls according to standard rules from  \cite{Gaiotto:2012rg}.
For convenience we collect in Appendix \ref{sec:traffic-rules} all the rules for each resolution.
In Appendix \ref{sec:soliton-computations} we present a technique for carrying out computations of soliton data in a simplified way, together with some examples. 
These techniques are especially useful for the critical graphs we study, although they can be of separate interest for handling complicated $\CK$-wall jumps, such as those considered in \cite{Galakhov:2013oja, Hollands:2016kgm}.

Taking the soliton data as given for both resolutions, we define two generating series for each two-way street $p\in\CW_c$
\be\label{eq:classical-Q-functions}
\begin{split}
	Q^{(\pm)}(p) = 1+\sum_{a\in\tGamma_{ij}(p)}\sum_{b\in\tGamma_{ji}(p)} \mu^{(\pm)}(a)\,\mu^{(\pm)}(b) \, X_{ab}
\end{split}
\ee
where the $\pm$ superscript denotes that the 2d-4d indices $\mu^{(\pm)}$ are computed in the American or British resolution of $\CW_c$.\footnote{For a standard $\CK$-wall jump as defined in \cite{Gaiotto:2012rg}, one would have $Q^{(+)}\equiv Q^{(-)}$. However this is not true anymore in our case, this is one of the features that distinguish the full KS monodromy jump from the $\CK$-wall jumps.} 
There exist a distinguished set of functions $\{Q^{(\pm)}_i\}_{i}$ such that every $Q^{(\pm)}(p)$ can be written as
\be\label{eq:Q-pm-factorization}
	Q^{(\pm)}(p) = \prod_i \left(  Q^{(\pm)}_i\right)^{\alpha^{\pm}_i(p)} 
\ee
with $\alpha_i^\pm(p)\in\IZ$.  In Section \ref{sec:main-proof} we provide a precise characterization of the $Q_i$.
To each $Q_i$ we associate a path $L_i$ on $\Sigma$, defined by the following formal sum
\be\label{eq:L-i-def}
	L_i := \sum_{p\in\CW_c} \alpha^{+}_i(p) \, p_{\tSigma}\,,
\ee
where $p_{\tSigma}$ is the canonical lift of a street $p$ to $\tSigma$ (we defined the canonical lift below  (\ref{eq:GMN-L})). 
In fact $L_i$ is closed $\partial L_i = \emptyset$, and we will denote by $\tilde\gamma_i$ the corresponding homology class. Note that we are focusing on the $\alpha_i^{+}$ at this point.
Using the $L_i$ we define a set of functions $S_\gamma$ as follows
\be\label{eq:S-gamma-def}
	S_{{\gamma}} = \prod_j \left(Q_j^{(+)}\right)^{\langle {\tilde\gamma}_j,{\tilde\gamma}\rangle}\,.
\ee
These are really functions of $\gamma\in\Gamma$, not $\tilde\gamma\in\tGamma$, since $H$ (the generator of the $\IZ$-extension of $\Gamma$) is in the annihilator of the intersection pairing as recalled in Section \ref{sec:review-4d-wc}.
Finally, we claim that the $S_\gamma$ encode precisely the action of the classical BPS monodromy (or spectrum generator) on generic formal variables $X_{\tilde\gamma}$ as 
\be\label{eq:KS-monodromy-formula}
	\IS (X_{\tilde\gamma}) = X_{\tilde\gamma}\, S_{\gamma}\,.
\ee

A proof of these claims will detailed below, in Section \ref{sec:main-proof}.
We insist on characterizing the classical monodromy by its action on generic formal variables $X_{\tilde\gamma}$, as opposed to giving a factorization of $\IS$ involving BPS indices, since the latter would not be manifestly wall crossing invariant.
An important point is whether the functions $S_\gamma$ \emph{completely} characterize $\IS$ or not. They almost do,  the exception being contributions from pure-flavor BPS states, since factors $\CK^{\Omega(\gamma_f)}_{\gamma_f}$ act trivially, and therefore their contribution is not contained in the functions $S_\gamma$. 
However, flavor states are automatically recorded by the $Q^{\pm}(p,y)$, thanks to the physical interpretation of framed 2d-4d wall crossing. We will illustrate this point in detail with an example in Section \ref{sec:T2}.
Therefore the set $\{\tilde \gamma_j,Q^{(+)}_j\}_j$ completely characterizes $\IS$, in fact it is possible to extract algorithmically a factorization of $\IS$ into $\CK$ transformations, from knowledge of the $S_{\gamma}$, see e.g. \cite[\S 5.1]{Galakhov:2013oja}.

\subsection{Quantum monodromy from critical graphs}\label{sec:quantum-monodromy-equations}
We now turn to explaining how the BPS monodromy $\mathbb{U}$ is encoded by $\CW_c$.
Now for each 2-way street $p\in\CW_c$ we again compute the soliton content in both American and British resolutions, using the standard soliton ``traffic rules" collected in Appendix \ref{sec:traffic-rules}. 
However, this time we don't just keep track of homology classes of solitons, instead we keep track of the \emph{actual path} of a soliton, which we define next. 
Recall that each street $p$ admits a canonical lift to $\tSigma$: for example a street of type $ij$ lifts to two segments on sheets $i$ and $j$ with opposite fixed orientations \cite{Gaiotto:2012rg}, the coordinate in the $S^1$ fiber of $\tilde\pi:\tSigma\to\Sigma$ is determined by the unit tangent vector to $p$. 
Using this information, we construct the actual path of each soliton by recording the ordered sequence of lifted street segments on $\tSigma$, along which the soliton path develops. 
The traffic rules collected in Appendix \ref{sec:traffic-rules} allow to do just this.
Then for any point $z\in p$ on a 2-way street $p$ of type $ij/ji$ we consider the sets
\be
	\CP^{(\pm)}_{ij}(z) = \{\text{actual soliton paths of $p$ from $z_i$ to $z_j$, in American/British resolution}\} \,,
\ee 
of \emph{actual} soliton paths of $ij$ type, and similarly for $\CP^{(\pm)}_{ji}(z)$.
Using this data we define the following two generating series
\be\label{eq:motivic-Q-def}
\begin{split}
	Q^{(-)}(p,y) & = 1 + \sum_{\alpha\in \CP^{(\pm)}_{ij}(z)}\sum_{\beta\in \CP^{(\pm)}_{ji}(z)} y^{\wr(\alpha\beta)-\iota_p(\alpha\beta)} \, \hY_{[\alpha\beta]} \\
	Q^{(+)}(p,y) & = 1 + \sum_{\alpha\in \CP^{(\pm)}_{ij}(z)}\sum_{\beta\in \CP^{(\pm)}_{ji}(z)} y^{-\wr(\beta\alpha)-\iota_p(\alpha\beta)} \, \hY_{[\beta\alpha]}\,.
\end{split}
\ee
Here $\alpha\beta$ denotes the concatenation of the two paths at ${\rm end}(\alpha)={\rm beg}(\beta)=z_j$, while $\beta\alpha$ is the concatenation at the opposite endpoint $z_i$.
$\wr{(\alpha\beta)}$ in $Q^{(-)}$ is the \emph{writhe} computed with respect to the basepoint ${\rm beg}(\alpha)={\rm end}(\beta)=z_i$, which differs in general from $\wr{(\beta\alpha)}$ appearing in $Q^{(+)}$.%
\footnote{More precisely, the writhe is computed by considering the projection of the actual path $\alpha\beta$ to $\Sigma$, i.e. this is the writhe of $\tilde\pi(\alpha\beta)$, see \cite{Galakhov:2014xba}.}
$\iota_p(\alpha\beta)$ is a positive integer, counting the number of times the actual path $\alpha\beta$ travels over street $p$, either on sheet $i$ or $j$ (the count is the same).
Finally $[\alpha\beta]\equiv [\beta\alpha]$ denotes the homology class of the closure of these concatenations in $H_1(\tSigma,\IZ)$, hence defining an element of $\tGamma$.

We stress that \emph{the functions $Q^{(\pm)}(p,y)$ are entirely determined by the topological data of the critical graph $\CW_c$}, the rest of the spectral network $\CW(\vartheta_c,u_c)$ being entirely negligible.
Given the whole set of $\{Q^{(\pm)}(p,y)\}_{p\in\CW_c}$,  the BPS monodromy is finally determined by the set of equations
\be\label{eq:U-eqn}
	\mathbb{U}\,Q^{(-)}(p,y) = Q^{(+)}(p,y) \, \mathbb{U}\,, \qquad \forall p\in\CW_c\,.
\ee
A remarkable feature of this characterization of $\mathbb{U}$ is the complete absence of information about vanilla BPS states, in this sense it is \emph{manifestly} wall crossing invariant.
In the next section we will present an algorithmic technique for solving these equations. 

Before moving on, a brief comment on the relation between (\ref{eq:U-eqn}) and its classical counterpart (\ref{eq:KS-monodromy-formula}) is in order.
First of all, note that definitions (\ref{eq:motivic-Q-def}) and (\ref{eq:classical-Q-functions}) are related by
\be\label{eq:classical-limit-Q}
	Q^{(\pm)}(p) = Q^{(\pm)}(p,y=1)\,.
\ee
The analogue of $S_{\gamma}$ would be given by $\hat R_{\tilde\gamma}$, defined by $ \mathbb{U} \hY_{\tilde\gamma} \mathbb{U}^{-1} = \hY_{\tilde\gamma} \, \hat R_{\tilde\gamma} $. 
We did not find a simple closed form for $\hat R_{\tilde\gamma}$. Instead, rewriting (\ref{eq:U-eqn})  as $ \mathbb{U} Q^{(-)} \mathbb{U}^{-1} = Q^{(-)} \, \big( {Q^{(-)}}^{-1}\,Q^{(+)}\big)$, we recognize ${Q^{(-)}}^{-1}\,Q^{(+)}$ as the analogue of acting with the classical monodromy $\IS$ on the classical generating function $Q^{(-)}$. In fact we will show in Section \ref{sec:main-proof} that $ \IS \left(Q^{(-)}(p)\right) = Q^{(+)}(p) $ for every $p\in\CW_c$.

\subsection{Solving the monodromy equations}\label{sec:solving-monodromy-eqs}

For practical purposes one is often interested in obtaining an explicit expression for $\mathbb{U}$, factorized into quantum dilogarithms as in (\ref{eq:factorized-U}), to compute BPS spectra.
In this section we present an algorithmic procedure for efficiently computing the $a_m(\gamma)$ (defined as  Laurent coefficients of the PSC in (\ref{eq:PSC-def})). 
Details on the derivation can be found in appendix \ref{sec:algorithmic-factorization}.

At $u_c$ all central charges group together at two opposite phases $\vartheta_c, \,\vartheta_c+\pi$. Without loss of generality, we choose  a half-plane $\mathbb{H}(\vartheta_c)$  in the complex plane of central charges centered around $e^{i \vartheta_c}$.
In turn this determines a half-lattice $\Gamma^+\subset\Gamma$ of all $\gamma$ with central charge $Z_\gamma$ contained in  $\mathbb{H}(\vartheta_c)$.
There is a unique positive integral basis for $\Gamma^+$, let $\gamma_k$ with $k=1,\dots, \text{rank}(\Gamma)$ be the generators.
Any charge $\gamma\in\Gamma^+$ admits then a unique expansion $\gamma = \sum_k r_k\gamma_k$, with non-negative integer coefficients $r_k$.
The basis therefore induces a filtration on $\Gamma^+$, we say that $\gamma$ is of level $|\gamma|=n$ if $\sum_k r_k = n$.

Now suppose we want to compute $\mathbb{U}$ at some $u\in\CB$, at which the basis central charges $Z_{\gamma_i}$ have moved away from the ray $e^{i\vartheta_c}$, but have not crossed the boundaries of $\mathbb{H}(\vartheta_c)$.\footnote{This is a simplifying assumption made for clarity of exposition. It can be relaxed by suitably rotating the half-plane, and (if necessary)  conjugating  $\mathbb{U}$ by the dilogarithms of those BPS rays which exit/enter the half-plane.}
Then both $\mathbb{U}$ and $Q^{(\pm)}(p,y)$ admit an expansion of the form
$F = \sum_{n\geq0} F_n$, with $F_n$ a finite sum of monomials $f_{\tilde\gamma} \hY_{\tilde\gamma}$ containing only charges of level $|\gamma|=n$ (here $\gamma$ is the pushforward of $\tilde\gamma$ by $\tilde\pi:\tSigma\to\Sigma$).
Let $U_i, Q_i^{(\pm)}(p) $ be the series coefficients of $\mathbb{U}$ and $Q^{(\pm)}(p,y) $ respectively.
Then define 
\be
\begin{split}
	\tilde R_{i+1}(p) & = \sum_{j=0}^{i-1} \left(  U_j \, Q^{(-)}_{i-j+1}(p) - Q^{(+)}_{i-j+1}(p) \, U_j \right) + \hat U_i \, Q^{(-)}_{1}(p) - Q^{(+)}_{1}(p) \, \hat U_i   \\
	& = \sum_{|{\tilde\gamma}|=i+1} \tilde r_{\tilde\gamma}(p) \hY_{\tilde\gamma}
\end{split}
\ee
where we introduced
\be\label{eq:hat-U-i-def}
	\hat U_i = \left[   \, \prod_{|\gamma|<i}^{\nwarrow}\prod_{m\in\IZ}   \Phi\left(  (-y)^m \hY_{\tilde\gamma}\right)^{a_m(\gamma)}  \,   \right]_{{\rm{lvl}} = i}\,,
\ee		
in which we keep monomials of level $i$, and $\tilde\gamma$ is the \emph{preferred lift} of $\gamma$ (see \cite{Gaiotto:2012rg} or Appendix \ref{sec:homology-conventions}).
A key property of the $\tilde r_\gamma$ defined in this way, is that they depend only on the $a_m(\gamma')$ for $|\gamma'|<i$. 
Moreover, the $a_m(\gamma)$ for $|\gamma|=i$ are related to the $\tilde r_\gamma$ by the linear equations
\be \label{eq:a-m-eqn}
	\tilde r_{\tilde\gamma}(p)  + \sum_{k=1}^{d} q_k(p) \, \frac{ y^{\langle \tilde\gamma,\tilde\gamma_k\rangle}- y^{-\langle \tilde\gamma,\tilde\gamma_k\rangle}  }{y-y^{-1}} \sum_{m\in\IZ} (-y)^m\, a_m(\gamma-\gamma_k)  =0 \,,
\ee
where $q_k(p)$ are the coefficients of the level-$1$ terms, namely $Q_1^{(+)}(p) = \sum_{k=1}^d q_k(p) \hY_{{\tilde\gamma}_k}$. 
Of course $a_m(\gamma)$ should depend on a choice of central charges, in fact this enters in the factorization of $\mathbb{U}$,  through $\hat U_i$ in (\ref{eq:hat-U-i-def}).
The  $a_m(\gamma)$ can therefore be computed recursively level-by-level in $|\gamma|$, by applying  equation (\ref{eq:a-m-eqn}) simultaneously  to all $p\in\CW_c$.
In Section \ref{sec:examples} we provide several examples of its application.

\subsection{Derivation of monodromy equations}\label{sec:main-proof}

In this section we give proofs for the constructions of classical and quantum monodromies presented in Sections \ref{sec:classical-monodromy-equations} and \ref{sec:quantum-monodromy-equations}.
Some familiarity with spectral networks will be assumed.

\begin{figure}[h!]
\begin{center}
\includegraphics[width=0.25\textwidth]{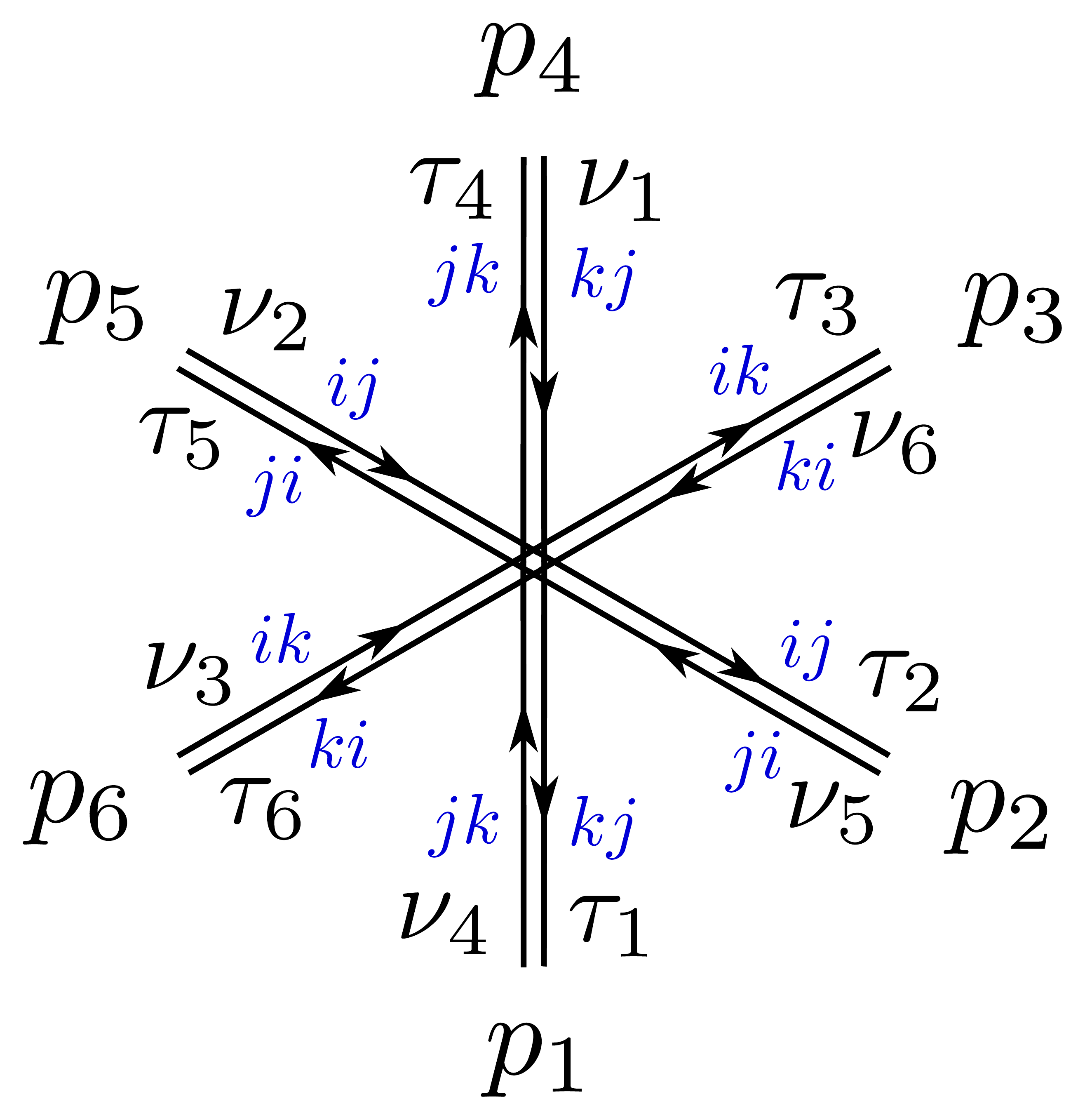}
\caption{A joint with six 2-way streets ending on it. Blue labels denote $\CS$-wall types.}
\label{fig:joint-conservation}
\end{center}
\end{figure}

Let us start from the classical monodromy.
First of all we must justify the claim of equation (\ref{eq:Q-pm-factorization}), by defining the distinguished set of factors $\{Q_i\}_i$.
To define these we cut the graph $\CW_c$ into disjoint sub-graphs $\CW_c^{(i)}$, by removing the branch point nodes.
Each $\CW_c^{(i)}$ will have its own set of $Q_i$, so we treat them separately.
If $\CW_c$ does not contain any joint nodes, then each $\CW_c^{(i)}$ consists of exactly one street $p_i$, in this case we are done: each $Q^{(+)}(p_i)$ is identified with a generator $Q_i$.
On the other hand if $\CW_c^{(i)}$ has joints, the $Q_i$ naturally emerge from the algebra of soliton traffic rules. 
For the sake of generality consider a six-way joint of 2-way street as in Figure \ref{fig:joint-conservation}, the generating functions $Q^{(+)}(p_i)$ for $i=1,\dots,6$ can be computed in full generality by repeated application of the rules in Appendix \ref{sec:traffic-rules}.

In British resolution they factorize as follows
\be\label{eq:joint-factorization}
\begin{split}
	Q^{(+)}(p_i) &= 1+\tau_{i}\nu_{i+3} = \frac{(1+\nu_{i+1}\nu_{i+3}\nu_{i+5})(1+\nu_{i}\nu_{i+3}(1+\nu_{i+2}\nu_{i+5}))}{1-(\nu_6\nu_2\nu_4)(\nu_5\nu_3\nu_1)}\,\qquad i=1,3,5\,\\
	Q^{(+)}(p_i) &= 1+\tau_{i}\nu_{i+3} = \frac{(1+\nu_{i+5}\nu_{i+3}\nu_{i+1})(1+\nu_{i}\nu_{i+3}(1+\nu_{i+2}\nu_{i+5}))}{1-(\nu_6\nu_2\nu_4)(\nu_5\nu_3\nu_1)}\,\qquad i=2,4,6\,
\end{split}
\ee
The $\nu_i$ denote generating functions of incoming solitons on the respective streets, they are determined by the global topology of the network. This formula covers all types of joints shown in Figure \ref{fig:all-nodes}, by just setting $\nu_i=0$ for those streets that are not 2-way.
All $Q(p_i)$ for streets ending on a generic 6-way joint $J$ can be thus expressed in terms of six factors:
\be\label{eq:joint-factors}
\begin{split}
	Q^{(J)}_1 & = 1+\nu_1\nu_4 (1+\nu_3\nu_6)%
	\qquad\qquad%
	Q^{(J)}_{\rm even} = 1+\nu_6\nu_2\nu_4\\
	Q^{(J)}_2 & = 1+\nu_2\nu_5 (1+\nu_1\nu_4)%
	\qquad\qquad%
	Q^{(J)}_{\rm odd} = 1+\nu_5\nu_3\nu_1\\
	Q^{(J)}_3 & = 1+\nu_3\nu_6 (1+\nu_2\nu_5)%
	\qquad\qquad%
	Q^{(J)}_{\rm all} = 1-(\nu_6\nu_2\nu_4)(\nu_5\nu_3\nu_1)
\end{split}	
\ee
Taking into account the factorization induced by the joint equations produces a candidate set of $Q_i$ for a whole sub-graph $\CW_c^{(j)}$.
There is one remaining issue: if a street $p$ has both endpoints on joints, one may wonder if the two factorizations (\ref{eq:joint-factorization}) produce the same factors or not.
If they do, we are done. 
If they do not, it means that some of the $Q^{(J)}_i$'s in (\ref{eq:joint-factors}) can be factorized further. 
In this case the $Q_i$ are defined by a choice of common multiplicative basis for all $Q^{(J)}_i$ and $Q^{(J')}_i$ that appear in $Q^{(+)}(p)$. 
This choice may or may not be unique a priori, but this will not affect the following.
What is important is that the $Q_i$ are independent from each other, i.e. that there are no nontrivial (multiplicative) relations among them, to ensure that the corresponding $\alpha_i(p)$ are \emph{uniquely defined}.

Next, we should support the claim, made below  (\ref{eq:L-i-def}), that $L_r$ is a closed cycle. The only contributions to $\partial L_r$ come from endpoints of the lifts of two-way streets. Endpoints above branch points obviously receive canceling contributions. 
At joints we consider instead the contributions on each sheet separately.
From (\ref{eq:joint-factorization}) it follows that
\be\label{eq:joint-conservation}
	\left(\prod_{r}Q_r^{\alpha_r(p_i)}\right)\left(\prod_{r}Q_r^{\alpha_r(p_{i+1})}\right) = \left(\prod_{r}Q_r^{\alpha_r(p_{i+3})}\right)\left(\prod_{r}Q_r^{\alpha_r(p_{i+4})}\right)\,,
\ee
with $i\in \IZ / 6\IZ$. Then referring to Figure \ref{fig:joint-conservation}, we have e.g. on sheet $j$
\be
	\partial L_r \supset z_j \,  \big( -\alpha_{r}(p_1) - \alpha_{r}(p_2) + \alpha_{r}(p_4) + \alpha_{r}(p_5)  \big) 
\ee
which vanishes as a consequence of (\ref{eq:joint-conservation}). Likewise on other sheets, relations among the $\alpha_r(p)$ of different streets ending on the same joint ensure canceling contributions from the joint to $\partial L_r$, establishing the claim.

To prove (\ref{eq:KS-monodromy-formula}) we will derive the following formula, which describes the jump of the framed 2d-4d degeneracies at the critical phase $\vartheta_c$ for $u_c\in\CB_c$
\be\label{eq:gen-K-wall-formula-F}
	F(\wp,\vartheta_c^+,u_c) = \IS \big( F(\wp, \vartheta_c^-,u_c) \big)\,,
\ee
where $\IS(\, \cdot \,) $ is here defined as the substitution rule on formal variables
\be\label{eq:gen-K-wall-formula}
	\IS(X_a) = X_a \prod_k \big(Q^{(+)}_k\big)^{\langle\tilde\gamma_k,a\rangle} \,.
\ee
We shall call this the \emph{generalized $\CK$-wall formula}, because its specialization to a generic $u$ recovers the $\CK$-wall formula of \cite{Gaiotto:2012rg}.
Establishing the generalized $\CK$-wall formula clearly implies (\ref{eq:KS-monodromy-formula}): simply specializing the relative homology class $a\in\tGamma_{ij}$ to be a closed cycle with basepoint $end(a) = beg(a)$ in homology class $\tilde\gamma\in\tGamma$ turns (\ref{eq:gen-K-wall-formula}) into (\ref{eq:KS-monodromy-formula}).


To prove (\ref{eq:gen-K-wall-formula-F}) we follow a strategy used in \cite[\S 6.6]{Gaiotto:2012rg}. 
A key observation from that paper is that $F(\wp, \vartheta, u)$ can be interpreted as a formal parallel transport on $C$, which is moreover \emph{flat}, i.e. it only depends on the relative homotopy class of $\wp$.
While this remains true in our case, there is an important novelty: there are now  \emph{two different} functions $Q^{(\pm)}(p)$ associated to each two-way street of $\CW_c$.
Let $\CF(\wp)$ denote either the LHS or RHS of (\ref{eq:gen-K-wall-formula-F}), both obviously satisfy the four following properties:
\begin{itemize}
 \item $\CF(\wp)$ is a homotopy invariant of $\wp$.
 \item If $\wp\cap\CW=\emptyset$, then $\CF(\wp) = D(\wp) \equiv \sum_{i}X_{\wp^{i}}$.
 \item If $\wp,\wp'$ have endpoints off $\CW_c$, then $ \CF(\wp) \CF(\wp') = \CF(\wp \wp') $.
 \item If $\wp\cap\CW=\{z\}$ on a one-way street of type ${ij}$, then 
\be\label{eq:f-one-way}
 \CF(\wp) = D(\wp_+) \left(1 + \sum_{a \in \tGamma_{ij}(p)} \mu(a) X_{a}\right) D(\wp_-),
\ee
for some $\mu(a) \in \IZ$.
\end{itemize}
where in the second point $\wp^{i}$ denotes the (tangent framing) lift of $\wp$ to the $i$-th sheet of $\pi\circ\tilde\pi:\tSigma\to C$.

\begin{figure}[h!]
\begin{center}
\includegraphics[width=0.35\textwidth]{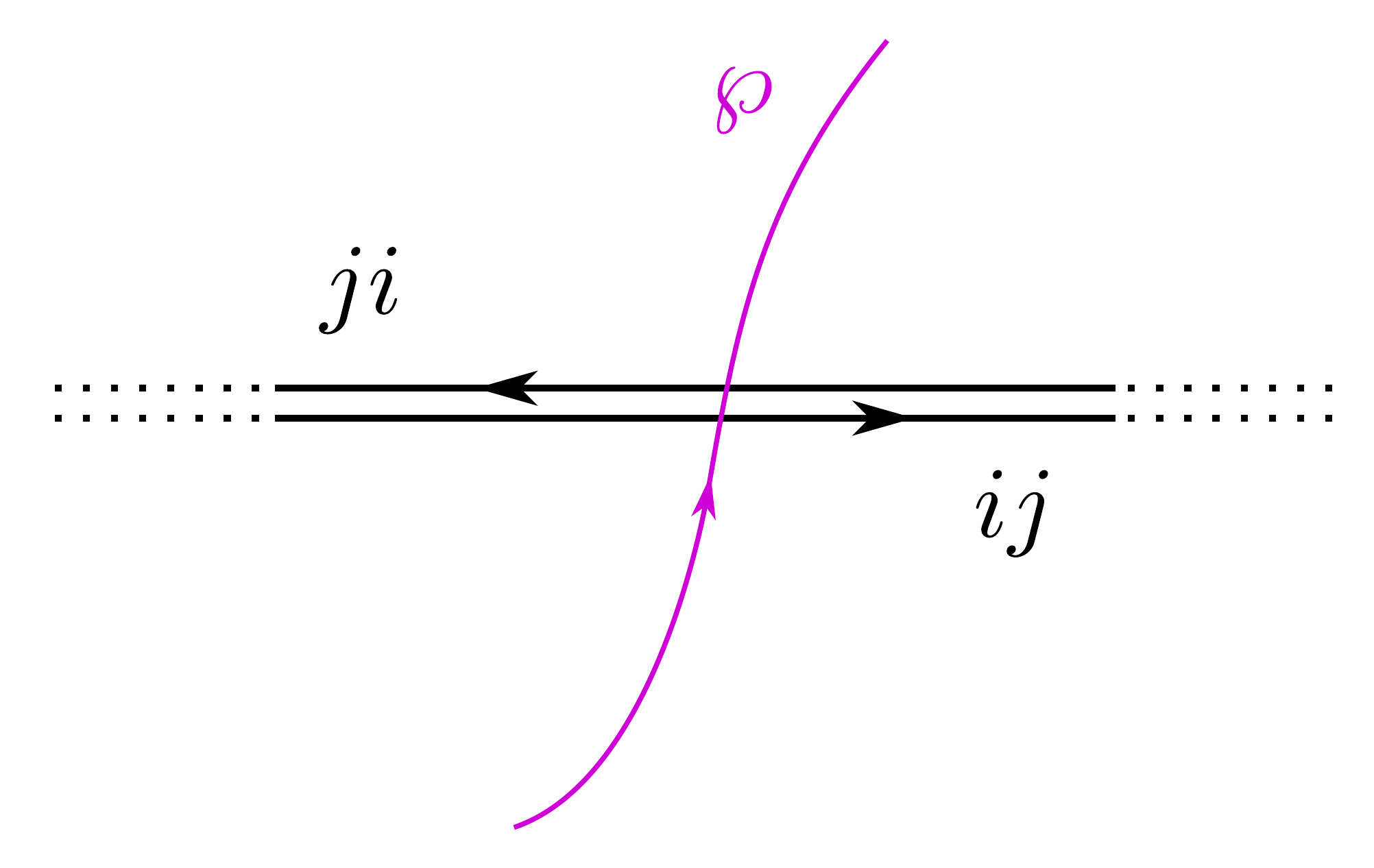}
\caption{
A path $\wp$ on $C$ crosses a 2-way street $p$ of type $ij/ji$, shown in  American resolution. 
The canonical lift of $p$ to $\Sigma$ is $p_\Sigma = p^{(i)}\cup p^{(j)}$, running to the left on sheet $i$ and to the right on sheet $j$. 
The lifts of $\wp$ to sheets $i$ and $j$ intersects $p_\Sigma$ with pairing $\langle\wp^{(i)}, p_\Sigma\rangle = 1 = - \langle\wp^{(j)}, p_\Sigma\rangle $.
}
\label{fig:two-way}
\end{center}
\end{figure}

In addition, the LHS of (\ref{eq:gen-K-wall-formula-F})  manifestly satisfies
\begin{itemize}
	\item If $\wp\cap\CW=\{z\}$ on a two-way street $p\in\CW_c$, and the intersection between $\wp$ and $p$
is positive\footnote{Computed with respect to the orientation of the complex curve $C$, with the orientation of $p$ fixed by the underlying $ij$ $\CS$-wall.},
\be \label{eq:f-two-way}
 \CF(\wp) = D(\wp_+) \left(1 + \sum_{ b \in \tGamma_{ji}(p)} \mu(b) X_{b}\right) \left(1 +  \sum_{ a \in \tGamma_{ij}(p)} \mu(a) X_{a} \right) D(\wp_-)
\ee
for some $\mu(a),\,\mu(b) \in \IZ$.
\end{itemize}
To prove that this last property also holds for the RHS of (\ref{eq:gen-K-wall-formula-F}), define 
\be
	\IT(X_a) = X_a \prod_k T_k^{\langle\tilde\gamma_k, a\rangle}\,.
\ee
for some generic $T_k$.
First of all note that associativity
\be
	\IT(X_a) \IT(X_b) = \IT(X_a X_b)
\ee
holds for any $X_a, X_b$ (here $a,b$ may be either open or closed homology classes, independently from each other).
Then consider $\wp$ crossing one of the streets of $\CW_c$ as shown in Figure \ref{fig:two-way}, and apply $\IT$ to $ F(\wp, \vartheta_c^-,u_c) $. 
We compute this by considering separately the components of different topological types:
\be\label{eq:KS-monodromy-direct-computation}
\begin{split}
	\IT\big( F(\wp,\vartheta_c^-,u_c)_{ii} \big) %
	& = X_{\wp^{i}} \, T(p)^{-1} \, \IT(Q^{(-)}(p)) \\
	\IT\big( F(\wp,\vartheta_c^-,u_c)_{ij} \big) %
	& = \sum_{a\in\tGamma_{ij}}\mu^{(-)}(a) \, X_{\wp^{i}_{+}} \,  \IT(X_a) \, X_{\wp^{j}_{-}}  \, \prod_k T_k^{\langle\tilde\gamma_k, \wp^{i}_+ + \wp^{j}_-\rangle}\\
	\IT\big( F(\wp,\vartheta_c^-,u_c)_{ji} \big) %
	& = \sum_{b\in\tGamma_{ji}}\mu^{(-)}(b) \, X_{\wp^{j}_{+}} \,  \IT(X_b) \, X_{\wp^{i}_{-}}  \, \prod_k T_k^{\langle\tilde\gamma_k, \wp^{j}_+ + \wp^{i}_-\rangle}\\
	\IT\big( F(\wp,\vartheta_c^-,u_c)_{jj} \big) %
	& = X_{\wp^{j}} \, T(p) \\
\end{split}
\ee
where $T(p) := \prod_k T_k^{\alpha^{+}_k(p)}$. This factor and its inverse arise from acting with $\IT$ on $X_{\wp^{j}}$ and $X_{\wp^{i}}$ respectively.%
\footnote{
For example consider $\IT \left(X_{\wp^{i}}\right) = X_{\wp^{i}}\prod_{k} T_k^{\langle\tilde\gamma_k, \wp^{i}\rangle}$. Recalling that $\tilde\gamma_k = [L_{k}]$ as defined below (\ref{eq:L-i-def}), this yields $\langle\tilde\gamma_k, \wp^{i}\rangle = \alpha_{k}^{+}(p) \langle p_{\tilde\Sigma} , \wp^{i}\rangle = - \alpha_{k}^{+}(p)$, since by assumption $\wp$ intersects only $p$, and in the last step we used the intersection pairing evaluated in Figure \ref{fig:two-way}.  
}
Now the desired property will be satisfied if these are the components of a parallel transport which factorizes into  the form
\be\label{eq:factorization-property}
	D(\wp_+) \, \left( 1+ \sum_{b} \mu^{(-)}(b) \IT(X_b) \right) \left( 1+ \sum_{a} \mu^{(-)}(a) \IT(X_a) \right)  D(\wp_-)
\ee
because the action of the operator $\IT$ acting on the $X_a$ can be reabsorbed into a change in the $\mu(a)$.
Note (\ref{eq:factorization-property}) will manifestly agree with (\ref{eq:KS-monodromy-direct-computation}) if we take
\be
\begin{split}
	T(p) & =1 + \sum_{a\in\tGamma_{ij}(p)}\sum_{b\in\tGamma_{ji}(p)} \, \mu^{(-)}(a)\mu^{(-)}(b)\, \IT(X_a)\, \IT(X_b) \\
	& = 1 + \sum_{a\in\tGamma_{ij}(p)}\sum_{b\in\tGamma_{ji}(p)} \, \mu^{(-)}(a)\mu^{(-)}(b)\, \IT(X_aX_b) \\
	& = \IT\big( Q^{(-)}(p) \big) \,.
\end{split}
\ee
So far we have proved that, if we take $T(p)$ defined in this way, then all five properties hold both for $F(\wp,\vartheta_c^+,u_c)$ and $\IT \big( F(\wp, \vartheta_c^-,u_c) \big)$.%
\footnote{
More precisely, we have just examined the validity of the fifth property. 
The first four properties are easily seen to hold as well, by the same arguments adopted for $\IS \big( F(\wp, \vartheta_c^-,u_c) \big)$. 
The only minor caveat is: for the fourth property to hold,  $T_k$ must be a polynomial involving only formal variables of type $X_\gamma$ associated with \emph{closed} cycles (as opposed to $X_a$). This would then ensure that $\IT(X_a) = \sum_b c_b X_b$ with $b\in\tGamma_{ij}(p)$ in (\ref{eq:f-one-way}).
}
Following \cite{Gaiotto:2012rg}, we then note that these five properties actually \emph{determine} $\CF(\wp)$ uniquely, therefore establishing the equality 
\be\label{eq:gen-K-wall-formula-F-T}
	F(\wp,\vartheta_c^+,u_c) = \IT \big( F(\wp, \vartheta_c^-,u_c) \big)\,.
\ee
The final step in proving (\ref{eq:gen-K-wall-formula-F}) is to note that (\ref{eq:gen-K-wall-formula-F-T}) now implies 
\be
	T(p) = Q^{(+)}(p)
\ee
by simply evaluating the LHS directly on the network in the British resolution, and comparing with (\ref{eq:KS-monodromy-direct-computation}).
In turn, by definition of $T(p)$ and recalling the factorization (\ref{eq:Q-pm-factorization}), this means that $T_k = Q^{}_k$.\footnote{By construction the $Q_k$ are independent from each other,  there are no nontrivial relations among them, as previously remarked.}
Therefore we also find that
\be
	\IT = \IS\,.
\ee
This completes the proof of (\ref{eq:gen-K-wall-formula-F}), and also establishes the relation
\be\label{eq:classical-Q-jump}
	Q^{(+)}(p) \equiv \IS \big(  Q^{(-)}(p)  \big)  \,, 
\ee
which fits naturally with the general formula.
Note that, in the case of a standard $\CK$-wall jump (e.g. if the whole spectrum consists of one BPS state plus its CPT conjugate), the above formula asserts that $Q^{(-)}(p)=Q^{(+)}(p)$, recovering the $\CK$-wall formula of \cite{Gaiotto:2012rg}.

We now turn to proving the quantum monodromy equations (\ref{eq:U-eqn}).
These correspond to the statement that the generating functions $Q^{(\pm)}(p,y)$ before and after the jump are related by conjugation by $\mathbb{U}$, which is a ``promotion'' of (\ref{eq:classical-Q-jump}) to the noncommutative setting.
The proof goes along the lines of the classical case, we highlight the few new key steps. 
First of all, the jump (\ref{eq:gen-K-wall-formula-F}) is replaced by%
\be\label{eq:motivic-U-conjugation}
	F(\wp,\vartheta_c^+,u_c;y) = \mathbb{U} \, F(\wp, \vartheta_c^-,u_c;y) \,  \mathbb{U}^{-1}
\ee
where $F(\wp,\vartheta,u_c;y)$ is a refinement of (\ref{eq:framed-2d-4d-generating-function}) defined in \cite{Galakhov:2014xba} (see equation (2.28), also see \cite{Gabella:2016zxu} for a recent application of this framework).
Let us look at the explicit form of the jump for some components of $F(\wp)$: recall that these count ``detours'', for example
\be
	F_{ii}(\wp, \vartheta_c^-,u_c;y) =\hY_{\wp^{(i)}} + \sum_{\alpha\in \CP^{(\pm)}_{ij}(z)}\sum_{\beta\in \CP^{(\pm)}_{ji}(z)} \, y^{\wr(\alpha\beta)}\hY_{\wp_+^{(i)} \alpha\beta \wp_-^{(i)}}
\ee
where we slightly abused notation and denoted by ${\wp_+^{(i)} \alpha\beta \wp_-^{(i)}}$ the relative homology class of the detour (whereas $\alpha$ and $\beta$ are really \emph{actual paths}).
As can be checked in Figure \ref{fig:two-way}, the path $\wp$ lifted to sheet $i$ intersects the lift of street $p$ to the same sheet once, with a positive intersection sign $\langle\wp^{(i)},p^{(i)}\rangle = 1$. 
Therefore we may rewrite the above as
\be
	F_{ii}(\wp, \vartheta_c^-,u_c;y) =\hY_{\wp^{(i)}} \left( 1+ \sum_{\alpha\in \CP^{(\pm)}_{ij}(z)}\sum_{\beta\in \CP^{(\pm)}_{ji}(z)} \, y^{\wr(\alpha\beta)-\iota_p(\alpha\beta)}\hY_{\alpha\beta} \right) \equiv \hY_{\wp^{(i)}} \,Q^{(-)}(p,y)
\ee
where $\iota_p(\alpha\beta)$ counts how many times the concatenated path $\alpha\beta$ runs over street $p$ on sheet $i$. 
Acting with $\mathbb{U}$ must yield $F_{ii}(\wp, \vartheta_c^+,u_c;y) = \hY_{\wp^{(i)}} $, therefore
\be\label{eq:def-of-CU}
	\mathbb{U}  \hY_{\wp^{(i)}}  \mathbb{U}^{-1} \, = \hY_{\wp^{(i)}} \CU^{-1}
\ee
with $\CU := \mathbb{U} \,Q^{(-)}(p,y)\, \mathbb{U}^{-1}$.
On the other hand, the jump of the $jj$ component takes  $F_{jj}(\wp, \vartheta_c^-,u_c;y) = \hY_{\wp^{(j)}} $ to
\be\label{eq:jump-wp-j}
\begin{split}
	\mathbb{U} \hY_{\wp^{(j)}} \mathbb{U}^{-1} %
	& = F_{jj}(\wp, \vartheta_c^+,u_c;y) \\
	& =\hY_{\wp^{(j)}} \left( 1+ \sum_{\alpha\in \CP^{(\pm)}_{ij}(z)}\sum_{\beta\in \CP^{(\pm)}_{ji}(z)} \, y^{\wr(\beta\alpha)+\iota_p(\alpha\beta)}\hY_{\alpha\beta} \right) \equiv \hY_{\wp^{(j)}} \,Q^{(+)}(p,y^{-1})\,,
\end{split}
\ee
where we used the fact that $\langle\wp^{(j)},p^{(j)}\rangle = -1$ and that $\iota_p(\alpha\beta)$ counts also the number of times $\alpha\beta$ runs over $p$ on sheet $j$.
Next we introduce a technical identity, define $R_{\tilde\gamma}(y)$ by 
\be\label{eq:y-invariance-eqn-1}
	\mathbb{U}\hY_{\tilde\gamma}\mathbb{U}^{-1} = \hY_{\tilde\gamma} \,R_{\tilde\gamma}(y)
\ee
then we claim that 
\be\label{eq:y-invariance-eqn-2}
	\mathbb{U}\hY_{-\tilde\gamma}\mathbb{U}^{-1} = \hY_{-\tilde\gamma} \,\left(R_{\tilde\gamma}(y^{-1})\right)^{-1}\,.
\ee
The dependence of $R_{\tilde\gamma}(y)$ on $y$ needs clarification due to the possibility of absorbing powers of $y$ through the ring relations of the formal variables. The clarification will come  in a moment, while proving the relation. 
First note that due to ${\mathfrak{so}}(3)$ rotational symmetry of the gauge theory, any factorization of $\mathbb{U}$ into dilogarithms must have $a_{m}(\gamma) \equiv a_{-m}(\gamma)$. Therefore $\mathbb{U}$ written as a factorization $\prod \Phi((-y)^m \hY_{\tilde\gamma})^{a_m(\gamma)}$ is invariant under the replacement $y\to y^{-1}$.
Then the whole LHS of (\ref{eq:y-invariance-eqn-1}), when expanded as a power series in $\hY_{\tilde\gamma}$ is still invariant under $y\to y^{-1}$. 
This follows easily by applying repeatedly the identity (\ref{eq:dilog-relation}), to write 
\be
\begin{split}
	\Phi(\hY_{\tilde\gamma})\, %
	\hY_{\tilde\gamma'}\,%
	\Phi(\hY_{\tilde\gamma})^{-1} \ %
	& = \ %
	\hY_{\tilde\gamma'}  \, %
	\Phi_{\langle\tilde\gamma',\tilde\gamma\rangle}(\hY_{\tilde\gamma})^{-{\rm sgn} \langle\tilde\gamma',\tilde\gamma\rangle\,} \\
	& = \sum_{\tilde\gamma''} c_{\tilde\gamma''}(y) \hY_{\tilde\gamma''}\,,
\end{split}
\ee
since the definition of the compact dilogarithm (\ref{eq:dilog-definition}) then implies
\be
	c_{\tilde\gamma''}(y) = c_{\tilde\gamma''}(y^{-1})\,.
\ee
Conjugation by $\Phi((-y)^m \hY_{\tilde\gamma})^{a_m}\Phi((-y)^{-m} \hY_{\tilde\gamma})^{a_{-m}}$ likewise gives a series with symmetric coefficients $c_{\tilde\gamma''}(y)=c_{\tilde\gamma''}(y^{-1})$, so overall 
\be
	\mathbb{U}\hY_{\tilde\gamma}\mathbb{U}^{-1} = \sum_{\tilde\gamma''} c_{\gamma''}(y) \hY_{\tilde\gamma''}
\ee
with $c_{\tilde\gamma''}(y)=c_{\tilde\gamma''}(y^{-1})$.
Expanding $R_{\tilde\gamma}(y)$ as follows 
\be
	R_{\tilde\gamma}(y) = \sum_{\tilde\gamma'}r_{\tilde\gamma'}(y)\hY_{\tilde\gamma'}
\ee
and plugging into its definition (\ref{eq:y-invariance-eqn-1}), gives
\be
	\sum_{\tilde\gamma''} c_{\gamma''}(y) \hY_{\tilde\gamma''} = \hY_{\tilde\gamma}R_{\tilde\gamma}(y)  = \sum_{\tilde\gamma'}y^{\langle\tilde\gamma,\tilde\gamma'\rangle}r_{\tilde\gamma'}(y)\hY_{\tilde\gamma'+\gamma}\,.
\ee
Taken term by term in the series, this means that 
\be
	y^{\langle\tilde\gamma,\tilde\gamma'\rangle}r_{\tilde\gamma'}(y) = y^{-\langle\tilde\gamma,\tilde\gamma'\rangle}r_{\tilde\gamma'}(y^{-1})\,.
\ee
Therefore, using the noncommutative product rule (\ref{eq:extended-formal-variables-motivic})
\be
	\hY_{\tilde\gamma} \, R_{\tilde\gamma}(y) =  R_{\tilde\gamma}(y^{-1}) \hY_{\tilde\gamma}\,,
\ee
finally this leads to
\be
\begin{split}
	\mathbb{U}  \hY_{-\tilde\gamma}  \mathbb{U}^{-1}  %
	& = \left(\mathbb{U}  \hY_{\tilde\gamma}  \mathbb{U}^{-1} \right)^{-1} 
	= \left(R_{\tilde\gamma}(y^{-1}) \hY_{\tilde\gamma} \right)^{-1}   
	= \hY_{-\tilde\gamma} \left(R_{\tilde\gamma}(y^{-1}) \right)^{-1}   \,,
\end{split}
\ee
which is the claim we wanted to prove in (\ref{eq:y-invariance-eqn-2}).

Now, since $\langle\wp^{(i)},\tilde\gamma\rangle=-\langle\wp^{(j)},\tilde\gamma\rangle$ for all $\tilde\gamma$, acting by conjugation by $\mathbb{U}$ on $\wp^{(j)}$ has the same effect it would have on $-\wp^{(i)}$. Then (\ref{eq:jump-wp-j}) and (\ref{eq:def-of-CU}) imply that
\be
	Q^{(+)}(p,y^{}) = \CU^{} = \mathbb{U} \,Q^{(-)}(p,y)\, \mathbb{U}^{-1}\,
\ee
establishing (\ref{eq:U-eqn}).

\section{Examples}\label{sec:examples}

In this section we present applications of our framework to several examples, deriving both classical and quantum BPS monodromies.
For pedagogical purposes we will begin by going in detail through the simplest example, that of the Argyres-Douglas $AD_3$ theory (see \cite{Gaiotto:2009hg} for nomenclature).
In more involved examples the computations of soliton generating functions can become rather unwieldy using the standard techniques of spectral networks, for this reason we developed a simplified formalism that greatly facilitates the computation of $Q^{(\pm)}(p)$. We make extensive use of these techniques, which are presented in Appendix \ref{sec:soliton-computations}.
On the other hand, for computing generating functions $Q^{(\pm)}(p,y)$ of the quantum monodromy, we instead developed a software which generates the soliton content of each street and computes self-intersections of paths, this is available at \cite{python-code}.
Useful tools for drawing spectral networks at critical loci include the mathematica package \cite{swn-plotter} and the web program \emph{loom} introduced in \cite{Longhi:2016rjt}.

\subsection{$AD_3$ Argyres-Douglas Theory}

The Coulomb branch of the $AD_3$ theory is one-dimensional, it is divided into two chambers by a pair of walls of marginal stability as shown in Figure \ref{fig:flavorless}. 
The critical locus $\CB_c$ coincides with the union of the two MS walls,  the spectral network at a generic  $u_c\in\CB_c$  is shown in Figure \ref{fig:AD3-MS}.
As shown there, the critical graph $\CW_c$ of this theory  consists of two 2-way streets, denoted by  $p_1, p_2$ in Figure \ref{fig:AD3} below. The symmetry group of $\CW_c$ is trivial, because exchanging the two edges would not preserve the cyclic ordering at the middle branch point.

\begin{figure}[h!]
\begin{center}
\includegraphics[width=0.45\textwidth]{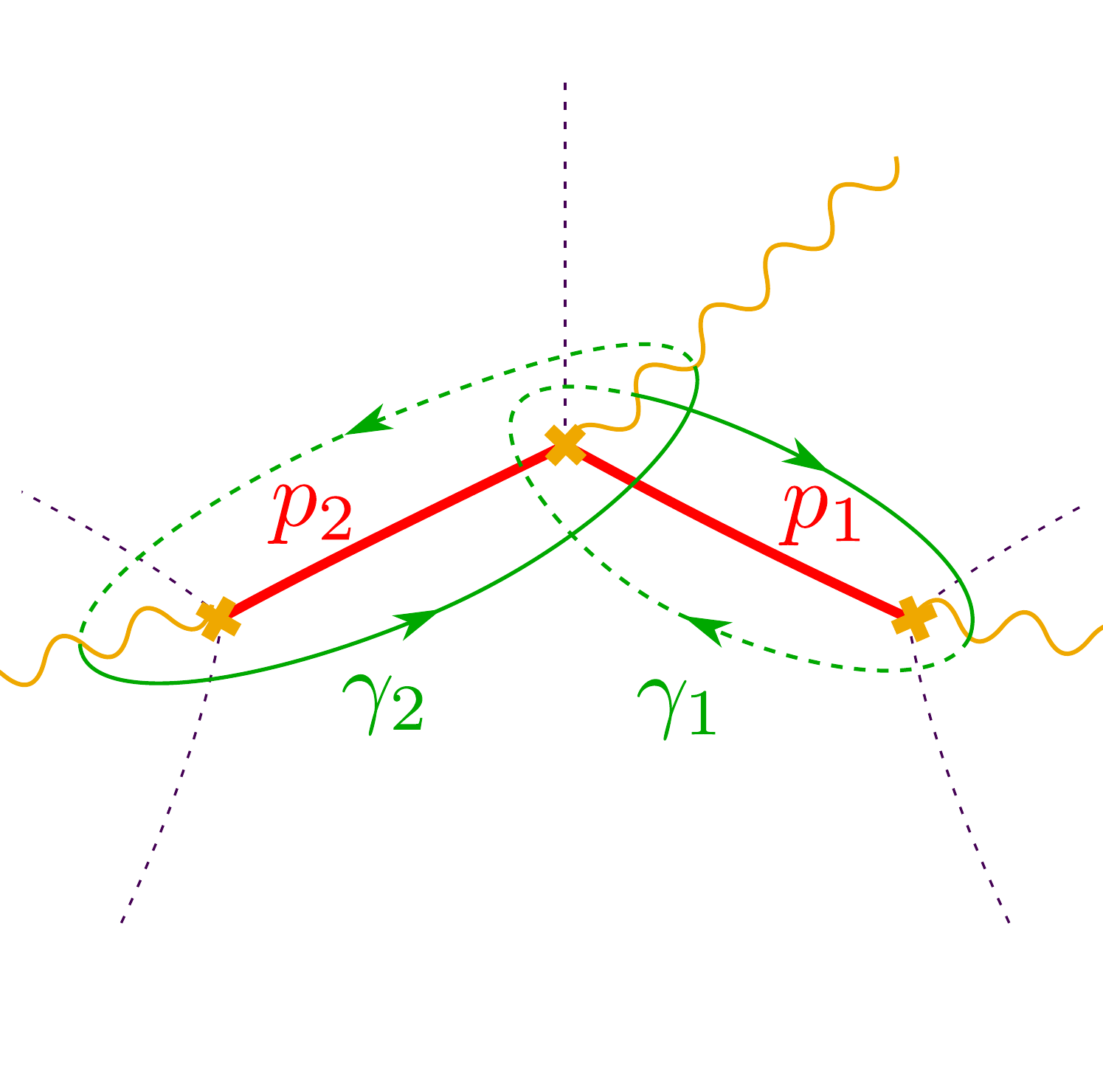}
\caption{
The critical graph $\CW_c$ for $AD_3$ theory, as obtained from the spectral network shown in Figure \ref{fig:AD3-MS}.
The graph has two edges and three nodes, 
each of them is a branch point. 
Homology cycles $\gamma_1$ and $\gamma_2$ are depicted in green. The covering $\Sigma\to C$ has two sheets (see \cite{Gaiotto:2009hg}), the solid/dashed green lines run on the upper/lower sheet respectively.  
}
\label{fig:AD3}
\end{center}
\end{figure}

We now wish to compute $Q^{(\pm)}(p_i)$ as defined in (\ref{eq:classical-Q-functions}), in particular we need to obtain the soliton data $\mu(a)$ for each of the two $\CS$-walls that make up $p_i$.
These can be obtained by an elementary application of the standard traffic rules of \cite{Gaiotto:2012rg} (collected for convenience in Appendix \ref{sec:traffic-rules}). 
In British resolution, the generating functions $\Upsilon(p_i)/ \Delta(p_i)$ of soliton data for up/down-going $\CS$-walls  are respectively 
\be\label{eq:AD3-up-down}
	\Upsilon(p_1) = X_{a_4} \,,\qquad %
	\Delta(p_1) = X_{a_3}\,,\qquad%
	\Upsilon(p_2) = X_{a_1} \,,\qquad %
	\Delta(p_2) = X_{a_2} + \Upsilon(p_1) \,,
\ee
where $a_1\in \tGamma_{ij}(z)$ is the relative homology in class with endpoints above $z\in p_2$, passing through the left-most ramification point, while $a_2\in \tGamma_{ji}(z)$ goes through the central ramification point; similarly for homology classes $a_3, a_4$ supported on $p_1$.
Note that in the expression for $\Delta(p_2)$ it is understood that one must ``parallel transport'' $a_4$ along street $p_1$, from the rightmost ramification point to the central one, to make it a homology class supported on $p_2$. Taking this into account we immediately obtain
\be
\begin{split}
	Q^{(+)}(p_1) & := 1+ \Upsilon(p_1)\Delta(p_1) =  1 +   X_{\tilde\gamma_1} \,,\\
	Q^{(+)}(p_2) & := 1+ \Upsilon(p_2)\Delta(p_2) =  1 + X_{\tilde\gamma_2} +  X_{\tilde\gamma_1+\tilde\gamma_2} \,,
\end{split}
\ee
where $\tilde\gamma_1$ (resp. $\tilde\gamma_2$) is the closure of the concatenation $a_3a_4$ (resp. $a_1 a_2$) in $\tGamma$.
The factors $Q^{(+)}_i$ defined in (\ref{eq:Q-pm-factorization}) are here $Q^{(+)}_1 = 1+X_{\tilde\gamma_1}$ and $Q^{(+)}_2 = 1 + X_{\tilde\gamma_2} +  X_{\tilde\gamma_1+\tilde\gamma_2}$, corresponding to $\alpha^+_{i}(p_j)=\delta_{ij}$.%
\footnote{A precise definition of $Q_i$ was given in Section \ref{sec:main-proof}.}
Then according to (\ref{eq:L-i-def}) the paths $L_i$ correspond to the basis homology cycles 
\be
	[L_i] = \left[\left(p_i\right)_\Sigma\right] = \tilde\gamma_i\,.
\ee
Now applying (\ref{eq:S-gamma-def}) gives, for example
\be\label{eq:AD3-S-gamma}
	S_{\gamma_{1}} = \left(Q_2^{(+)}\right)^{-1}\,,\qquad 
	S_{\gamma_{2}} = Q_1^{(+)}\,,
\ee
where we used $\langle\gamma_1,\gamma_2\rangle=1$, which can be checked in Figure \ref{fig:AD3}.

Since the BPS monodromy of this theory is well-known, let us compare it with our prediction.
Acting with
\be
\IS = \CK_{\gamma_1}\CK_{\gamma_2} = \CK_{\gamma_2}\CK_{\gamma_1+\gamma_2}\CK_{\gamma_1}\,
\ee
on the ring generators $X_{\tilde\gamma_i}$ gives
\be
	\IS\,( X_{\tilde\gamma_1} )= X_{\tilde\gamma_1}  (1+ X_{\tilde\gamma_2} (1+ X_{\tilde\gamma_1}))^{-1} ,\qquad  \IS\,( X_{\tilde\gamma_2} )= X_{\tilde\gamma_2}   (1+  X_{\tilde\gamma_1}) \,,
\ee
in agreement with (\ref{eq:AD3-S-gamma}), details of this computation can be found in Appendix \ref{sec:active-passive}.
Next we turn to the quantum monodromy $\mathbb{U}$.
We first compute the $Q^{(\pm)}(p,y)$ as described above (\ref{eq:motivic-Q-def}), a straightforward computation gives (all writhes vanish)
\be\label{eq:AD3-motivic-Q}
\begin{split}
	&Q^{(-)}(p_1,y) = 1+y^{-1}\hY_{\tilde\gamma_1}+y^{-1}\hY_{\tilde\gamma_1+\tilde\gamma_2}\\
	&Q^{(-)}(p_2,y) = 1+y^{-1}\hY_{\tilde\gamma_2} \\
	&Q^{(+)}(p_1,y) = 1+y^{-1}\hY_{\tilde\gamma_1}\\
	&Q^{(+)}(p_2,y) = 1+y^{-1}\hY_{\tilde\gamma_2}+y^{-1}\hY_{\tilde\gamma_1+\tilde\gamma_2} \,.
\end{split}
\ee
The BPS monodromy  $\mathbb{U}$ is determined by these through equation (\ref{eq:U-eqn}).
To show this explicitly, let us derive a factorization for $\mathbb{U}$ by applying formula (\ref{eq:a-m-eqn}). 
Using the filtration introduced in Section \ref{sec:solving-monodromy-eqs}, we split the generating functions into terms of levels $0,1,2$
\be
\begin{split}
	& Q_0^{(\pm)}(p_1) = 1 \,, \qquad Q_1^{(\pm)}(p_1) = y^{-1}\hY_{\tilde\gamma_1}  \,, \qquad Q_2^{(+)}(p_1) = 0\,,  \,\qquad  Q_2^{(-)}(p_1) =y^{-1} \hY_{\tilde\gamma_1+\tilde\gamma_2} \,,\\
	& Q_0^{(\pm)}(p_2) = 1 \,, \qquad Q_1^{(\pm)}(p_2) =y^{-1}\hY_{\tilde\gamma_2} \,, \qquad Q_2^{(+)}(p_2) = y^{-1}\hY_{\tilde\gamma_1+\tilde\gamma_2} \,,  \qquad Q_2^{(-)}(p_2) =  0\,,
\end{split}
\ee
giving $q_i(p_j) =y^{-1} \delta_{ij}$ for $i,j=1,2$.
From the generic form (\ref{eq:factorized-U}) and the expansion of quantum dilogarithms (see e.g. (\ref{eq:dilog-expansion})) it follows that $U_0 = 1$.
Moreover by definition $\hat U_1 = 0$, therefore 
\be
	\tilde R_2(p) = U_0 \, Q_2^{(-)}(p) - Q_2^{(+)}(p) \, U_0 = \left\{ \begin{array}{lc}
	y^{-1}\hY_{\tilde\gamma_1+\tilde\gamma_2} & (p=p_1) \\
	-y^{-1}\hY_{\tilde\gamma_1+\tilde\gamma_2} & (p=p_2) 
	\end{array}\right.
\ee
giving $\tilde r_{\tilde\gamma_1+\tilde\gamma_2}(p_1) = y^{-1} = -\tilde r_{\tilde\gamma_1+\tilde\gamma_2}(p_2)$.
Equation (\ref{eq:a-m-eqn}) becomes
\be
\begin{split}
	& 1 = \sum_{m}a_m(\gamma_2) (-y)^m \,\qquad  \text{for }\tilde\gamma = \tilde\gamma_1+\tilde\gamma_2\,, p=p_1\\
	& 1 = \sum_{m}a_m(\gamma_1) (-y)^m \,\qquad \text{for }\tilde\gamma = \tilde\gamma_1+\tilde\gamma_2\,, p=p_2
\end{split}
\ee
the unique solution is 
\be
	a_m(\gamma_1) =a_m(\gamma_2) = \delta_{m,0} \,.
\ee 
Going to the next level, we must compute $U_1$ and $\hat U_2$ from knowledge of the $a_m(\gamma_i)$. 
A choice of phase-ordering for the factorization of $\mathbb{U}$ must be made: choosing $\arg Z_{\gamma_1} > \arg Z_{\gamma_2}$ gives
\be
\begin{split}
	U_1 & = \frac{1}{y-y^{-1}} \left(\hY_{\tilde\gamma_1}+\hY_{\tilde\gamma_2}\right) \\
	\hat U_2 & = \frac{1}{(y-y^{-1})^2} \left(\frac{1}{1+y^2}\left(\hY_{2\tilde\gamma_1}+\hY_{2\tilde\gamma_2}\right)  +y \hY_{\tilde\gamma_1+\tilde\gamma_2} \right)  \\
\end{split}
\ee
from which it follows that $\tilde R_3(p_1)=\tilde R_3(p_2) = 0$ and therefore that $a_m(\gamma) = 0$ for all $|\gamma|=2$.
Likewise, proceeding to higher orders in $|\gamma|$ gives vanishing $a_m(\gamma)$, recovering  the expected factorization 
\be\label{eq:AD3-strong-coupling-motivic}
	\mathbb{U} = \Phi(\hY_{\tilde\gamma_1}) \Phi(\hY_{\tilde\gamma_2})\,.
\ee

An equivalently good choice for central charges would have been to pick $\arg Z_{\gamma_1} < \arg Z_{\gamma_2}$, in this case we obtain
\be
	\hat U_2 = \frac{y}{(1-y^2)^2} \left(\frac{y}{1+y^2}\left(\hY_{2\tilde\gamma_1}+\hY_{2\tilde\gamma_2}\right)  + \hY_{\tilde\gamma_1+\tilde\gamma_2} \right)  
\ee
which gives
\be
	\tilde R_3(p_1) = y^{-1}\hY_{2\tilde\gamma_1+\tilde\gamma_2} \qquad \tilde R_3(p_2) = -y^{-1}\hY_{\tilde\gamma_1+2\tilde\gamma_2} \,.
\ee
The only nonzero coefficients are then $\tilde r_{2\tilde\gamma_1+\tilde\gamma_2}(p_1) = -\tilde r_{\tilde\gamma_1+2\tilde\gamma_2}(p_2) = y^{-1} $, plugging these into (\ref{eq:a-m-eqn}) yields
\be
\begin{split}
	& 1 = \sum_{m}a_m(\gamma_1+\gamma_2) (-y)^m \,\qquad  \text{for }\gamma = 2\gamma_1+\gamma_2,\, p=p_1\\
	& 1 = \sum_{m}a_m(\gamma_2+\gamma_2) (-y)^m \,\qquad \text{for }\gamma = \gamma_1+2\gamma_2,\, p=p_2
\end{split}
\ee
whose unique solution is
\be
	a_m(\gamma_1+\gamma_2) = \delta_{m,0}\,.
\ee
Higher levels with $|\gamma|>2$ simply give $a_m(\gamma)=0$. We have thus recovered the other side of the pentagon identity 
\be
	\mathbb{U} = \Phi(\hY_{\tilde\gamma_2})\Phi(\hY_{\tilde\gamma_1+\tilde\gamma_2})\Phi(\hY_{\tilde\gamma_1})\,.
\ee

\subsection{SU(2) SYM}
\begin{figure}[h!]
\begin{center}
\includegraphics[width=0.35\textwidth]{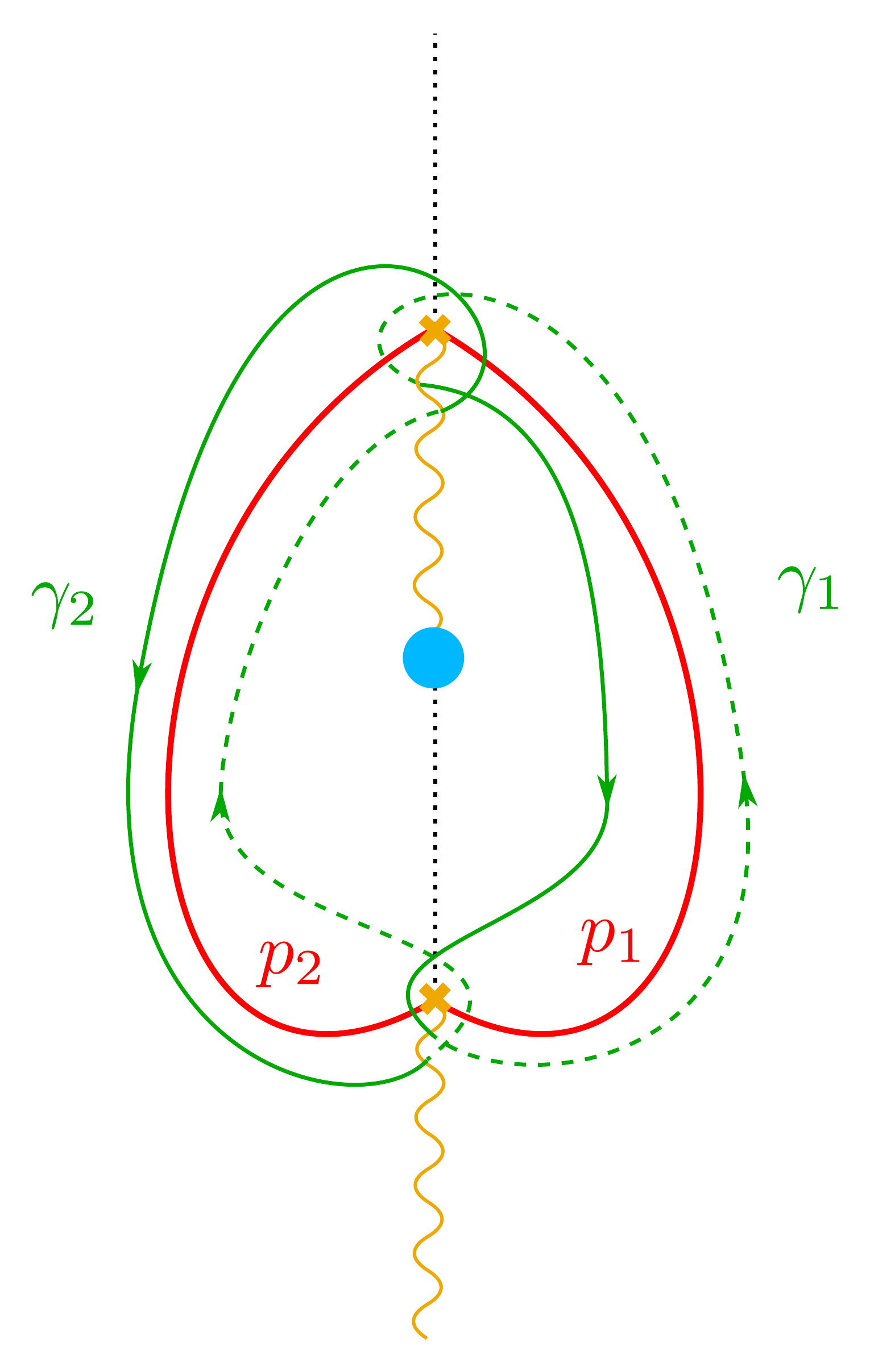}
\caption{
The critical graph $\CW_c$ for SU(2) SYM theory.
The two edges of the graph are depicted in red, the two nodes corresponding to branch points are marked by orange crosses. The light blue disc marks an irregular puncture.
Homology cycles $\gamma_1$ and $\gamma_2$ are depicted in green, solid/dashed green lines run on the upper/lower sheet of the covering $\Sigma\to C$.
Black dotted lines denote ordinary $\CS$-walls, which are not part of the critical graph.
}
\label{fig:SU2}
\end{center}
\end{figure}

The Coulomb branch of this theory bears similarities to that of the previous example, as it is one-dimensional and divided in two regions by a pair of walls of marginal stability. 
The critical locus $\CB_c$ corresponds to the union of these walls, the critical spectral network for generic $u_c\in\CB_c$ is shown in Figure \ref{fig:SU2}.
The critical graph $\CW_c$ consists of two 2-way streets $p_1$ and $p_2$, each of these lifts to a closed homology cycle  $\tilde\gamma_i\in\tGamma$, with intersection $\langle \tilde\gamma_1,\tilde \gamma_2\rangle = 2$.
The symmetry group of $\CW_c$ is trivial, because exchanging $p_1, p_2$ would not preserve the cyclic ordering at the branch points.

Let $a_1$, $a_2$ the solitons emanating from the lower ramification point, and stretching upwards along streets $p_1, p_2$ respectively. Similarly let $b_1, b_2$ be the solitons sourced by the upper ramification point.
The soliton generating functions $\Upsilon,\Delta$  of upwards/downwards solitons can readily be computed from the standard traffic rules (see Appendix \ref{sec:traffic-rules}). 
In British resolution we find
\be
\begin{split}
	\Upsilon(p_1) = X_{a_1} \,,\quad \Delta(p_1) = X_{b_1} \,, \quad \Upsilon(p_2) = X_{a_2}+ \Delta(p_1) \,, \quad \Delta(p_2) = X_{b_2} + \Upsilon(p_1)\,.
\end{split}
\ee
It's straightforward to solve these equations to obtain $\Upsilon,\Delta$ as functions of $X_{a_i},X_{b_i}, X_{\tilde\gamma_i}$, which immediately yield
\footnote{As in the previous section, due care must be taken in parallel transporting solitons across streets. This explains the extra factors of $\gamma_1$ and $\gamma_1+\gamma_2$ in some of the monomials.}
\be
	Q^{(+)}(p_1) = 1+X_{\tilde\gamma_1}\,,\qquad Q^{(+)}(p_2) = 1+X_{\tilde\gamma_2}(1+X_{\tilde\gamma_1})^2\,.
\ee
These characterize the BPS monodromy, in fact applying formula (\ref{eq:S-gamma-def}) we can compute the action of $\IS$ on the ring generators:
\be\label{eq:classical-SU2-monodromy}
\begin{split}
	S_{\gamma_1} & = Q(p_2)^{-2} = (1+X_{\tilde\gamma_2}(1+X_{\tilde\gamma_1})^2)^{-2}   \\
	S_{\gamma_2} & = Q(p_1)^{2} = (1+X_{\tilde\gamma_1})^2\,.
\end{split}
\ee
As a check, let us compare with the well-known monodromy operator $\IS=\CK_{\gamma_1}\CK_{\gamma_2}$, whose action is  
\be
\begin{split}
	\CK_{\gamma_1} \CK_{\gamma_2} \hat X_{\gamma_1} & = \CK_{\gamma_1} \, \hat X_{\gamma_1} (1- \hat X_{\gamma_2})^{-2} = \hat X_{\gamma_1} (1- \hat X_{\gamma_2}(1- \hat X_{\gamma_1})^2)^{-2}  \\
	\CK_{\gamma_1} \CK_{\gamma_2} \hat X_{\gamma_2} & = \CK_{\gamma_1} \, \hat X_{\gamma_2}  = \hat X_{\gamma_2} (1- \hat X_{\gamma_1})^2  \,.
\end{split}
\ee
Switching to twisted variables $\hX_{\gamma_i}$ in (\ref{eq:classical-SU2-monodromy}) (with $\rho(\tilde\gamma_i)=-1$ in (\ref{graph:variables-relations})) we find exact agreement.

Turning to the quantum monodromy, a simple computation gives
\be\label{eq:SU2-motivic-Q}
\begin{split}
	& Q^{(-)}(p_1,y) = 1+y^{-1}\hY_{\tilde\gamma_1}+(1+y^{-2}) \hY_{\tilde\gamma_1+\tilde\gamma_2} + y^{-1}\hY_{\tilde\gamma_1+2\tilde\gamma_2} \,,\\
	& Q^{(-)}(p_2,y) = 1+y^{-1}\hY_{\tilde\gamma_2}\,,\\
	& Q^{(+)}(p_1,y) = 1+y^{-1}\hY_{\tilde\gamma_1} \,,\\
	& Q^{(+)}(p_2,y) = 1+y^{-1}\hY_{\tilde\gamma_2} +(1+y^{-2}) \hY_{\tilde\gamma_1+\tilde\gamma_2} +y^{-1} \hY_{2\tilde\gamma_1+\tilde\gamma_2}\,,
\end{split}
\ee
which completely characterize $\mathbb{U}$.
In fact, using (\ref{eq:a-m-eqn}) to derive a factorization of $\mathbb{U}$ for $\arg Z_{\gamma_1}>\arg Z_{\gamma_2}$ yields the well known expression
\be\label{eq:SU2-motivic-U}
	\Phi(\hY_{\tilde\gamma_1})\Phi(\hY_{\tilde\gamma_2})\,.
\ee

\subsection{Wild Wall Crossing}\label{sec:wwc}

\begin{figure}[h!]
\begin{center}
\includegraphics[width=0.55\textwidth]{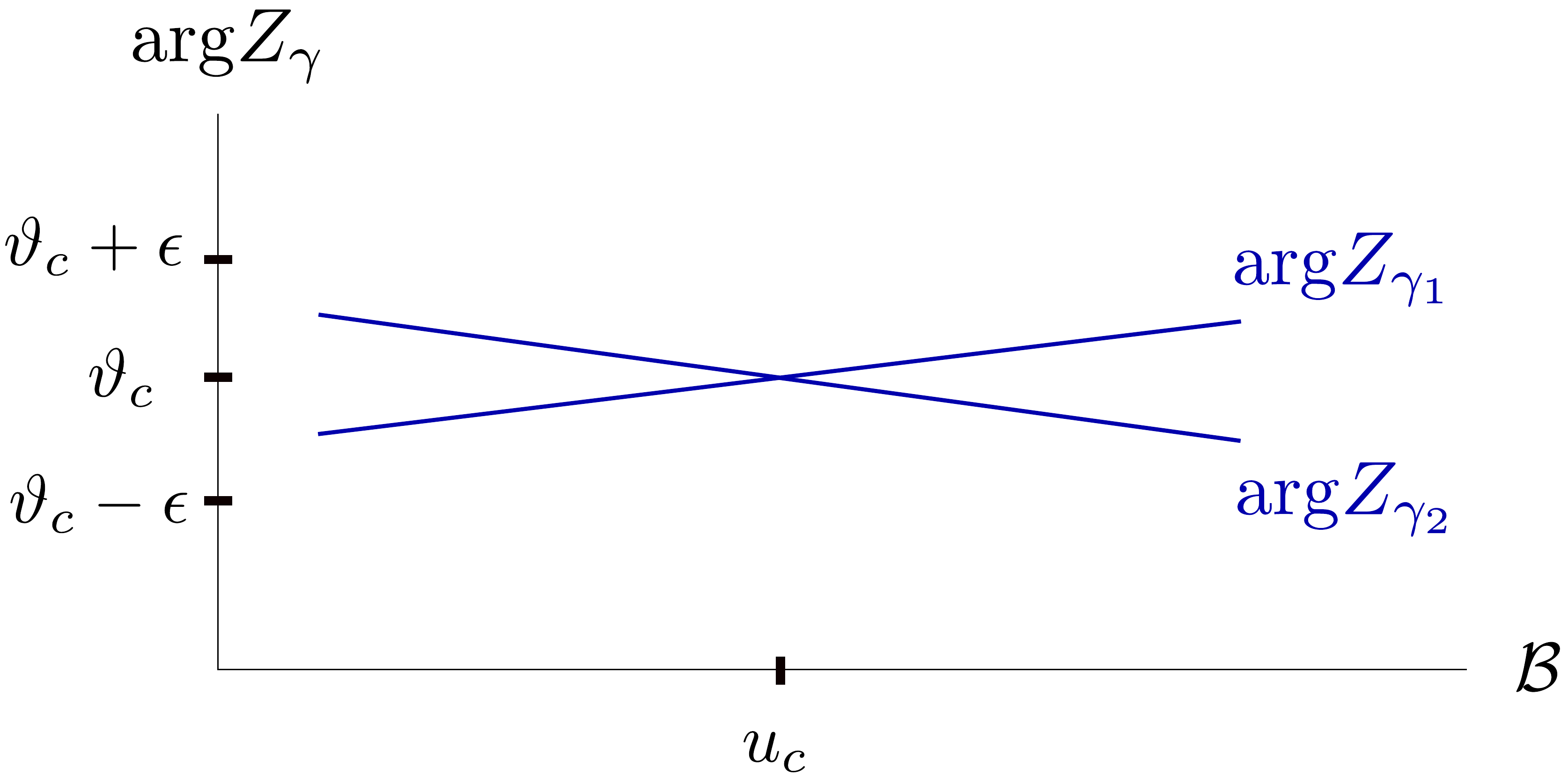}
\caption{
Schematic picture of phases of central charges $Z_{\gamma_1}, Z_{\gamma_2}$ near a point $u_c$ on the MS wall of $\gamma_1$ and $\gamma_2$.
}
\label{fig:m-herd-phases}
\end{center}
\end{figure}

Up to now we have applied our formalism to the computation of whole BPS monodromies, however the same ideas readily generalize to sub-sectors of the BPS spectrum.
Let us consider a generic Coulomb branch $\CB$ with a wall of marginal stability $MS(\gamma_1,\gamma_2)$.
Choosing $u_c$ on the wall causes the central charges $Z_{\gamma_1}, Z_{\gamma_2}$ to align in the complex plane, on the ray of phase $e^{i\vartheta_c}$. At the same time, by a genericity assumption, the central charges of all other BPS states must have phases located at finite distance from $\vartheta_c$.
Within an arbitrarily small neighborhood of $u_c$ there is an angular sector $A = (\vartheta_c-\epsilon, \vartheta_c+\epsilon)$ containing both $\arg Z_{\gamma_1}, \arg Z_{\gamma_2}$, as illustrated in Figure \ref{fig:m-herd-phases}

We wish to study the jump in the spectrum of framed 2d-4d degeneracies by comparing $F(\wp,\vartheta_c\pm\epsilon,u_c)$. 
This is described by a natural generalization of (\ref{eq:gen-K-wall-formula-F})
\be
	F(\wp,\vartheta_c +\epsilon,u_c) = \IA \big( F(\wp, \vartheta_c-\epsilon,u_c) \big)\,,
\ee
where 
\be
	\IA = \prod^{\nwarrow}_{n_1,n_2\geq 0} \CK^{\Omega(n_1\gamma_1+n_2\gamma_2)}_{n_1\gamma_a+n_2\gamma_2}\,.
\ee
The frameworks developed in Sections \ref{sec:classical-monodromy-equations} and \ref{sec:quantum-monodromy-equations} can be applied \emph{mutatis mutandis} to the computation of $\IA$, and of its ``quantum'' counterpart. The main difference is that, instead of a full-rank half-lattice for $\Gamma$, we will instead restrict to the slice generated by $\gamma_1,\gamma_2$.
The critical graph $\CW_c$ will be replaced by the sub-network of 2-way streets within $\CW(\vartheta_c,u_c)$.

\begin{figure}[h!]
\begin{center}
\includegraphics[width=0.25\textwidth]{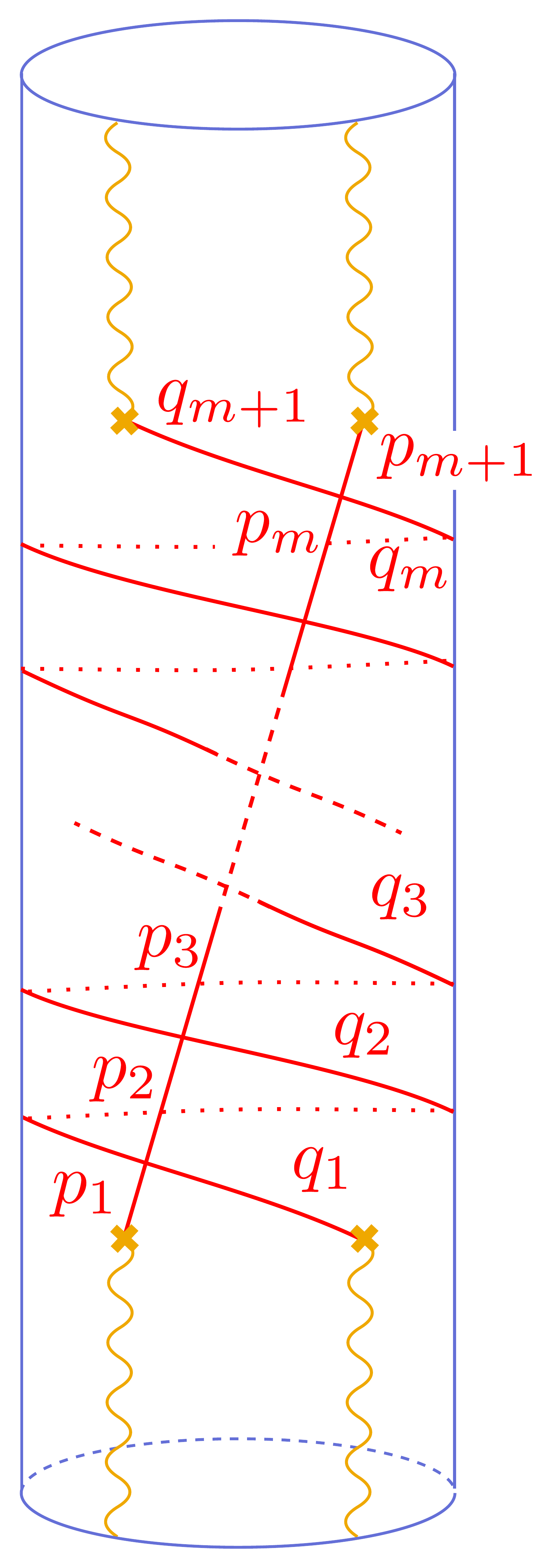}
\caption{A degenerate limit of an $m$-herd network, corresponding to $\CW_c$ on the wall of marginal stability of two charges with intersection pairing $m$. The network wraps a cylinder, 2-way streets are depicted in red, regular $\CS$-walls are omitted.}
\label{fig:wild-m}
\end{center}
\end{figure}

As an interesting toy example, consider two charges with pairing $\langle\gamma_1,\gamma_2\rangle = m > 0$. 
A possible critical graph for this example is the degenerate limit of an "$m$-Herd" shown in Figure \ref{fig:wild-m} (see also Figure 2 of \cite{Galakhov:2013oja}). 

The graph consists of $2m+2$ edges, four branch point nodes and $m$ joint nodes.
Working in British resolution, the $r$-th joint node yields the following equations for the edges attached to it
\be\label{eq:m-herd-up-down}
\begin{split}
	\Delta(p_r) & = \Delta(p_{r+1}) + \Delta(p_{r+1}) \Delta(q_{r+1}) \Upsilon(q_r) \\
	\Delta(q_r) & = \Delta(q_{r+1}) \\
	\Upsilon(p_r) & = \Upsilon(p_{r-1}) + \Delta(q_{r})  \Upsilon(q_{r-1}) \Upsilon(p_{r-1}) \\
	 \Upsilon(q_r)&=  \Upsilon(q_{r-1})\,,
\end{split}
\ee
where we denoted as usual by $\Upsilon$ and $\Delta$ the generating functions of upwards/downwards-flowing solitons on each street.
This set of equations is supplemented by boundary conditions, provided by the branch point nodes. Each branch point contributes an equation for the attached terminal street
\be
	\Upsilon(p_1) =  X_{a_1}\,,\qquad %
	\Upsilon(q_1) =  X_{a_2}\,,\qquad %
	\Delta(p_{m+1}) =  X_{a_3}\,,\qquad %
	\Delta(q_{m+1}) =  X_{a_4}\,. %
\ee
Homology classes $\tilde\gamma_1,\tilde\gamma_2$ arise from concatenations of soliton paths as follows
\be
	cl(a_1 a_3) =\left[ \sum_{r=1}^{m+1}{\pi^{-1}(p_r)}\right]= \tilde\gamma_2\qquad %
	cl(a_2 a_4) =\left[ \sum_{r=1}^{m+1}{\pi^{-1}(q_r)} \right]= \tilde\gamma_1\,,
\ee 
where it is understood that $a_1, a_3$ (respectively $a_2, a_4$) are joined upon being transported along the lifts of streets $p_j$ (respectively $q_j$).
Equations (\ref{eq:m-herd-up-down}) have a simple unique solution
\be
\begin{split}
	\Upsilon(p_r) & =  X_{a_1}(1+ X_{\tilde\gamma_1})^{r-1}\\
	\Upsilon(q_r) & =  X_{a_2}\\
	\Delta(p_r) & =  X_{a_3}(1+ X_{\tilde\gamma_1})^{m+1-r}\\
	\Delta(q_r) & =  X_{a_4}\,.
\end{split}
\ee
From this we obtain the 2-way street generating functions
\be
	Q^{(+)}(p_r) = 1+ X_{\tilde\gamma_2}(1+  X_{\tilde\gamma_1})^m\qquad Q^{(+)}(q_r) = 1+ X_{\tilde\gamma_1}\qquad r=1,\dots,m+1\,.
\ee
Identifying factors (\ref{eq:Q-pm-factorization}) with $Q^{(+)}_1 = 1+X_{\tilde\gamma_1}$ and $Q^{(+)}_2 = 1+ X_{\tilde\gamma_2}(1+  X_{\tilde\gamma_1})^m$ yields
\be
	\alpha_1^{+}(q_r) = 1\,,\qquad \alpha_2^{+}(q_r) = 0\,,\qquad \alpha_1^{+}(p_r) = 0\,,\qquad \alpha_2^{+}(p_r) = 1\,. 
\ee
Hence we have $[L_1]=\tilde\gamma_1$ and $[L_2]=\tilde\gamma_2$, and applying (\ref{eq:gen-K-wall-formula}) we find the action of the classical BPS monodromy to be
\be\label{eq:m-herd-S-gamma}
	S_{\gamma_1} =  \left(Q^{(+)}_2\right)^{-m} \,,\qquad 
	S_{\gamma_2} = \left(Q^{(+)}_1\right)^{m}\,.
\ee
Indeed the operator $\IA$ in this example is  $\IA = \CK_{\gamma_1}\CK_{\gamma_2}$, it acts on the ring generators as
\be
\begin{split}
	\IA  \left(\hX_{\gamma_1} \right) & = \hX_{\gamma_1}  (1-\hX_{\gamma_2} (1-\hat X_{\gamma_1})^m )^{-m}\\
	\IA  \left(\hX_{\gamma_2}\right)   & =\hX_{\gamma_2} (1-\hX_{\gamma_1})^m \,.
\end{split}
\ee
Upon switching to twisted variables $\hX_{\gamma_i} $ in (\ref{eq:m-herd-S-gamma}) we find perfect agreement.

Turning to the quantum monodromy, an explicit computation gives\footnote{We checked these expressions explicitly for $m=3,4,5$ using \cite{python-code}.}
\be\label{eq:wwc-motivic-Q}
\begin{split}
	Q^{(-)}(p_r,y) &= 1+y^{-1}\hY_{\tilde\gamma_2} \,,\\
	Q^{(-)}(q_r,y) &= 1+y^{-1}\sum_{k=0}^{m} \left[\begin{array}{c}m\\k\end{array}\right]_y \,\hY_{\tilde\gamma_1+k\tilde\gamma_2} \,,\\
	Q^{(+)}(p_r,y) &= 1+y^{-1}\sum_{k=0}^{m} \left[\begin{array}{c}m\\k\end{array}\right]_y \,\hY_{k \tilde\gamma_1+\tilde\gamma_2} \,,\\
	Q^{(+)}(q_r,y) &= 1+y^{-1}\hY_{\tilde\gamma_1} \,.
\end{split}
\ee
where 
\be
	\left[\begin{array}{c}m\\k\end{array}\right]_y := \prod_{j=1}^{k-1}\frac{y^{m-j}-y^{-m+j}}{y^{j+1}-y^{-j-1}}\,
\ee
is the symmetrized $q$-binomial.
Using formula (\ref{eq:a-m-eqn}) with $\arg Z_{\gamma_1}>\arg Z_{\gamma_2}$ we recover the expected factorization \cite{Galakhov:2013oja}
\be\label{eq:wild-motivic-U}
	\Phi(\hY_{\tilde\gamma_1})\Phi(\hY_{\tilde\gamma_2})\,.
\ee

%
%
%

\subsection{$AD_k$ Argyres-Douglas Theories}
\begin{figure}[h!]
\begin{center}
\includegraphics[width=0.3\textwidth]{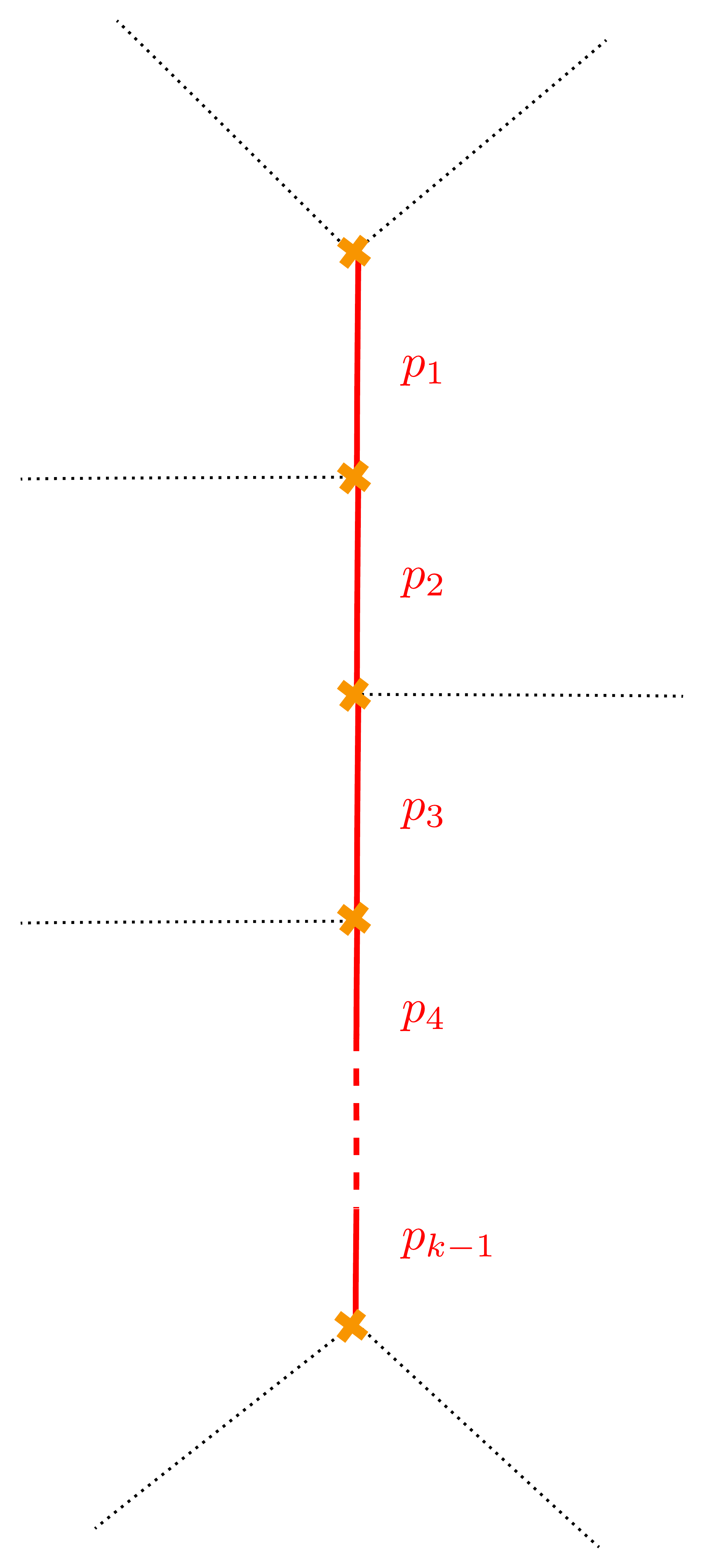}
\caption{
The critical graph of $AD_k$ theory includes $k-1$ edges, depicted in red. All nodes are branch points, marked by crosses. Dotted black lines are regular $\CS$-walls, which are not part of the critical graph.
}
\label{fig:ADk}
\end{center}
\end{figure}

$AD_k$ Argyres-Douglas theories are engineered as class $\CS$ theories of type $A_1$ by taking $C$ to be a sphere with an irregular singularity at infinity, the label $k$ is related to the degree of the singularity \cite{Gaiotto:2009hg, Xie:2012hs}. 
The critical spectral network $\CW(\vartheta_c,u_c)$ is shown in Figure \ref{fig:ADk}, it contains $k-1$ edges and $k$ nodes of branch point type.

For each street $p_i$, let $a_i$ be the solitons sourced from the branch point attached to the lower end of $p_i$ and stretching upwards. Likewise $b_i$ will denote the downards-stretching soliton sourced at the upper branch point. 
Each $p_i$ admits a canonical oriented lift to a closed path on $\tSigma$, whose homology class we denote by $\tilde\gamma_i$ (we defined the canonical lift below (\ref{eq:GMN-L})).
The only non-zero intersections of these cycles are  $\langle\tilde \gamma_{2n-1}, \tilde\gamma_{2n}\rangle = \langle\tilde\gamma_{2n+1} , \tilde\gamma_{2n}\rangle = 1$.
Applying the standard traffic rules (collected in Appendix \ref{sec:traffic-rules}) in British resolution, we obtain the following equations for the generating functions $\Upsilon/\Delta$ of upwards/downwards-flowing solitons
\be\label{eq:ADk-up-down}
\begin{array}{ll}
	\Upsilon(p_{2n+1}) = X_{a_{2n+1}}  \, \qquad & \Delta(p_{2n+1}) = X_{b_{2n+1}}  \\
	\Upsilon(p_{2n}) = X_{a_{2n}} + \Upsilon_{2n+1}  \, \qquad \qquad & \Delta(p_{2n}) = X_{b_{2n}} + \Delta_{2n-1}  
\end{array}\,\qquad %
n=0,1,2,\dots
\ee
As usual, due care must be taken in parallel transporting solitons across streets to properly make sense of these equations. Concretely, an effective way of taking this into account is to use equivalences $a_{m+1} \sim  \gamma_{m+1}+a_{m}$ and $b_{m-1} \sim  \gamma_{m-1}+b_{m}$. 
Equations (\ref{eq:ADk-up-down}) are then easily solved, and we obtain the 2-way street generating functions 
\be
\begin{split}
	Q^{(+)}(p_{2n}) & = 1+ X_{\tilde\gamma_{2n}}(1+ X_{\tilde\gamma_{2n-1}})(1+ X_{\tilde\gamma_{2n+1}}) \,,\\
	Q^{(+)}(p_{2n+1}) & = 1+ X_{\tilde\gamma_{2n+1}}\,.
\end{split}
\ee
The factors defined in (\ref{eq:Q-pm-factorization}) are identified with $Q_i^{} \equiv Q^{(+)}(p_i)$, therefore 
\be
	\alpha_i(p_j) = \delta_{ij}\,.
\ee
The corresponding homology cycles then correspond directly to lifts of each street
\be
	\left[L_i\right] = \left[\left(p_i\right)_\Sigma\right] = \tilde\gamma_i\,,
\ee
this completes the characterization of the classical monodromy.

Applying formula (\ref{eq:gen-K-wall-formula}) we find the action of $\mathbb{S}$ on $X_{\tilde\gamma_{2n}},X_{\tilde\gamma_{2n+1}}$:
\be
\begin{split}
	S_{\gamma_{2n}} & = Q_{2n-1}^{\langle\gamma_{2n-1},\gamma_{2n}\rangle}  Q_{2n+1}^{\langle\gamma_{2n+1},\gamma_{2n}\rangle} \\
	& = (1+ X_{\tilde\gamma_{2n-1}})  (1+ X_{\tilde\gamma_{2n+1}}) \\
	S_{\gamma_{2n+1}} & = Q_{2n}^{\langle\gamma_{2n},\gamma_{2n+1}\rangle}  Q_{2n+2}^{\langle\gamma_{2n+2},\gamma_{2n+1}\rangle} \\
	& = \left[  1+ X_{\tilde\gamma_{2n}}(1+ X_{\tilde\gamma_{2n-1}})(1+ X_{\tilde\gamma_{2n+1}})   \right]^{-1} \, \left[  1+ X_{\tilde\gamma_{2n+2}}(1+ X_{\tilde\gamma_{2n+1}})(1+ X_{\tilde\gamma_{2n+3}})   \right]^{-1}
\end{split}
\ee
except for
\be
\begin{split}
	S_{\gamma_1} & = \left[  1+ X_{\tilde\gamma_{2}}(1+ X_{\tilde\gamma_{1}})(1+ X_{\tilde\gamma_{3}})   \right]^{-1} \\
	S_{\gamma_{k-1}} & = \left\{ \begin{array}{lr} 
	\left[  1+ X_{\tilde\gamma_{k-2}}(1+ X_{\tilde\gamma_{k-3}})(1+ X_{\tilde\gamma_{k-1}})   \right]^{-1} & (k\in2\IZ) \\
	1+X_{\tilde\gamma_{k-2}}& (k\in 2\IZ+1)
	\end{array}\right.
\end{split}
\ee
It can be checked that these agree in fact with the action of
\be
	\IS = \left(\prod_{\text{odd}} \, \CK_{\gamma_{2n+1}}\right) \,  \left(\prod_{\text{even}} \, \CK_{\gamma_{2n}}\right)\,.
\ee

Turning to the quantum monodromy, it's easy to compute by hand the generating functions
\be
\begin{split}
        Q^{(-)}(p_{2n+1},y) & = 1+y^{-1}Y_{\tilde\gamma_{2n+1}} +y^{-1}Y_{\tilde\gamma_{2n+1}+\tilde\gamma_{2n}}+y^{-1}Y_{\tilde\gamma_{2n+1}+\tilde\gamma_{2n+2}}+y^{-1}Y_{\tilde\gamma_{2n+1}+\tilde\gamma_{2n}+\tilde\gamma_{2n+2}} \\
	Q^{(-)}(p_{2n},y) & = 1+y^{-1}Y_{\tilde\gamma_{2n}}\,\\
	Q^{(+)}(p_{2n+1},y) & = 1+y^{-1}Y_{\tilde\gamma_{2n+1}} \\
	Q^{(+)}(p_{2n},y) & = 1+y^{-1}Y_{\tilde\gamma_{2n}}+y^{-1}Y_{\tilde\gamma_{2n}+\tilde\gamma_{2n-1}}+y^{-1}Y_{\tilde\gamma_{2n}+\tilde\gamma_{2n+1}}+y^{-1}Y_{\tilde\gamma_{2n}+\tilde\gamma_{2n+1}+\tilde\gamma_{2n-1}} \,
\end{split}
\ee
which determine $\mathbb{U}$ through relations (\ref{eq:U-eqn}). 
A factorization can be given using formula (\ref{eq:a-m-eqn}), for example choosing all even/odd-indexed charges to have same central charges respectively, with $\arg(Z_{\text{odd}})>\arg(Z_{\text{even}})$, we find
\be
	\mathbb{U} = \left(\prod_{\text{odd}} \, \Phi(Y_{\tilde\gamma_{2n+1}})\right) \,  \left(\prod_{\text{even}} \, \Phi(Y_{\tilde\gamma_{2n}})\right)\,.
\ee

\subsection{$SU(2)$ $\CN=2^*$ theory}
\label{sec:N2-star}

\begin{figure}[h!]
\begin{center}
\includegraphics[width=0.33\textwidth]{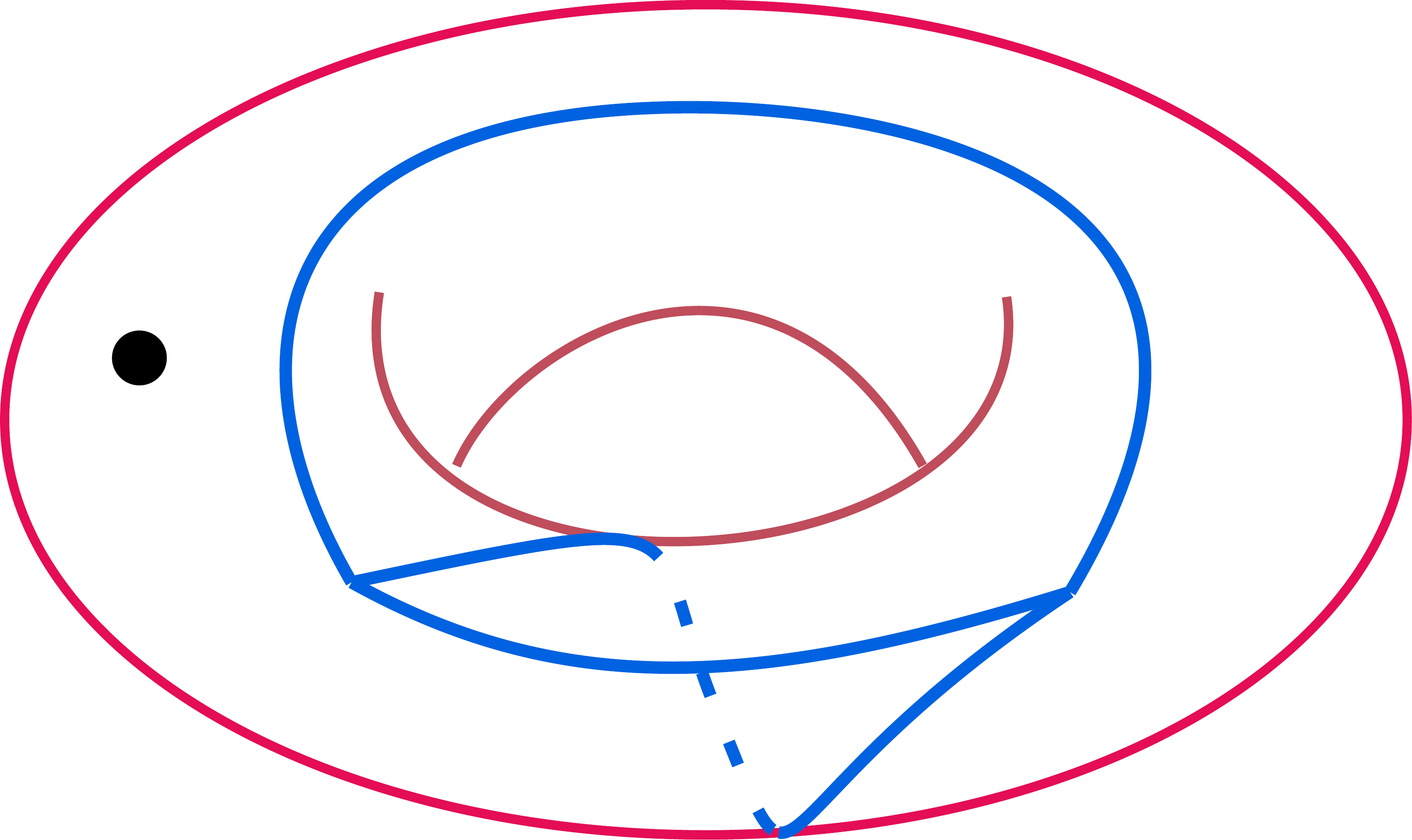}
\hspace{30pt}
\includegraphics[width=0.28\textwidth]{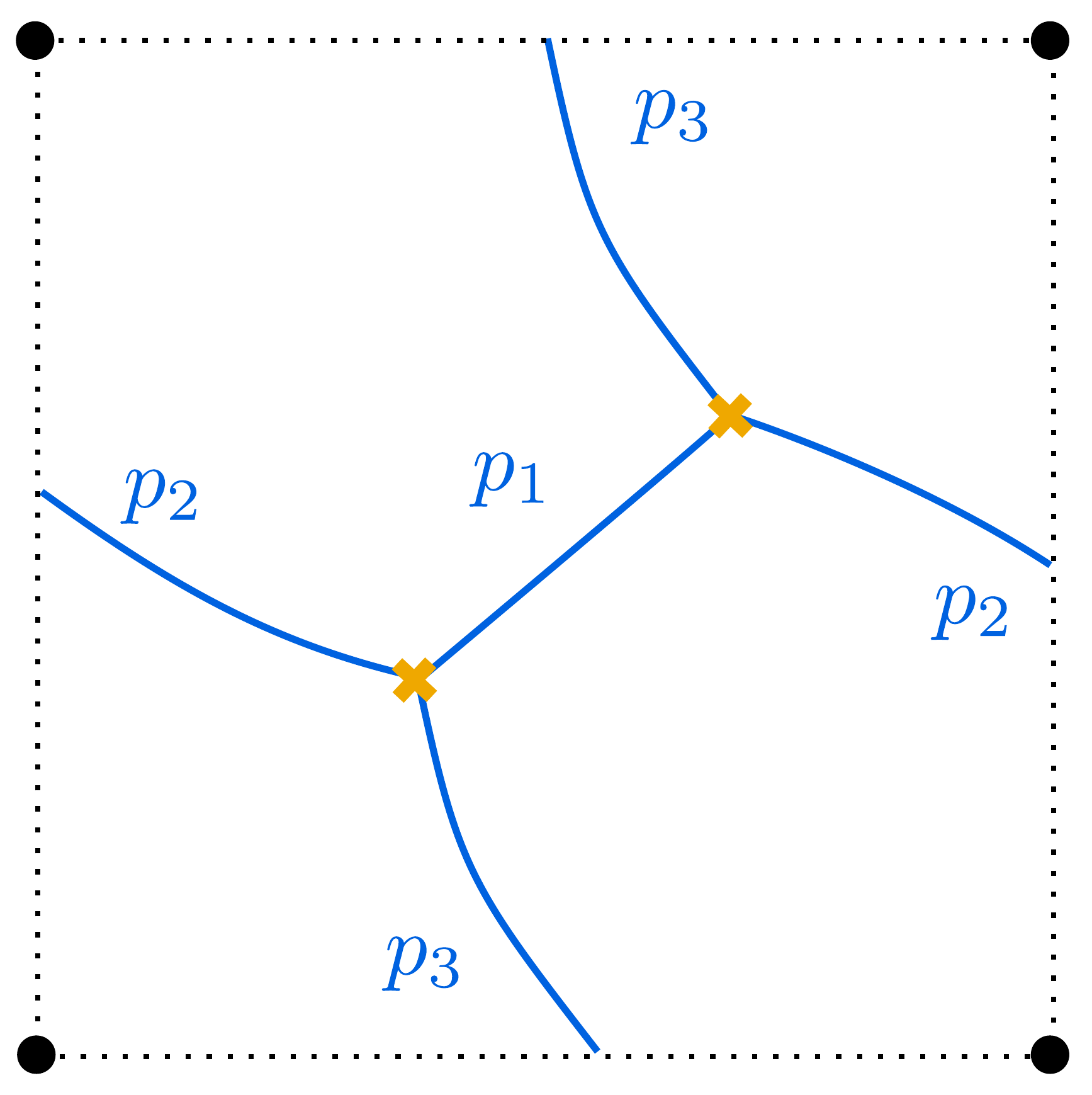}
\caption{The critical graph for SU(2) $\CN=2^*$ theory contains 3 edges and two nodes, each is a branch point. On the right: labels and up/down orientations of streets, as employed in computations. The graph has a manifest $\IZ_3$ symmetry, which is inherited by the soliton generating functions.}
\label{fig:N2star}
\end{center}
\end{figure}

This theory can be engineered as a class $\CS$ theory of $A_1$ type, with $C$ a punctured torus with a regular singularity \cite{Gaiotto:2009we, Gaiotto:2009hg}.
The critical graph $\CW_c$ is shown in Figure \ref{fig:N2star}, it consists of three two-way streets $p_1, p_2, p_3$ connecting two branch points. Note that $\CW_c$ has a $\IZ_3$ cyclic symmetry.
This $\IZ_3$ preserves both the topology and the cyclic ordering of edges at the nodes, therefore it is an honest symmetry, as we will see it is realized on the generating functions that determine the monodromy.

Let $a_i$ be the solitons sourced at the NE (top-right) branch point on streets $p_i$, and $b_i$ the solitons sourced at the SW (bottom-left) branch point.
As usual  $\Upsilon_i,\Delta_i$ denote the generating functions of $\CS$-wall soliton data for walls  running upwards/downwards on each street.

In British resolution, the network equations are determined by the two branch points
\be\label{eq:N2-star-equations}
\begin{array}{c}
NE \\
\hline
\Delta_1 = X_{a_1} + \Upsilon_2 \\
\Delta_2 = X_{a_2} + \Delta_3 \\
\Upsilon_3 = X_{a_3} + \Upsilon_1 \\
\end{array} \qquad %
\begin{array}{c}
SW \\
\hline
\Upsilon_1 = X_{b_1} + \Delta_2 \\
\Upsilon_2 = X_{b_2} + \Upsilon_3 \\
\Delta_3 = X_{b_3} + \Delta_1 \\
\end{array}  
\ee
In solving these equations, it is crucial to keep track of the parallel transport of solitons along streets of the network, as the identities are repeatedly applied. 
This is handled easily by using the techniques of Appendix \ref{sec:soliton-computations}, where the details of this computation can be found. The result is 
\be
\begin{split}
	\Delta_1 & = X_{a_1} \frac{1+X_{\tilde\gamma_2}+X_{\tilde\gamma_2+\tilde\gamma_3} +X_{\tilde\gamma_1+\tilde\gamma_2+\tilde\gamma_3}+X_{\tilde\gamma_1+2\tilde\gamma_2+\tilde\gamma_3} +X_{\tilde\gamma_1+2\tilde\gamma_2+2\tilde\gamma_3}}{1 - X_{2\tilde\gamma_1+2\tilde\gamma_2+2\tilde\gamma_3}} \\
	\Upsilon_1 & = X_{b_1} \frac{1+X_{\tilde\gamma_2}+X_{\tilde\gamma_2+\tilde\gamma_3} +X_{\tilde\gamma_1+\tilde\gamma_2+\tilde\gamma_3}+X_{\tilde\gamma_1+2\tilde\gamma_2+\tilde\gamma_3} +X_{\tilde\gamma_1+2\tilde\gamma_2+2\tilde\gamma_3}}{1 - X_{2\tilde\gamma_1+2\tilde\gamma_2+2\tilde\gamma_3}} 
\end{split}
\ee
with similar expressions for $\Delta_{2},\Delta_{3}, \Upsilon_{2}, \Upsilon_{3}$ obtained by cyclic permutations of the indices. Here $\tilde\gamma_i\in\tGamma$ denote the homology classes of canonical lifts of streets $p_i$.
The 2-way street generating functions are therefore
\be
\begin{split}
	Q^{(+)}(p_1) & = 1 + X_{\tilde\gamma_1} \left(  \frac{1+X_{\tilde\gamma_2}+X_{\tilde\gamma_2+\tilde\gamma_3} +X_{\tilde\gamma_1+\tilde\gamma_2+\tilde\gamma_3}+X_{\tilde\gamma_1+2\tilde\gamma_2+\tilde\gamma_3} +X_{\tilde\gamma_1+2\tilde\gamma_2+2\tilde\gamma_3}}{1 - X_{2\tilde\gamma_1+2\tilde\gamma_2+2\tilde\gamma_3}}  \right)^2 \\
	& \equiv 1+A_1
\end{split}
\ee
where the last equality defines $A_1$, and similar expressions hold for $Q^{(+)}(p_2), Q^{(+)}(p_3)$.
The factors defined in (\ref{eq:Q-pm-factorization}) are identified with $Q_i^{} \equiv Q^{(+)}(p_i)$.\footnote{
Actually, one may factorize the denominator separately. However we will see in subsequent examples that denominators don't play a role in computing spectrum generators. In fact in the present case the denominators simply drop out in equation (\ref{eq:N2star-Sgamma1}).
}  Therefore $\alpha_i(p_j)=\delta_{ij}$, and the corresponding homology cycles are precisely the $\tilde\gamma_i$
\be
	\left[L_i\right] = \left[\left(p_i\right)_\Sigma\right] = \tilde\gamma_i\,.
\ee
These cycles have pairing $\langle\tilde\gamma_i,\tilde\gamma_{i-1}\rangle = 2$ with $i\in\IZ/3\IZ$, therefore applying (\ref{eq:gen-K-wall-formula}) we obtain e.g.
\be\label{eq:N2star-Sgamma1}
	S_{\gamma_1} = \left(\frac{1+A_2}{1+A_3}\right)^2\qquad S_{\gamma_2} = \left(\frac{1+A_3}{1+A_1}\right)^2 \qquad S_{\gamma_3} = \left(\frac{1+A_1}{1+A_2}\right)^2\,.
\ee
As a check, this agrees with eqs (5.16) and (5.17) of \cite{Longhi:thesis}, also note that the result is $\IZ_3$-symmetric as expected.
Factorization of $\IS$ is highly nontrivial, because the spectrum appears to be infinite in all chambers of $\CB$. 
An explicit factorization of $\IS$ for a particular configuration of central charges can however be found in \cite[\S\S 1.5 \& 5]{Longhi:thesis}.\footnote{This factorization was found in joint work with Greg Moore.} 
We will reproduce it shortly, in the context of the quantum monodromy.

Turning to the quantum monodromy, with the aid of \cite{python-code} we obtain 
\be
\begin{split}
	Q^{(-)}(p_1,y) & = \Big[1+y^{-1}\hY_{\tilde\gamma_1}+\left(1+y^{-2}\right) \hY_{\tilde\gamma_1+\tilde\gamma_3}  + y^{-1} \hY_{\tilde\gamma_1+2\tilde\gamma_3} + \left(1+y^{-2}\right) \hY_{\tilde\gamma_1+\tilde\gamma_2+2\tilde\gamma_3} \\
	&+ y^{-1} \hY_{\tilde\gamma_1+2\tilde\gamma_2+2\tilde\gamma_3} +y^{-2} \hY_{2\tilde\gamma_1+2\tilde\gamma_2+2\tilde\gamma_3} \Big] \, \left(1-y^{-2}\hY_{2\tilde\gamma_1+2\tilde\gamma_2+2\tilde\gamma_3}\right)^{-2}	\\
	Q^{(+)}(p_1,y) & = \Big[1+y^{-1}\hY_{\tilde\gamma_1}+\left(1+y^{-2}\right) \hY_{\tilde\gamma_1+\tilde\gamma_2}  +y^{-1}\hY_{\tilde\gamma_1+2\tilde\gamma_2} + \left(1+y^{-2}\right) \hY_{\tilde\gamma_1+2\tilde\gamma_2+\tilde\gamma_3} \\
	& +y^{-1} \hY_{\tilde\gamma_1+2\tilde\gamma_2+2\tilde\gamma_3} +y^{-2}\hY_{2\tilde\gamma_1+2\tilde\gamma_2+2\tilde\gamma_3} \Big] \left(1-y^{-2}\hY_{2\tilde\gamma_1+2\tilde\gamma_2+2\tilde\gamma_3}\right)^{-2}	
\end{split}
\ee
$Q^{(\pm)}(p_2,y) \ \& \ Q^{(\pm)}(p_3,y)$ are obtained by cyclic shifts of $\tilde\gamma_1,\tilde\gamma_2,\tilde\gamma_3$.
Finally, an explicit factorization can be obtained from (\ref{eq:U-eqn}), using formula (\ref{eq:a-m-eqn})%
\footnote{Comparing with \cite[\S 1.5]{Longhi:thesis} the roles of $\gamma_1$ and $\gamma_2$ appear switched. This is because the pairing matrix with our conventions is the opposite of the one in the reference (The three charges in the center mutually commute). 
}
\be
\begin{split}
	\mathbb{U} & = %
	\left(\prod_{n\geq 0}^{\nearrow} \Phi\left(\hY_{\tilde\gamma_1 + n(\tilde\gamma_1+\tilde\gamma_2)}\right)\right) \\%
	& \qquad\qquad\quad  \times \Phi\left(\hY_{\tilde\gamma_3}\right)%
	\Phi\left((-y)\hY_{\tilde\gamma_1+\tilde\gamma_2}\right)^{-1} \Phi\left((-y)^{-1}\hY_{\tilde\gamma_1+\tilde\gamma_2}\right)^{-1}
	\Phi\left(\hY_{2\tilde\gamma_1+2\tilde\gamma_2+\tilde\gamma_3}\right) \\
	&\qquad\qquad\qquad\qquad \qquad\qquad\qquad\qquad \qquad\qquad\qquad\qquad \times \left(\prod_{n\geq 0}^{\searrow} \Phi\left(\hY_{\tilde\gamma_2 + n(\tilde\gamma_1+\tilde\gamma_2)}\right)\right)\,,
\end{split}
\ee
where we took $\arg Z_{\gamma_1}>\arg Z_{\gamma_3}=\arg Z_{\gamma_1+\gamma_2}>\arg Z_{\gamma_2}$. 
This does not correspond to a MS wall, since $\langle\gamma_1+\gamma_2,\gamma_3\rangle=0$, therefore the BPS spectrum is unambiguously defined.

\subsection{$T_2$ theory}\label{sec:T2}
The $T_2$ theory presents an interesting test of our framework, because its Coulomb branch is trivial and its BPS spectrum consists entirely of gauge-neutral states. 
In particular this means that the BPS monodromy acts trivially, since all flavor charges are mutually local.

\begin{figure}[h!]
\begin{center}
\includegraphics[width=0.9\textwidth]{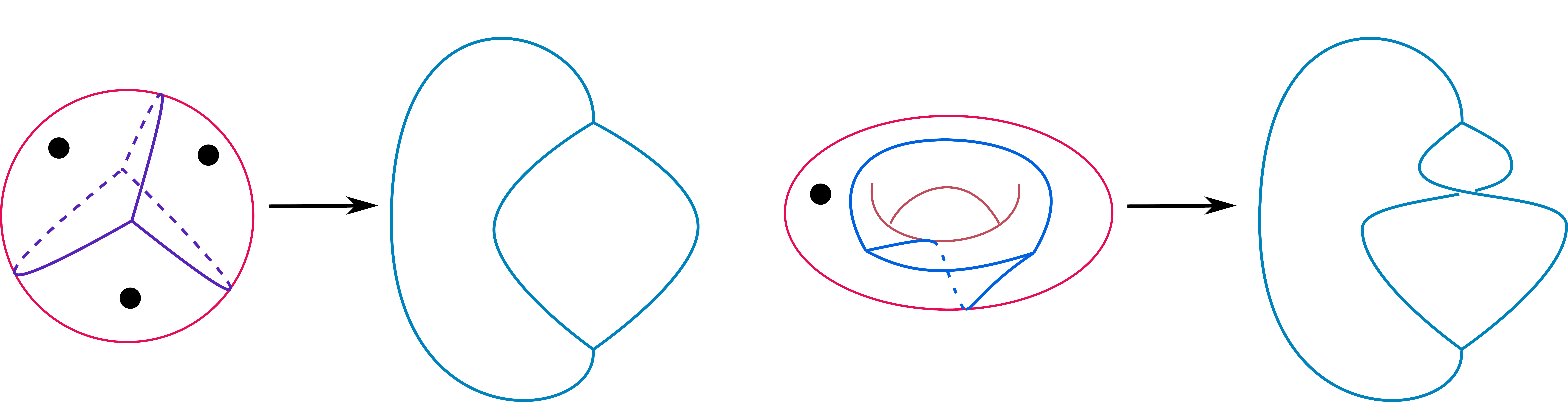}
\caption{Left: the critical graph of $T_2$. Right: the one for $\CN=2^*$. Both are class $\CS$ theories of type $A_1$, whose networks cannot involve any joints. All nodes in the two graphs are branch points.}
\label{fig:framing}
\end{center}
\end{figure}

\begin{figure}[h!]
\begin{center}
\includegraphics[width=0.33\textwidth]{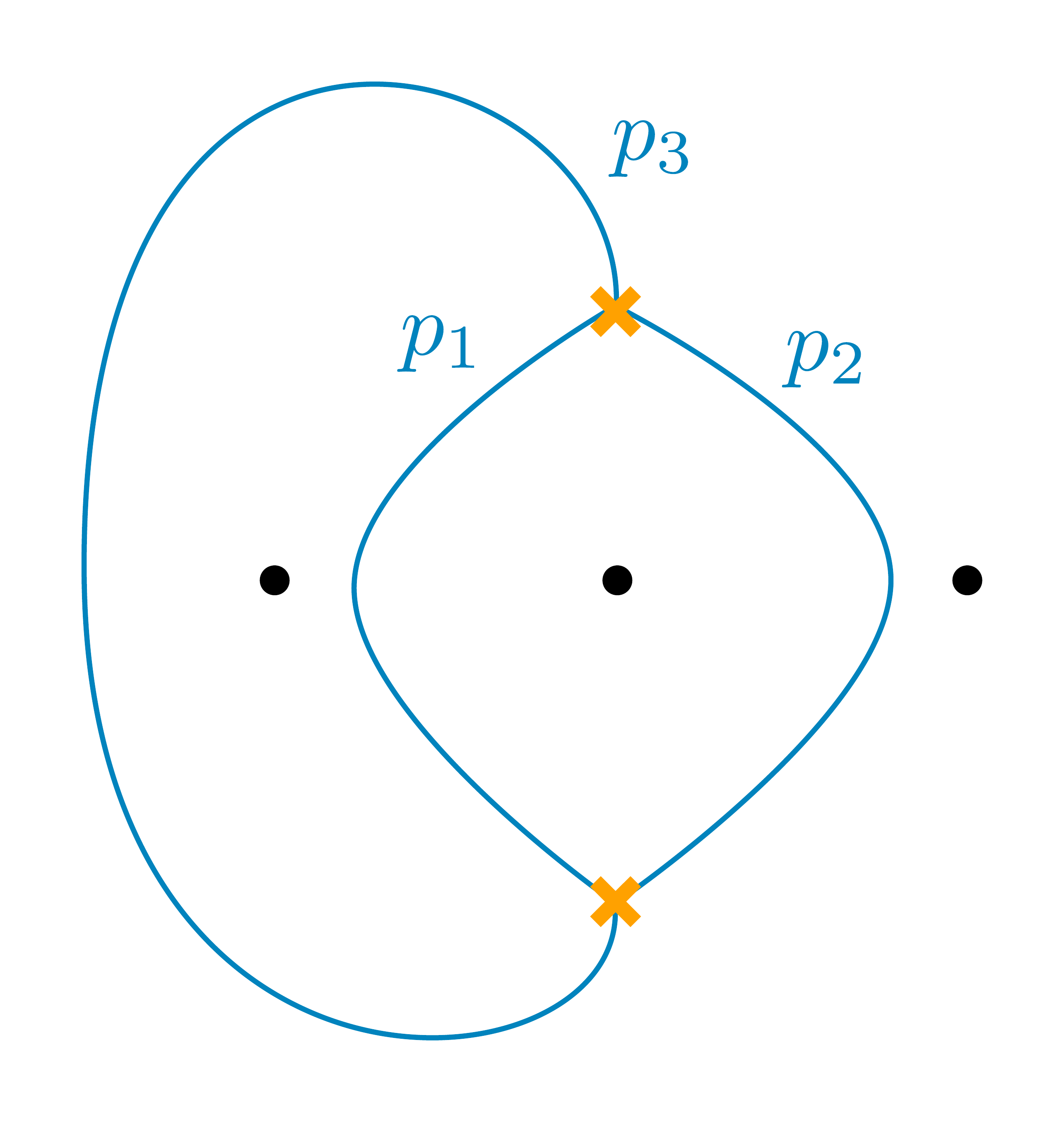}
\caption{The critical graph of the $T_2$ theory involves three edges and two nodes, each corresponding to a branch point. The graph has a manifest $S_3$ symmetry, which is inherited by the soliton generating functions: its generators are cyclic shifts of the three edges, and a rotation of the plane by an angle $\pi$.}
\label{fig:T2}
\end{center}
\end{figure}

Nevertheless there are some interesting aspects to the critical graph of this theory, which is shown in Figure \ref{fig:T2}.
For one thing, the generating functions associated to two way streets are still non-trivial, and related by an $S_3$ symmetry of the graph.
Another curiosity is that the topology of the network is essentially identical to that of $\CN=2^*$, 
the BPS monodromies however are quite different. 
This is because the monodromy does not just depend on the graph's topology, but also on extra ``framing data'', as detailed in Section \ref{sec:monodromy-graphs}.
The difference between the $T_2$ and $\CN=2^*$ critical graphs is an exchange in the cyclic ordering of edges at one of the branch points, see Figure \ref{fig:framing}.

The class $\CS$ spectral curve $\Sigma$ is a sphere with 6 punctures, The physical charge lattice $\Gamma$ has rank  3, due to the projection on $H_1(\Sigma,\IZ)$ explained in \cite{Gaiotto:2009hg, Longhi:2016rjt}.
Generators of $\Gamma$ can be taken to be $\gamma_L,\gamma_C,\gamma_R$, corresponding to $\IZ_2 $ anti-invariant combinations of small circles around the left, central and right puncture of figure \ref{fig:T2}.
In terms of these, the cycles obtained by canonically lifting $p_1, p_2, p_3$ to $\Sigma$ are $\gamma_1 = (\gamma_C-\gamma_L+\gamma_R)/2$,  $\gamma_2 =  (\gamma_C+\gamma_L-\gamma_R)/2$,  $\gamma_3 =  -(\gamma_C+\gamma_L+\gamma_R)/2$.\footnote{These relations can be obtained by choosing branch cuts and trivializing the 2-sheeted covering $\pi:\Sigma\to C$. Here $\gamma_{C,L,R}$ denote a linear combination of a counter-clockwise path around the lift of the $C,L,R$ puncture on the upper sheet and a clockwise path around the lift of the puncture on the lower sheet.} 

Let $a_i$ be the solitons on street $p_i$ sourced by the Northern branch point, and $b_i$ those by the Southern one. Also denote as usual $\Upsilon/\Delta$ the soliton generating functions for up/down-going $\CS$-walls which make up each 2-way street (for street $p_3$ the direction is understood to be considered in the neighborhood of each branch point.). 
The network equations, determined by each branch point in British resolution, are
\be
\begin{array}{c}
N \\
\hline
\Delta_1 = X_{a_1} + \Upsilon_2 \\
\Delta_2 = X_{a_2} + \Delta_3 \\
\Upsilon_3 = X_{a_3} + \Upsilon_1 \\
\end{array} \qquad %
\begin{array}{c}
S \\
\hline
\Upsilon_1 = X_{b_1} + \Upsilon_3 \\
\Upsilon_2 = X_{b_2} + \Delta_1 \\
\Delta_3 = X_{b_3} + \Delta_2 \\
\end{array}  
\ee
Comparing with (\ref{eq:N2-star-equations}) we note that, while the first set of equations is identical,  the effect of the twisting is to rotate the $\Delta_i, \Upsilon_i$ on the RHS in the second set of equations. 
Solving these equations with the techniques of Appendix \ref{sec:soliton-computations} is straightforward. We find
\be
	\Delta(p_1) = \frac{X_{a_1} + X_{b_2+\tilde\gamma_2}}{1-X_{\tilde\gamma_1+\tilde\gamma_2}}\,,
	\qquad %
	\Upsilon(p_1) = \frac{X_{b_1} + X_{a_3+\tilde\gamma_3}}{1-X_{\tilde\gamma_1+\tilde\gamma_3}}\,,
\ee
and similarly for $p_2, p_3$, by a cyclic permutation of the indices.
Hence the soliton generating functions for 2-way streets are
\be\label{eq:T2-Q-factor}
	Q^{(+)}(p_1) = \frac{(1+X_{\tilde\gamma_1})(1+X_{\tilde\gamma_1+\tilde\gamma_2+\tilde\gamma_3})}{(1-X_{\tilde\gamma_1+\tilde\gamma_2})(1-X_{\tilde\gamma_1+\tilde\gamma_3})}\,.
\ee
The $S_{\gamma}$ are all trivial, because all intersection pairings vanish. 
However we still find nontrivial $Q^{(+)}(p_i)$, what are they counting? 
The $Q^{(\pm)}(p_i)$ contain information on framed 2d-4d wall-crossing, which can occur even if the 4d theory contains only flavor charges, as can be seen from the explicit computation in (\ref{eq:KS-monodromy-direct-computation}).

In fact, in this example the jump of framed 2d-4d degeneracies at $\vartheta_c$ is actually of the standard ``$\CK$-wall'' type described in \cite{Gaiotto:2012rg}, because we are working at a regular point on the Coulomb branch (due to the complete lack of walls of marginal stability). 
In particular, $Q^{(-)}(p) = Q^{(+)}(p) $ coincide with the $Q(p)$ defined in \cite{Gaiotto:2012rg} for this theory. 
Therefore the exponent of each factor in (\ref{eq:T2-Q-factor}) carries a precise physical meaning in terms of the halo picture of wall-crossing: each factor comes in as $(1\pm X_{\tilde\gamma})^{\omega(\tilde\gamma,a)}$ where $\omega(\gamma,a)$ counts contributions to framed 2d-4d wall crossing both from 4d states and from 2d-4d states with the same charge $\tilde\gamma$.

Now this example has the special feature of containing flavor states, as mentioned in Section \ref{sec:classical-monodromy-equations} we should slightly modify the rules to capture their information.
The essential observation is that pure-flavor BPS states always contribute factors of the $\CK$-wall type $(1+X_{\tilde\gamma_f})^{\Omega}$, because $\CK_{\gamma_f}$ commutes with all other $\CK_{\gamma}$ in the classical monodromy $\IS$. 
The modification to the rules of Section \ref{sec:classical-monodromy-equations} is that one should introduce $Q_i$ for the flavor contributions, which always factorize.
In this example every state is gauge-neutral, so there are only flavor factors to consider as $Q_i$ into (\ref{eq:Q-pm-factorization}). They are obtained naturally from the factorization of $Q(p_i)$\footnote{As explained in \cite{Gaiotto:2012rg}, uniqueness of the factorization of $Q(p)$ (for a $\CK$-wall type jump) is guaranteed by using $X_{\tilde\gamma_f}$ corresponding to the \emph{preferred lift} of a path.}
\be
\begin{array}{lll}
	Q_1 = 1+X_{\tilde\gamma_1}  & Q_2 = 1+X_{\tilde\gamma_2}  \\
	Q_3 = 1+X_{\tilde\gamma_3} & Q_4 = 1+X_{\tilde\gamma_1+\tilde\gamma_2+\tilde\gamma_3} \qquad \\
	Q_5 = 1+X_{\tilde\gamma_1+\tilde\gamma_2} \qquad\qquad &Q_6 = 1+X_{\tilde\gamma_2+\tilde\gamma_3} &\qquad Q_7 = 1+X_{\tilde\gamma_1+\tilde\gamma_3}
\end{array}
\ee
the corresponding homology cycles are
\be\label{eq:T2-L}
\begin{array}{ll}
	[L_1] = [(p_1)_{\tSigma} ]= \tilde\gamma_1 \qquad\qquad\qquad\qquad & [L_2] = [(p_2)_{\tSigma} ] = \tilde\gamma_2 \\
	{[L_3]} = [(p_3)_{\tSigma}] = \tilde\gamma_3 &[L_4] = [(p_1+p_2+p_3)_{\tSigma}] = \tilde\gamma_1+\tilde\gamma_2+\tilde\gamma_3 \\
	{[L_5]} = [(-p_1 -p_2)_{\tSigma}] = -\tilde\gamma_1-\tilde\gamma_2  &[L_6] = [(-p_2-p_3)_{\tSigma}] = -\tilde\gamma_2-\tilde\gamma_3 \\
	{[L_7]} = [(-p_1-p_3)_{\tSigma}] = -\tilde\gamma_1-\tilde\gamma_3 
\end{array}
\ee
There is an important distinction between cycles $L_1,\dots,L_4$ and $L_{5}, L_6, L_7$.
When $\omega(\tilde\gamma,a)=-1$, and the corresponding charge $\gamma$ is a cycle winding around a single puncture, this is usually interpreted as halo contributions from 2d particles, not 4d BPS states (see discussion below equation (8.11) in \cite{Gaiotto:2012rg}).\footnote{Also see studies of 2d $\IC\IP^n$ sigma model spectra using spectral networks in \cite{Gaiotto:2011tf, Longhi-Park-2d}, where states of this kind appear as 2d massive quanta.} 
In fact the factors corresponding to $L_5, L_6, L_7$ correspond precisely to this description, their homology classes are indeed
\be
	-(\gamma_1+\gamma_2) = -\gamma_C\,, \qquad%
	-(\gamma_1+\gamma_3) = \gamma_L\,, \qquad%
	-(\gamma_2+\gamma_3) = \gamma_R\,.
\ee
The other four factors appearing in the $Q(p_i)$ are instead due to honest 4d BPS states, their BPS indices are all $1$, hence they are hypermutiplets.
The key point is that, despite not contributing to $S_\gamma$ through equation (\ref{eq:gen-K-wall-formula}),  they are nevertheless detected by 2d-4d wall crossing.

We therefore propose a natural enhancement of our framework of Section \ref{sec:classical-monodromy-equations}, by taking into account such pure 4d flavor states by including in $\IS$ a factor $\CK^{\Omega(\gamma)}_\gamma$ for each of them. Thus the BPS monodromy for $T_2$ reads
\be
	\CK_{\gamma_1}\CK_{\gamma_2}\CK_{\gamma_3}\CK_{\gamma_1+\gamma_2+\gamma_3}\,.
\ee

From this theory we have learned that some extra care is needed to capture pure flavor states within BPS monodromies, since the functions $S_{\gamma}$ defined in (\ref{eq:S-gamma-def}) are blind to factors of the form $\CK_{\gamma_f}$. Happily, the generating functions $Q^{(\pm)}(p)$ do detect such states, thanks to the mechanism of 2d-4d wall crossing.
The procedure adopted for this example generalizes naturally to other theories whose BPS spectrum contains pure flavor states (possibly, in addition to charged states): every time there is a gauge-neutral BPS state in the 4d spectrum, it will appear in the jumps of framed 2d-4d degeneracies. Moreover, since its charge has zero pairing with all other 4d charges, it will appear precisely as a factor $(1+ X_{\tilde\gamma_f})^{\Omega(\gamma_f)}$.
In addition to pure flavor states, there can also be contributions from purely 2d particles, whose contributions are somewhat similar. The rules we discussed for discerning 2d particles from 4d flavor states apply in full generality. Below in Section \ref{sec:SU2-Nf4} we will encounter a more involved example containing 2d particles.

Turning to quantum monodromies, the story is very similar.
Computing the generating functions with the aid of \cite{python-code} we find
\be\label{eq:T2-motivic-Q}
\begin{split}
	Q^{(\pm)}(p_1, y) & = %
	\frac{(1+ y^{-1}\hY_{\tilde\gamma_1}) (1+y^{-1}\hY_{\tilde\gamma_1+\tilde\gamma_2+\tilde\gamma_3})}{\left(1- y^{-2} {\hY_{\tilde\gamma_1+\tilde\gamma_2}}\right) (1- \hY_{\tilde\gamma_1+\tilde\gamma_3} )} \,,\\
	Q^{(\pm)}(p_2, y) & = %
	\frac{(1+y^{-1}\hY_{\tilde\gamma_2}) (1+ y^{-1}\hY_{\tilde\gamma_1+\tilde\gamma_2+\tilde\gamma_3})}{(1- \hY_{\tilde\gamma_1+\tilde\gamma_2} ) \left(1- y^{-2} {\hY_{\tilde\gamma_2+\tilde\gamma_3}}\right)}\,,\\
	Q^{(\pm)}(p_3, y) & = %
	\frac{(1+y^{-1}\hY_{\tilde\gamma_3}) (1+ y^{-1}\hY_{\tilde\gamma_1+\tilde\gamma_2+\tilde\gamma_3})}{\left(1-y^{-2} {\hY_{\tilde\gamma_1+\tilde\gamma_3}} \right) (1- \hY_{\tilde\gamma_2+\tilde\gamma_3} )}\,,\\
\end{split}
\ee
as expected $Q^{(+)}(p_i,y) = Q^{(-)}(p_i,y)$, because $\mathbb{U}$ commutes with $Q^{(\pm)}(p_i,y)$. Also the $\IZ_3$ symmetry of the graph is again manifest. 
As for the classical monodromy, also in the case of $\mathbb{U}$  equation (\ref{eq:U-eqn}) does not impose any constraint. 
However, by the same reasoning outlined above, it is natural to include the four factors appearing in the numerators into $\mathbb{U}$
\be
	\mathbb{U} = \Phi(\hY_{\tilde\gamma_1})\Phi(\hY_{\tilde\gamma_2})\Phi(\hY_{\tilde\gamma_3})\Phi(\hY_{\tilde\gamma_1+\tilde\gamma_2+\tilde\gamma_3}) \,.
\ee
The reason for having $a_m(\gamma)=\delta_{m,0}$ in each of these contributions can be understood by recalling that the relation between $\mathbb{U}$ and factors in $Q^{(\pm)}(p,y)$ is through the relation (\ref{eq:dilog-relation}) between finite-type and noncompact quantum dilogarithms. The relevant intersection pairing which fixes $m$ is that between $L_i$ and a small path $\wp$  (more properly, its lift to $\Sigma$) crossing a 2-way street $p$ which contributes to $L_i$. Concretely, the relevant relations are of the type $\Phi(\hY_{\gamma_1}) \, \hY_{\wp^{(i)}} \, \Phi(\hY_{\gamma_1})^{-1} = \hY_{\wp^{(i)}} \, \Phi_{\langle\wp^{(i)},L_i\rangle}(\hY_{\gamma_1}) $ with ${\langle\wp^{(i)},L_i\rangle}=1$ and therefore $\Phi_{1}(\hY_{\gamma_1}) = (1+ y^{-1} \hY_{\gamma_1})$ as appears in $Q(p_1,y)$ in (\ref{eq:T2-motivic-Q}).
For more details we refer the reader to \cite[\S 2]{Galakhov:2014xba}.

\subsection{SU(2) $N_f=4$ SQCD}\label{sec:SU2-Nf4}
\begin{figure}[h!]
\begin{center}
\includegraphics[width=0.35\textwidth]{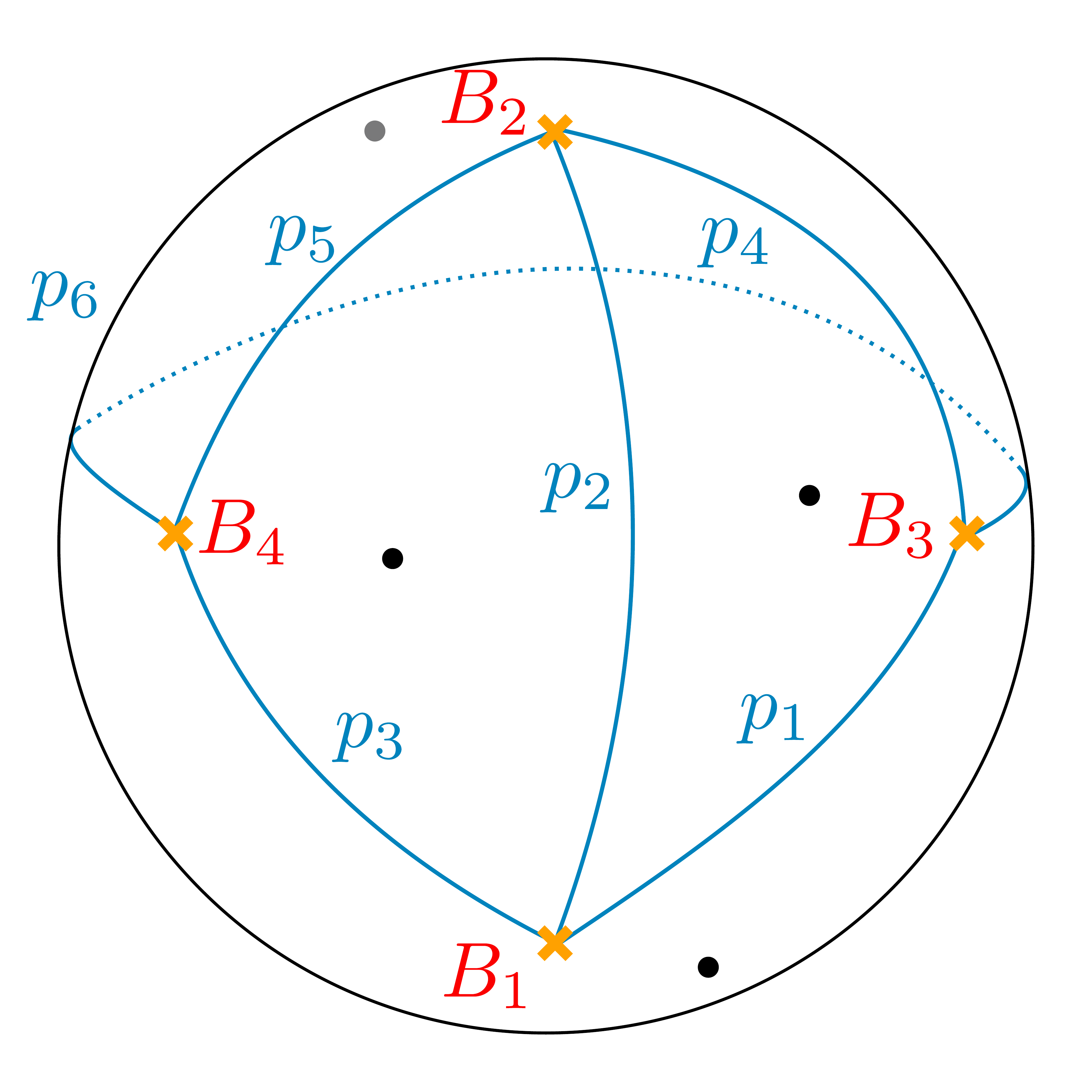}
\caption{Critical graph of the SU(2) $N_f=4$ theory. The graph has a manifest tetrahedral rotation symmetry, which preserves both its topology and cyclic ordering of edges at each node. The symmetry is inherited by the soliton generating functions $Q^{(\pm)}$ that determine the BPS monodromy.}
\label{fig:SU2_Nf4}
\end{center}
\end{figure}

This theory is realized in class $\CS$ by taking $C$ to be a four-punctured sphere, the critical graph is tetrahedral as shown in figure \ref{fig:SU2_Nf4}, each node is a branch point.
Symmetries of the graph which preserve both topology and cyclic ordering of edges at nodes correspond to rotations of the tetrahedron about any one of its vertices by $2\pi/3$.
Let $\tilde\gamma_i$ be the homology cycle obtained from the canonical oriented lift of street $p_i$.
One way of computing the intersection matrix of these cycles would be to choose branch cuts and draw representatives of cycles, using them to compute pairings. 
However a simple trick is to notice that intersections of these cycles can be computed by counting  shared branch points of the $p_i$ with signs.
At a branch point, if $p_2$ sits counter-clockwise to $p_1$, then $\langle\tilde\gamma_1,\tilde\gamma_2\rangle = -1$.\footnote{This can be seen by simply trivializing the neighborhood of a generic branch point, and counting intersections of canonical lifts of the streets ending there.} 
The pairing matrix is 
\be\label{eq:Nf4-pairing}
\langle\tilde\gamma_i,\tilde\gamma_j\rangle = %
\left(
\begin{array}{cccccc}
 0 & -1 & 1 & 1 & 0 & -1 \\
 1 & 0 & -1 & -1 & 1 & 0 \\
 -1 & 1 & 0 & 0 & -1 & 1 \\
 -1 & 1 & 0 & 0 & -1 & 1 \\
 0 & -1 & 1 & 1 & 0 & -1 \\
 1 & 0 & -1 & -1 & 1 & 0 \\
\end{array}
\right)
\ee
with the convention that e.g. $\langle\tilde\gamma_1,\tilde\gamma_2\rangle  = -1$.

Let $\Upsilon_1$ be the soliton generating function for the upward-flowing $\CS$-wall underlying the 2-way street $p_1$. 
Similarly, $\Delta_1$ will denote the generating function of downward-flowing solitons on $p_1$.
Both are easily obtained using the technique of Appendix  \ref{sec:soliton-computations}, to apply these we switch for convenience to variables $\tilde\Upsilon_1,\tilde \Delta_1$ defined by
\be
	\Upsilon_1 = X_{a_1}\tilde\Upsilon_1\qquad \Delta_1 = X_{b_1}\tilde\Delta_1
\ee
where $a_1$ is the soliton charge sourced at branch point $B_1$, and $b_1$ is the soliton sourced at branch point $B_2$. 
Then in British resolution the equations read
\be
\begin{split}	
	\tilde\Upsilon_1 &=1+X_{\tilde\gamma_2}(1+X_{\tilde\gamma_4}(1+X_{\tilde\gamma_1}\,\tilde\Upsilon_1)) \\
	\tilde\Delta_1 &= 1+X_{\tilde\gamma_6}(1+X_{\tilde\gamma_3}(1+X_{\tilde\gamma_1}\,\tilde\Delta_1)) 
\end{split}
\ee
where $a_1$ is the soliton charge sourced at branch point $B_1$, and $b_1$ is the soliton sourced at branch point $B_2$. 
These are easily solved to yield
\be
	\Upsilon_1 = X_{a_1}\frac{1+X_{\tilde\gamma_2}(1+X_{\tilde\gamma_4})}{1-X_{\tilde\gamma_1+\tilde\gamma_2+\tilde\gamma_4}} \,,\qquad %
	\Delta_1 = X_{b_1}\frac{1+X_{\tilde\gamma_6}(1+X_{\tilde\gamma_3})}{1-X_{\tilde\gamma_1+\tilde\gamma_3+\tilde\gamma_6}} \,,
\ee
which immediately give the 2-way street generating function
\be
\begin{split}
	Q^{(+)}(p_1) & :=1+\Upsilon_1 \Delta_1 = %
	1+X_{\tilde\gamma_1}%
	\frac{\left(1+X_{\tilde\gamma_6}(1+X_{\tilde\gamma_3})\right)%
	\left(1+X_{\tilde\gamma_2}(1+X_{\tilde\gamma_4})\right)}%
	{(1-X_{\tilde\gamma_1+\tilde\gamma_2+\tilde\gamma_4})%
	(1-X_{\tilde\gamma_1+\tilde\gamma_3+\tilde\gamma_6})} \\
	& = \frac{1+X_{\tilde\gamma_1}%
	(1 + X_{\tilde\gamma_2} + X_{ \tilde\gamma_6} + X_{ \tilde\gamma_2+\tilde\gamma_6} %
	(1 + X_{\tilde\gamma_3} + X_{\tilde\gamma_4} + X_{\tilde\gamma_3+\tilde\gamma_4} %
	(1+ X_{\tilde\gamma_1})) )}%
	{(1-X_{\tilde\gamma_1+\tilde\gamma_2+\tilde\gamma_4})%
	(1-X_{\tilde\gamma_1+\tilde\gamma_3+\tilde\gamma_6})} 
\end{split}
\ee
The rest of the generating functions can be obtained by simply exploiting the symmetries of the graph. For example to obtain $Q^{(+)}(p_2)$ we rotate the tetrahedron by $2\pi/3$ about the vertex $B_1$, and so on:
\be
\begin{split}
Q^{(+)}(p_2) %
	& = \frac{1+X_{\tilde\gamma_2}%
    	(1 + X_{\tilde\gamma_3} + X_{ \tilde\gamma_4} + X_{ \tilde\gamma_3+\tilde\gamma_4} %
    	(1 + X_{\tilde\gamma_1} + X_{\tilde\gamma_5} + X_{\tilde\gamma_1+\tilde\gamma_5} %
    	(1+ X_{\tilde\gamma_2})) )%
	}%
    	{%
	(1-X_{\tilde\gamma_1+\tilde\gamma_2+\tilde\gamma_4}) %
	(1-X_{\tilde\gamma_2+\tilde\gamma_3+\tilde\gamma_5})}	\\
Q^{(+)}(p_3) %
	& = \frac{1+X_{\tilde\gamma_3}%
    	(1 + X_{\tilde\gamma_1} + X_{ \tilde\gamma_5} + X_{ \tilde\gamma_1+\tilde\gamma_5} %
    	(1 + X_{\tilde\gamma_2} + X_{\tilde\gamma_6} + X_{\tilde\gamma_2+\tilde\gamma_6} %
    	(1+ X_{\tilde\gamma_3})) )%
	}%
    	{%
	(1-X_{\tilde\gamma_1+\tilde\gamma_3+\tilde\gamma_6}) %
	(1-X_{\tilde\gamma_2+\tilde\gamma_3+\tilde\gamma_5})}\\
Q^{(+)}(p_4) %
	& = \frac{1+X_{\tilde\gamma_4}%
    	(1 + X_{\tilde\gamma_1} + X_{ \tilde\gamma_5} + X_{ \tilde\gamma_1+\tilde\gamma_5} %
    	(1 + X_{\tilde\gamma_2} + X_{\tilde\gamma_6} + X_{\tilde\gamma_2+\tilde\gamma_6} %
    	(1+ X_{\tilde\gamma_4})) )%
	}%
    	{%
	(1-X_{\tilde\gamma_1+\tilde\gamma_2+\tilde\gamma_4}) %
	((1-X_{\tilde\gamma_4+\tilde\gamma_5+\tilde\gamma_6})}\\
Q^{(+)}(p_5) & = \frac{1+X_{\tilde\gamma_5}%
	(1 + X_{\tilde\gamma_2} + X_{ \tilde\gamma_6} + X_{ \tilde\gamma_2+\tilde\gamma_6} %
	(1 + X_{\tilde\gamma_3} + X_{\tilde\gamma_4} + X_{\tilde\gamma_3+\tilde\gamma_4} %
	(1+ X_{\tilde\gamma_5})) )
	}%
	{%
	(1-X_{\tilde\gamma_2+\tilde\gamma_3+\tilde\gamma_5}) %
	(1-X_{\tilde\gamma_4+\tilde\gamma_5+\tilde\gamma_6})}\\
Q^{(+)}(p_6) %
	& = \frac{1+X_{\tilde\gamma_6}%
    	(1 + X_{\tilde\gamma_3} + X_{ \tilde\gamma_4} + X_{ \tilde\gamma_3+\tilde\gamma_4} %
    	(1 + X_{\tilde\gamma_1} + X_{\tilde\gamma_5} + X_{\tilde\gamma_1+\tilde\gamma_5} %
    	(1+ X_{\tilde\gamma_6})) )%
	}%
	{%
	(1-X_{\tilde\gamma_1+\tilde\gamma_3+\tilde\gamma_6}) %
	(1-X_{\tilde\gamma_4+\tilde\gamma_5+\tilde\gamma_6})}
\end{split}
\ee
The factors defined in (\ref{eq:Q-pm-factorization}) are 10 in this case: there are the six numerators of $Q^{(+)}(p_i)$, which will be denoted $Q_i,\dots, Q_6$, as well as the four extra factors
\be	\label{eq:flavor-Q-Nf-4}
\begin{split}
	& Q_7 = (1-X_{\tilde\gamma_1+\tilde\gamma_3+\tilde\gamma_6}) \,,\qquad %
	Q_8 = (1-X_{\tilde\gamma_2+\tilde\gamma_3+\tilde\gamma_5}) \,,\\
	& Q_9 = (1-X_{\tilde\gamma_4+\tilde\gamma_5+\tilde\gamma_6})\,,\qquad 
	Q_{10} = (1-X_{\tilde\gamma_1+\tilde\gamma_2+\tilde\gamma_4}) \,,
\end{split}
\ee
which come from the denominators.
The reason for separating the four extra factors is due to the presence of pure-flavor charges in the theory, which must be taken into account separately as illustrated in Section \ref{sec:T2}. 
Generators of the flavor symmetry correspond to small cycles around punctures, which sit on faces of the tetrahedron. 
These flavor cycles correspond in fact to sums of lifted edges bounding each face: 
for example lifting $p_1\cup p_2 \cup p_4$ gives a pair of cycles winding around the puncture to the right, whose homology class is therefore $\tilde\gamma_1+\tilde\gamma_2+\tilde\gamma_4$. The 3 other charges are obtained by rotations of the tetrahedron, completing the set in (\ref{eq:flavor-Q-Nf-4}).
Based on this observation, according to the rules proposed in the previous section we separate precisely \emph{four} extra $Q_i$, each of the specific type $ (1-X_{\tilde\gamma_f})$.

The cycles corresponding to the $Q_i$ are
\be
\begin{split}
	& {[}L_{i}{]} = {[}(p_i){{}_{\tSigma}}{]} = \tilde\gamma_i \qquad i=1,\dots, 6 \\
	& [L_7] = [-(p_1)_{\tSigma}-(p_3)_{\tSigma}-(p_6)_{\tSigma}] = -(\tilde\gamma_1+\tilde\gamma_3+\tilde\gamma_6)\\
	& [L_8] = [-(p_2)_{\tSigma}-(p_3)_{\tSigma}-(p_5)_{\tSigma}] = -(\tilde\gamma_2+\tilde\gamma_3+\tilde\gamma_5)\\
	& [L_9] = [-(p_4)_{\tSigma}-(p_5)_{\tSigma}-(p_6)_{\tSigma}] = -(\tilde\gamma_4+\tilde\gamma_5+\tilde\gamma_6)\\
	& [L_{10}] = [-(p_1)_{\tSigma}-(p_2)_{\tSigma}-(p_4)_{\tSigma}] = -(\tilde\gamma_1+\tilde\gamma_2+\tilde\gamma_4)
\end{split}
\ee
Flavor charges $L_7,\dots, L_{10}$ have zero intersection pairing with all 4d charges, as can also be checked directly in (\ref{eq:Nf4-pairing}). 
Therefore they will not contribute at all to $S_\gamma$ in equation (\ref{eq:S-gamma-def}).
As discussed below (\ref{eq:T2-L}) these are contributions from 2d particles, which automatically decouple from $\IS$, and should not be reintroduced since they are not 4d flavor states. 

Applying formula (\ref{eq:gen-K-wall-formula}) we can write down the functions $S_\gamma$: for example 
\be
\begin{split}
	S_{\gamma_1} & = %
	\frac{Q_2 Q_6}{Q_3 Q_4}\,.
\end{split}
\ee
We can compare this with equation (11.54) of \cite{Gaiotto:2009hg}: the dictionary between our streets and those in figure 83 of the reference is as follows 
\be
	p_1: (12),\ p_2:(24),\ p_3:(23),\ p_4:(14),\ p_5:(34), \ p_6:(13)
\ee
the first factor in the denominator (i.e. $Q_3$) thus becomes
\be
\begin{split}
	& 1+X_{23}%
	+X_{23} X_{34}+X_{12} X_{23}%
	+X_{12} X_{23} X_{34}%
	+X_{12} X_{13} X_{23} X_{34}+X_{12} X_{23} X_{24} X_{34}\\
	& +X_{12} X_{13} X_{23} X_{24} X_{34} + X_{12} X_{13} X_{23}^2 X_{24} X_{34}
\end{split}
\ee
which should be compared to equation (11.43) of that paper, 
we find exact agreement.\footnote{Actually they almost coincide, except for the  summand $X_{12} X_{23} X_{34}$. This appears to be a typo in \cite{Gaiotto:2009hg}.}

Turning to the quantum monodromy, we computed the full motivic generating functions for all streets with the aid of \cite{python-code}.
To write them explicitly let us introduce the  convenient notation 
\be
\begin{split}
	\CQ^{(-)}(\xi;\alpha_1,\alpha_2;\beta_1,\beta_2)  & = 1 + y^{-1} \hY_\xi%
    	\frac{%
	\big(	1 	+ 	y^{-1} \hY_{\alpha_1} (1  +  y \hY_{\alpha_2} )	\big)%
	\big(	1	+	y^{-1} \hY_{\beta_1} (1  + y^{-1} \hY_{\beta_2})	\big)%
	}%
	{%
    	\big(	1 	- 	\hY_{\xi+\alpha_1+\alpha_2}	\big) %
	\big(	1	-	y^{-2} \hY_{\xi+\beta_1+\beta_2}	\big) %
	}%
	\\%
	\CQ^{(+)}(\xi;\alpha_1,\alpha_2;\beta_1,\beta_2)  & = 1 + y^{-1}\hY_\xi%
    	\frac{%
	\big(	1	+	y^{} \hY_{\beta_2} (1  + y^{-1} \hY_{\beta_1})	\big)%
	\big(	1 	+ 	y^{} \hY_{\alpha_2} (1  + y^{}  \hY_{\alpha_1} )	\big)%
	}%
	{%
    	\big(	1 	- 	\hY_{\xi+\alpha_1+\alpha_2}	\big) %
	\big(	1	-	y^{-2} \hY_{\xi+\beta_1+\beta_2}	\big)  %
	}\,.%
\end{split}
\ee
In this notation the generating functions are
\be
\begin{split}
	Q^{(\pm)}(p_1,y) & = \CQ^{(\pm)}(%
	\tilde\gamma_1	;\,  %
	\tilde\gamma_3 ,\, \tilde\gamma_6 ;\,  %
	\tilde\gamma_4 ,\, \tilde\gamma_2 %
	) \,,\\
	Q^{(\pm)}(p_2,y) & = \CQ^{(\pm)}(%
	\tilde\gamma_2	;\,  %
	\tilde\gamma_1 ,\, \tilde\gamma_4 ;\,  %
	\tilde\gamma_5 ,\, \tilde\gamma_3 %
	) \,,\\
	Q^{(\pm)}(p_3,y) & = \CQ^{(\pm)}(%
	\tilde\gamma_3	;\,  %
	\tilde\gamma_2 ,\, \tilde\gamma_5 ;\,  %
	\tilde\gamma_6 ,\, \tilde\gamma_1 %
	) \,,\\
	Q^{(\pm)}(p_4,y) & = \CQ^{(\pm)}(%
	\tilde\gamma_4	;\,  %
	\tilde\gamma_2 ,\, \tilde\gamma_1 ;\,  %
	\tilde\gamma_6 ,\, \tilde\gamma_5 %
	) \,,\\
	Q^{(\pm)}(p_5,y) & = \CQ^{(\pm)}(%
	\tilde\gamma_5	;\,  %
	\tilde\gamma_4 ,\, \tilde\gamma_6 ;\,  %
	\tilde\gamma_3 ,\, \tilde\gamma_2 %
	) \,,\\
	Q^{(\pm)}(p_6,y) & = \CQ^{(\pm)}(%
	\tilde\gamma_6	;\,  %
	\tilde\gamma_1 ,\, \tilde\gamma_3 ;\,  %
	\tilde\gamma_5 ,\, \tilde\gamma_4 %
	) \,.\\
\end{split}
\ee
Symmetry under rotations of the tetrahedron is manifest,
these functions completely determine $\mathbb{U}$ through relations (\ref{eq:U-eqn}) .

\subsection{\texorpdfstring{$T_3$}{T3} Theory}\label{sec:T3}
\begin{figure}[h!]
\begin{center}
\includegraphics[width=0.8\textwidth]{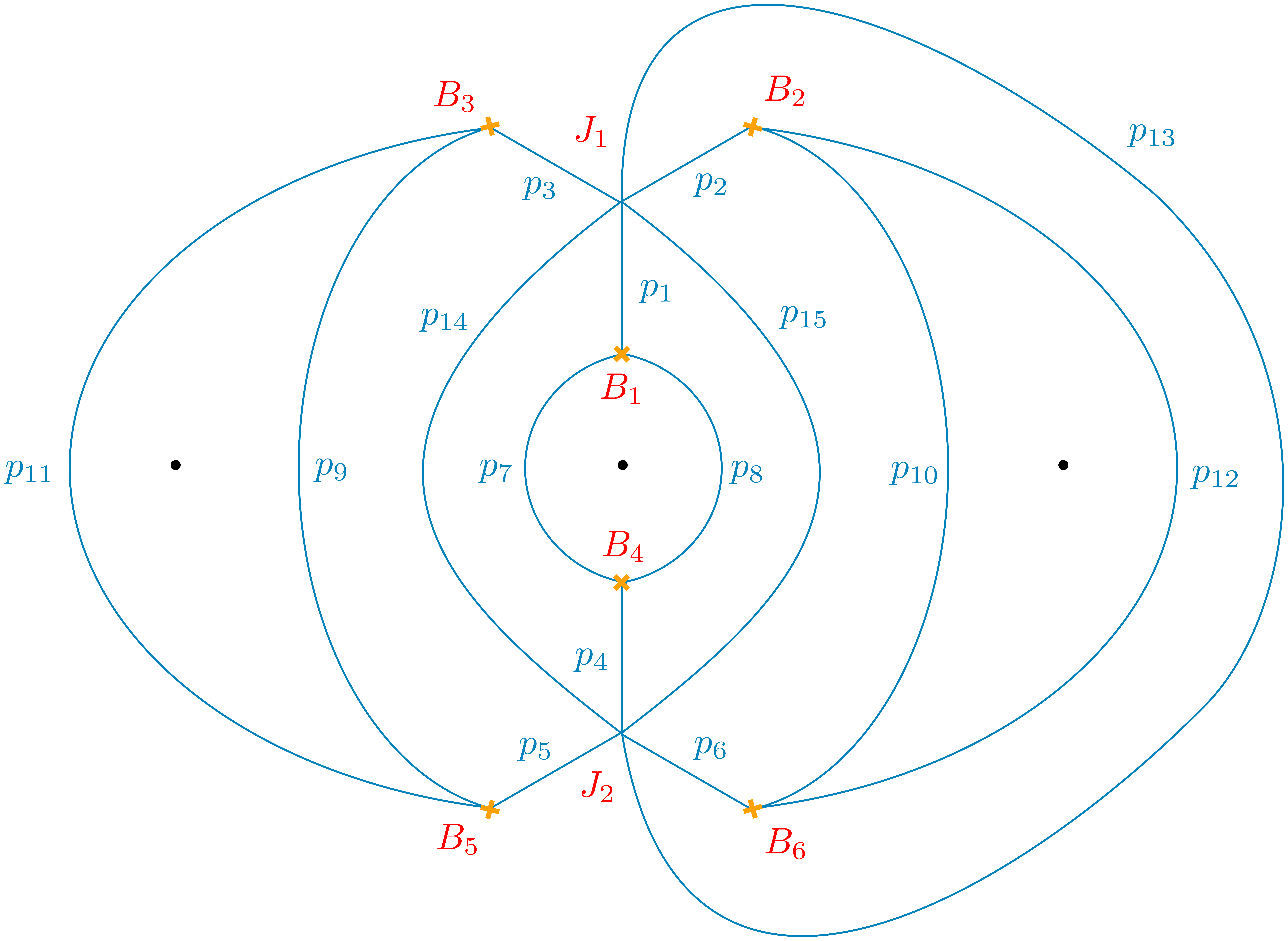}
\caption{Critical graph of the $T_3$ theory. The graph has a manifest $S_3$ symmetry, which is reflected in the soliton generating functions.}
\label{fig:T3}
\end{center}
\end{figure}

$T_3$ is a class $\CS$ theory of type $A_2$, engineered by taking $C$ to be a sphere with three ``{full}'' punctures \cite{Gaiotto:2009we}. This theory is also known as the $E_6$ superconformal theory of Minahan and Nemeschansky \cite{Minahan:1996fg}.
The critical graph was obtained in upcoming joint work \cite{network-quiver} and is shown in Figure  \ref{fig:T3}. 
The graph contains 15 edges $p_1,\dots, p_{15}$, six branch points $B_1 \dots B_6$ and two joints $J_{1}, J_2$. 

The physical charge lattice $\Gamma$ has rank 8, let  $\gamma_1,\dots,\gamma_8$ be the generators of the  positive half-lattice $\Gamma^+$ determined by the critical phase $\vartheta_c$. 
These cycles correspond to canonical lifts of the graph's edges, we defined them as follows\footnote{It is understood that $\gamma_i$ is the projection of $\tilde\gamma_i$ by $\tilde\pi_*$, see Section \ref{sec:review-4d-wc} or appendix \ref{sec:homology-conventions}.}
\be\label{eq:T3-basis-cycles}
\begin{split}
	&%
	\quad %
	\tilde\gamma_1 = \left[\sum_{i=1,2,3} (p_i)_{\tSigma}\right] \,,\qquad %
	\tilde\gamma_2 = \left[\sum_{i=4,5,6} (p_i)_{\tSigma}\right] \,,\\
	&%
	\tilde\gamma_3 = [(p_{7})_{\tSigma}]\,,\qquad %
	\ \tilde\gamma_4 = [(p_{8})_{\tSigma}]\,,\qquad %
	\ \tilde\gamma_5 = [(p_{9})_{\tSigma}]\,,\\
	&%
	\tilde\gamma_6 = [(p_{10})_{\tSigma}]\,,\qquad %
	\tilde\gamma_7 = [(p_{11})_{\tSigma}]\,,\qquad %
	\tilde\gamma_8 = [(p_{12})_{\tSigma}]\,.
\end{split}
\ee

The critical graph has a $\IZ_3$ symmetry preserving both its topology and  framing, which is generated by a simultaneous cyclic permutation of the following triples
\be\label{eq:T3-Z3}
	(p_1, p_2, p_3)\,,\quad (p_4, p_6, p_5)\,,\quad (p_9, p_8, p_{12})\,,\quad (p_7, p_{10}, p_{11})\,,\quad (p_{13}, p_{14}, p_{15})\,.
\ee
An additional $\IZ_2$ symmetry acts by exchanging simultaneously elements within the following pairs 
\be\label{eq:T3-Z2}
	(p_1,p_4)\,,\ %
	(p_2,p_5)\,,\ %
	(p_3,p_6)\,,\ %
	(p_7,p_8)\,,\ %
	(p_9,p_{10})\,,\ %
	(p_{11},p_{12})\,,\ %
	(p_{14}, p_{15})\,.
\ee
The soliton data on the critical graph is encoded by 30 soliton generating functions: one for each direction, on each 2-way street. 
The standard soliton traffic rules produce 30 coupled algebraic equations which uniquely determine each of these generating functions.
Here we shall sketch how these equations can be disentangled into a more manageable sub-system, leaving a more detailed analysis to the interested reader (the soliton data can also be obtained using the software distributed with this paper \cite{python-code}). 

We will work in the British resolution, it turns out that some of the generating functions can actually be obtained in closed form. 
For example, let $\Delta_{7}$ be the generating function of 2d-4d solitons flowing downward on street $p_{7}$, i.e. $\Delta_{7} = X_{a_{7}}+ \dots$ with $a_{7}$ sourcing from branch point $B_1$. 
To write the traffic rule for $\Delta_{7}$ it is convenient to introduce the reduced generating function $\tilde\Delta_{7}$ defined by $\Delta_{7} = X_{a_{7}}\tilde\Delta_{7}$. Then an elementary application of the rules in appendix \ref{sec:soliton-computations} gives
\be
	\tilde\Delta_{7} = 1+ X_{\tilde\gamma_4} (1+X_{\tilde\gamma_3} \tilde\Delta_{7})\,,
\ee
from which we immediately obtain
\be
	\Delta_{7} = X_{a_{7}} \frac{1+X_{\tilde\gamma_4}}{1-X_{\tilde\gamma_3+\tilde\gamma_4}}\,.
\ee
Acting on this solution with the symmetries (\ref{eq:T3-Z3}) and (\ref{eq:T3-Z2}) yields immediately  $\Delta_{10}, \Delta_{11}$ (downward-flowing solitons on streets $p_{10}, p_{11}$) and $\Upsilon_{8}, \Upsilon_{9}, \Upsilon_{12}$ (upward-flowing solitons on streets $p_8, p_9, p_{12}$).
Other generating functions are more difficult to compute. 
By direct inspection, it turns out that generating functions  $\Upsilon_k, \Delta_k$ of upward/downward-flowing solitons on street $p_k$ can be expressed entirely in terms of those of streets $p_{13},p_{14},p_{15}$.
Therefore the initial system of 30 coupled algebraic equations can be reduced to a coupled system of six equations for the six variables $\Delta_n,\Upsilon_n$ with $n=13,14,15$. The form of these equations is rather involved, and we were not able to find a closed-form solution, therefore we shall omit them.

Even in the absence of closed-form expressions for the $\Delta_n,\Upsilon_n$, we can still make progress with the analysis of the spectrum generator $\mathbb{S}$. 
Employing (\ref{eq:joint-factorization}) it is possible to factorize the generating functions $Q(p_n)=1+\Upsilon_n \Delta_n$ of those streets $p_n$ which terminate on at least one joint. 
For streets ending at joint $J_1$ we have
\be
\begin{split}
	& Q(p_1) = \frac{Q_{1}^{(J_1)}Q_{{\rm even}}^{(J_1)}}{Q_{{\rm all}}^{(J_1)}} \,, \qquad %
	 Q(p_2) = \frac{Q_{3}^{(J_1)}Q_{{\rm even}}^{(J_1)}}{Q_{{\rm all}}^{(J_1)}}\,, \qquad %
	 Q(p_3) = \frac{Q_{2}^{(J_1)}Q_{{\rm even}}^{(J_1)}}{Q_{{\rm all}}^{(J_1)}}\,, \\
	& Q(p_{13}) = \frac{Q_{1}^{(J_1)}Q_{{\rm odd}}^{(J_1)}}{Q_{{\rm all}}^{(J_1)}} \,,\qquad %
	 Q(p_{14}) = \frac{Q_{3}^{(J_1)}Q_{{\rm odd}}^{(J_1)}}{Q_{{\rm all}}^{(J_1)}} \,,\qquad %
	 Q(p_{15}) = \frac{Q_{2}^{(J_1)}Q_{{\rm odd}}^{(J_1)}}{Q_{{\rm all}}^{(J_1)}}\,. 
\end{split}
\ee
Likewise, for streets ending at joint $J_2$ we have the following factorization
\be
\begin{split}
	& Q(p_4) = \frac{Q_{1}^{(J_2)}Q_{{\rm even}}^{(J_2)}}{Q_{{\rm all}}^{(J_2)}}\,, \qquad %
	 Q(p_5) = \frac{Q_{3}^{(J_2)}Q_{{\rm even}}^{(J_2)}}{Q_{{\rm all}}^{(J_2)}} \,,\qquad %
	 Q(p_6) = \frac{Q_{2}^{(J_2)}Q_{{\rm even}}^{(J_2)}}{Q_{{\rm all}}^{(J_2)}}\,, \\
	& Q(p_{13}) = \frac{Q_{1}^{(J_2)}Q_{{\rm odd}}^{(J_2)}}{Q_{{\rm all}}^{(J_2)}} \,,\qquad %
	 Q(p_{14}) = \frac{Q_{2}^{(J_2)}Q_{{\rm odd}}^{(J_2)}}{Q_{{\rm all}}^{(J_2)}} \,,\qquad %
	 Q(p_{15}) = \frac{Q_{3}^{(J_2)}Q_{{\rm odd}}^{(J_2)}}{Q_{{\rm all}}^{(J_2)}} \,.
\end{split}
\ee
The factors given in (\ref{eq:joint-factors}) can be expressed in terms of 2d-4d generating functions as follows 
\be
\begin{split}
	& Q_1^{(J_1)} = 1 + \Upsilon_{1} \Upsilon_{13} (1 + \Upsilon_2 \Upsilon_{14})\,, \qquad%
	Q_{\rm odd}^{(J_1)} = 1 + \Upsilon_{13} \Upsilon_{15} \Upsilon_{14}\,, \\ %
	& Q_2^{(J_1)} = 1 + \Upsilon_{3} \Upsilon_{15} (1 + \Upsilon_{1} \Upsilon_{13}) \,,\qquad %
	Q_{\rm even}^{(J_1)} = 1 + \Upsilon_{1} \Upsilon_{2} \Upsilon_{3} \,, \\  %
	& Q_3^{(J_1)} = 1 + \Upsilon_{2} \Upsilon_{14} (1 + \Upsilon_{3} \Upsilon_{15})  \,,\qquad %
	Q_{\rm all}^{(J_1)} = 1 - \Upsilon_{1} \Upsilon_{2} \Upsilon_{3} \Upsilon_{13} \Upsilon_{15} \Upsilon_{14}\,, %
\end{split}
\ee
for those defined by the factorization at joint $J_1$, while for joint $J_2$ we have:
\be
\begin{split}
	& Q_1^{(J_2)} = 1 + \Delta_{4} \Delta_{13}(1 + \Delta_{5} \Delta_{15}) \,,\qquad%
	Q_{\rm odd}^{(J_2)} = 1 + \Delta_{13}\Delta_{14} \Delta_{15} \,, \\ %
	& Q_2^{(J_2)} = 1 + \Delta_{6} \Delta_{14}(1 + \Delta_{4} \Delta_{13}) \,,\qquad %
	Q_{\rm even}^{(J_2)} = 1 + \Delta_4 \Delta_5 \Delta_6 \,,\\ %
	& Q_3^{(J_2)} = 1 + \Delta_{5} \Delta_{15}(1 + \Delta_{6} \Delta_{14})  \,,\qquad 
	Q_{\rm all}^{(J_2)} = 1 - \Delta_4 \Delta_5 \Delta_6  \Delta_{13} \Delta_{14} \Delta_{15} \,.%
\end{split}
\ee
By direct inspection, it turns out that $Q_1^{(J_1)} = Q_1^{(J_2)}$, $Q_2^{(J_1)} = Q_3^{(J_2)}$ and $Q_3^{(J_1)} = Q_2^{(J_2)}$.\footnote{We checked this by expanding these functions as series in $X_{\gamma}$, with grading fixed by the generators of the charge lattice $\gamma_1\dots\gamma_8$, such that the degree of $X_{\sum n_i\gamma_i}$ is $\sum_i {n_i}$.}
The factors $Q^J_{{\rm all}}$ can be dropped, as we already saw in Sections \ref{sec:T2} and \ref{sec:SU2-Nf4}. 
Therefore we are left with the following twelve cycles
\be
\begin{split}
	& \qquad \qquad\qquad\qquad \qquad [L_i] = \left[(p_i)_{\tilde\Sigma}\right] \qquad i=7,\dots, 12\\
	& [L_{1}] = \left[\sum_{i=1,2,3}(p_i)_{\tilde\Sigma}\right] \quad \ \  %
	[L_{2}] = \left[\sum_{i=4,5,6}(p_i)_{\tilde\Sigma}\right]   \quad \ \ %
	[L_{{\rm odd}}] = \left[\sum_{i=13,14,15}(p_i)_{\tilde\Sigma}\right] \\
	& [L_{13}] = \left[\sum_{i=1,13,4}(p_i)_{\tilde\Sigma}\right]  \quad %
	[L_{14}] = \left[\sum_{i=2,14,6}(p_i)_{\tilde\Sigma}\right]\quad 
	[L_{15}] = \left[\sum_{i=3,15,5}(p_i)_{\tilde\Sigma}\right]  %
\end{split}
\ee
Among these, we recognize the eight basis cycles defined previously in (\ref{eq:T3-basis-cycles}).
The remaining four $L_{13,14,15}$ and $L_{{\rm odd}}$ are pure-flavor, and are positive-integer combinations of the basis ones.

We provide truncated expressions for the generating functions $Q^{(\pm)}(p,y)$ in appendix \ref{sec:T3-Qs}. Taking their classical limit (\ref{eq:classical-limit-Q}) gives the building blocks to construct the spectrum generator $\IS$ using (\ref{eq:KS-monodromy-formula}), in conjunction with the $L_i$ obtained here. 
In addition, they can be used to obtain directly the quantum monodromy $\mathbb{U}$ from (\ref{eq:U-eqn}).

%
%

\section{Future directions}\label{sec:coda}
Our work raises a number of questions which would be interesting to explore further, here we collect a selection of them.
\begin{itemize}
	\item The fact that $\mathbb{U}$ is encoded by a graph $\CW_c$ on the UV curve of the theory suggests a constructive approach for obtaining these graphs (and the monodromy $\mathbb{U}$), via Gaiotto's gluing \cite{Gaiotto:2009we}. 
	In particular, critical graphs of $A_n$ trinion theories $T_n$ are essentially already classified as duals of ``$K$-triangulations'' \cite{Gaiotto:2012db, network-quiver}. The main remaining question regards the details of gluing graphs from different trinions. 
	In the context of $A_1$ theories, this construction has essentially already appeared in the literature: the critical graph $\CW_c$ appears to coincide with the ``contracted Fenchel-Nielsen networks'' of \cite{Hollands:2013qza}.\\
	A constructive approach for $\mathbb{U}$ would also provide a powerful tool for testing the conjectural correspondence between BPS monodromy and specializations of the superconformal index, for a large class of theories.
	\item	A key element in our construction is the locus $\CB_c$, which can be characterized as a maximal intersection of marginal stability walls. It would be important to clarify the existence conditions for this locus, in order to assess the range of applicability of our construction of BPS monodromies.\footnote{For \emph{complete} 4d $\CN=2$ theories  the locus is guaranteed to exist \cite{Cecotti:2011rv}.} 
	In particular, it would be interesting to understand whether it can always be engineered to appear on a Coulomb branch, by shaping the geometry of the latter through a tuning of UV moduli.\footnote{Or, in some cases, by embedding the theory into a ``larger'' one, such as for the Argyres-Douglas $AD_3$ theory. I would like to thank Thomas Dumitrescu for raising this point.}
	\item Uniqueness of the critical graph $\CW_c$ of a theory is far from obvious. It is plausible that studying the spectral network at different points on $\CB_c$ would produce graphs with different topologies (see e.g. \cite{Hollands:2013qza, 1998math.ph..11024M}). On physical grounds, they should lead to the same monodromy. 
	For example the graph of pure SU(2) gauge theory and that of $m=2$ wild-wall crossing have different topologies, but encode the same monodromy $\mathbb{U}$. 
	It would be very interesting to classify equivalence relations of graphs under topological transitions.
	\item Another application of our construction would be to study the \emph{spin purity} conjecture recently formulated in \cite{Hollands:2016kgm}.
	 The key step is to find nice closed forms for the $Q^{(\pm)}(p_i)$ for the critical graph of Figure \ref{fig:T3}. Once these are known, it should not be too hard to obtain their non-commutative generalization $Q^{(\pm)}(p_i,y)$: this would amount to inserting appropriate powers of $y$, which is simple to accomplish by expanding $Q^{(\pm)}(p)$ in powers of $X_{\tilde\gamma_i}$ and comparing to the perturbative expansion obtained using the software \cite{python-code}. 
	 These exact expressions would allow to study the realization of the $E_6$ symmetry on $\mathbb{U}$, and hopefully provide a useful tool for testing the {spin purity} conjecture.\footnote{I would like to thank Andy Neitzke for raising this point.}
	\item Can our construction be extended beyond theories of class $\CS$? An important step in this direction would be to find a field theoretic interpretation of the critical graph $\CW_c$. Guidance for this is provided by the physical interpretation in terms of 2d-4d wall crossing outlined in Section \ref{sec:2d-4d-physics}. Edges of the graph are values of 2d couplings for which $\arg Z_{a} = \vartheta_c$ for some soliton charges $a\in\tGamma_{ij}$. Joint nodes are marginal stability points where 2d-4d solitons have central charges of equal phases $\arg Z_a = \arg Z_b$, while branch point nodes are singular points where some 2d-4d solitons become massless.
	The topology of $\CW_c$ thus makes sense as a (possibly higher-dimensional) locus embedded into the parameter space of a surface defect, even when abstracted from the  UV curve of a class $\CS$ theory. 
	More puzzling is the notion of framing for $\CW_c$, it would be quite interesting to understand its meaning from a field theoretic viewpoint. A potential approach would be to extend the analysis of \cite{Tong:2014yla, Moore:2015szp, Moore:2015qyu, Moore:2014jfa, Moore:2014gua, Brennan:2016znk} to the study of supersymmetric interfaces between surface defects.\footnote{An interesting class of toy models was considered in \cite[App. D]{Gaiotto:2011tf}.}
	\item The authors of \cite{1998math.ph..11024M} established a correspondence between Strebel differentials and dessins d'enfants, uncovering a relation of the differentials to the Belyi map.
	This correspondence is clearly at play in class $\CS$ theories  of type $A_1$, where the Strebel differential corresponds (in the class $\CS$ description via Hitchin systems) to the Coulomb vacuum at which the critical graph emerges. 
	In higher rank theories, ``maximally critical'' spectral networks can be viewed as a generalization of the notion of a Strebel quadratic differential \cite{network-quiver}, which involves several multi-differentials $\phi_k$ of different degrees $k\geq 2$. 
	It is natural to wonder whether the correspondence discovered in \cite{1998math.ph..11024M} admits an extension, relating the critical graphs $\CW_c$ and the corresponding $\phi_k$ to a Belyi map.%
	\footnote{Another relation between BPS spectra and dessins was found by the authors of \cite{Cecotti:2015qha}, in the context of BPS quivers. It would be interesting to elucidate the relation between the two frameworks.}
\end{itemize}

\section*{Acknowledgements}
\addcontentsline{toc}{section}{Acknowledgements}
I am indebted to my collaborators Chan Y. Park, Masahito Yamazaki, and especially Maxime Gabella, who drew my attention to the importance of critical graphs in connection with BPS quivers.
I am grateful to Thomas Dumitrescu, Davide Gaiotto, Joe Minahan, and Luigi Tizzano for interesting discussions, and to Maxime Gabella, Greg Moore, Andy Neitzke, Chan Y. Park and Masahito Yamazaki for helpful comments on the draft of this paper. 
I am grateful to the organizers of the IHP String-Math trimester 2016 for hospitality at Institut Henri Poincar\'e, and to the organizers of the DESY workshop ``Rethinking Quantum Field Theory'' for hospitality during completion of this work, and for giving me the opportunity to present the results of this paper at the workshop.
This work was partially supported by the Carl Tryggers foundation. The research of PL is supported by the grants ``Geometry and Physics'' and  ``Exact Results in Gauge and String Theories'' from the Knut and Alice Wallenberg foundation.

\appendix

\section{Notational conventions on Wall Crossing identities}
\label{sec:active-passive}

\subsection{Classical}
%

The classical version of the BPS monodromy, denoted $\IS$ in the main body of the paper, involves operators $\CK_{\gamma'}$ acting on functions of the formal variables $\hX_\gamma$ as follows 
\be
	\CK_{\gamma'}\, F\big(\hat X_\gamma\big) = F\big(\hat X_\gamma(1-\hat X_{\gamma'})^{\langle\gamma',\gamma\rangle}\big)
\ee
Classical wall crossing identities are most conveniently presented using the twisted variables $\hX_\gamma$. 
It is always possible to switch to these variables from the un-twisted $X_{\tilde\gamma}$ using the relation in (\ref{graph:variables-relations}), also reviewed in greater detail in Appendix \ref{sec:homology-conventions}.

As an illustration of our conventions, let us review in some detail the pentagon identity (here $\langle\gamma_1,\gamma_2\rangle=1$)
\be
\begin{split}
	& \CK_{\gamma_1}\CK_{\gamma_2} = \CK_{\gamma_2} \CK_{\gamma_1+\gamma_2} \CK_{\gamma_1} 
\end{split}	
\ee
by studying the action on the ring generators $\hX_{\gamma_1}, \hX_{\gamma_2}$. The LHS reads
\be
\begin{split}
	\CK_{\gamma_1}\CK_{\gamma_2}  \hat X_{\gamma_1} & = \CK_{\gamma_1}\hat X_{\gamma_1} (1-\hat X_{\gamma_2})^{-1} =   \hat X_{\gamma_1} (1-\hat X_{\gamma_2} (1-\hat X_{\gamma_1}))^{-1} \\
	\CK_{\gamma_1}\CK_{\gamma_2}  \hat X_{\gamma_2} & = \CK_{\gamma_1}\hat X_{\gamma_2} = \hat X_{\gamma_2} (1-\hat X_{\gamma_1})
\end{split}	
\ee	
while the RHS is
\be
\begin{split}
	\CK_{\gamma_2} \CK_{\gamma_1+\gamma_2} \CK_{\gamma_1} \hat X_{\gamma_1}& = %
	\CK_{\gamma_2} \CK_{\gamma_1+\gamma_2}  \hat X_{\gamma_1} \\%
	& = \CK_{\gamma_2}  \hat X_{\gamma_1} (1- \hat X_{\gamma_1+\gamma_2} )^{-1} \\ %
	& = \hat X_{\gamma_1}(1- \hat X_{\gamma_2} )^{-1}  (1- \hat X_{\gamma_1+\gamma_2}(1- \hat X_{\gamma_2} )^{-1}  )^{-1} \\
	& = \hat X_{\gamma_1}  (1 -\hat X_{\gamma_2}- \hat X_{\gamma_1+\gamma_2} )^{-1}  \\
	& = \hat X_{\gamma_1}  (1 -\hat X_{\gamma_2}( 1 - \hat X_{\gamma_1} ))^{-1}  \\
	\CK_{\gamma_2} \CK_{\gamma_1+\gamma_2} \CK_{\gamma_1} \hat X_{\gamma_2}& = %
	\CK_{\gamma_2} \CK_{\gamma_1+\gamma_2}  \hat X_{\gamma_2}(1-\hat X_{\gamma_1}) \\%
	& = \CK_{\gamma_2}  \hat X_{\gamma_2} (1- \hat X_{\gamma_1+\gamma_2} )(1-\hat X_{\gamma_1} (1- \hat X_{\gamma_1+\gamma_2} )^{-1}) \\%
	& = \CK_{\gamma_2}  \hat X_{\gamma_2}  (1-\hat X_{\gamma_1} - \hat X_{\gamma_1+\gamma_2} ) \\%
	& = \CK_{\gamma_2}  \hat X_{\gamma_2}  (1-\hat X_{\gamma_1}(1- \hat X_{\gamma_2}) ) \\%
	& = \hat X_{\gamma_2}  (1-\hat X_{\gamma_1} ) \\%
\end{split}	
\ee
notice that the use of the \emph{twisted} product law $\hX_{\gamma_1}\hX_{\gamma_2}=-\hX_{\gamma_1+\gamma_2}$ is fundamental in checking the identity.

\subsection{Motivic}
The quantum monodromy is expressed in terms of the noncommutative variables $\hY_{\tilde\gamma}$
\be
	\hY_{\tilde\gamma} \hY_{\tilde\gamma'} = y^{\langle\tilde\gamma,\tilde\gamma'\rangle} \hY_{\tilde\gamma+\tilde\gamma'}\,,
\ee
and in terms of quantum dilogarithms, which are defined as follows:
\be\label{eq:dilog-definition}
\begin{split}
	\Phi(\xi) & = \prod_{k\geq1}(1+y^{2k-1} \xi)^{-1} \,,\\
	\Phi_n(\xi) & = \prod_{s=1}^{|n|}(1+y^{-sgn(n)(2s-1)} \xi)\,.
\end{split}
\ee
An important relation between the first (a.k.a. non-compact) and the second (a.k.a. compact) type of dilogarithm is 
\be\label{eq:dilog-relation}
\begin{split}
	\Phi(\hY_{\tilde\gamma})\, %
	\hY_{\tilde\gamma'}\,%
	\Phi(\hY_{\tilde\gamma})^{-1} \ %
	= \ %
	\hY_{\tilde\gamma'}  \, %
	\Phi_{\langle\tilde\gamma',\tilde\gamma\rangle}(\hY_{\tilde\gamma})^{-{\rm sgn} \langle\tilde\gamma',\tilde\gamma\rangle}
\end{split}
\ee
For convenience let us define the following operators
\be\label{eq:dilog-relation}
\begin{split}
	\hat \CK_{\gamma}^{\Omega(\gamma,y)} \, \hY_{\tilde\gamma'} & := \left(\prod_m \left[\Phi((-y)^m  \hY_{\tilde\gamma})\right]^{a_m(\gamma)} \right)\, %
	Y_{\tilde\gamma'}\,%
	\left(\prod_m  \left[\Phi((-y)^m  \hY_{\tilde\gamma})\right]^{-a_m(\gamma)} \right) 
\end{split}
\ee
where $\Omega(\gamma,y)$ is related to $a_m(\gamma)$ by $\Omega(\gamma,y) = \sum_{m\in\IZ} a_m(\gamma) (-y)^{m}$, and ${\tilde\gamma}$ is the \emph{preferred lift} of $\gamma$ appearing on the LHS (the preferred lift was introduced in \cite{Gaiotto:2012rg}, see also Appendix \ref{sec:homology-conventions}).

As an illustration of our conventions , let us review the motivic pentagon identity
\be
\begin{split}
	&\hat \CK_{\gamma_1} \hat \CK_{\gamma_2} = \hat \CK_{\gamma_2} \hat \CK_{\gamma_1+\gamma_2} \hat \CK_{\gamma_1} \,,
\end{split}
\ee
by studying the action of either side on the ring generator $\hY_{\tilde\gamma_1}$
This is equivalent to the following identity on quantum dilogarithms
\be
	\Phi(\hY_{\tilde\gamma_1}) \Phi(\hY_{\tilde\gamma_2}) = \Phi(\hY_{\tilde\gamma_2}) \Phi(\hY_{\tilde\gamma_1+\tilde\gamma_2}) \Phi(\hY_{\tilde\gamma_1})\,.
\ee
The LHS acts as
\be
\begin{split}
	\hat \CK_{\gamma_1} \hat \CK_{\gamma_2}  \hY_{\gamma_1} & = %
	 \hat \CK_{\gamma_1}\, \hY_{\tilde\gamma_1} \, \Phi_1(\hY_{\tilde\gamma_2})^{-1} \\
	 & =\hY_{\tilde\gamma_1} \,\Phi_1(\hY_{\tilde\gamma_2} \,\Phi_{-1}(\hY_{\tilde\gamma_1}) )^{-1} \\
	 & = \hY_{\tilde\gamma_1} \, (1+y^{-1} \hY_{\tilde\gamma_2} + y^{-1} \hY_{\tilde\gamma_1+\tilde\gamma_2})^{-1}
\end{split}
\ee
The RHS of the identity instead gives
\be
\begin{split}
	 \hat \CK_{\gamma_2} \hat \CK_{\gamma_1+\gamma_2} \hat \CK_{\gamma_1}  \hY_{\tilde\gamma_1} & = %
	\hat \CK_{\gamma_2} \hat \CK_{\gamma_1+\gamma_2}  \hY_{\tilde\gamma_1} \\
	& = \hat \CK_{\gamma_2}   \hY_{\tilde\gamma_1}  \Phi_1(\hY_{\tilde\gamma_1+\tilde\gamma_2})^{-1}  \\
	& = \big( \hY_{\tilde\gamma_1}\Phi_1(\hY_{\tilde\gamma_2})^{-1}  \big) \big( \Phi_1(\hY_{\tilde\gamma_1+\tilde\gamma_2}\Phi_1(\hY_{\tilde\gamma_2})^{-1} )^{-1}   \big) \\
	& = \hY_{\tilde\gamma_1} (1+y^{-1} \hY_{\tilde\gamma_2})^{-1} \, (1+y^{-1} \hY_{\tilde\gamma_1+\tilde\gamma_2}(1+y^{-1} \hY_{\tilde\gamma_2})^{-1})^{-1} \\
	& = \hY_{\tilde\gamma_1}  \, ((1+y^{-1} \hY_{\tilde\gamma_1+\tilde\gamma_2}(1+y^{-1} \hY_{\tilde\gamma_2})^{-1}) \, (1+y^{-1} \hY_{\tilde\gamma_2}) )^{-1} \\
	 & = \hY_{\tilde\gamma_1} \, (1+y^{-1} \hY_{\tilde\gamma_2} + y^{-1} \hY_{\tilde\gamma_1+\tilde\gamma_2})^{-1}
\end{split}
\ee
where, in the second to last step we used relation for non-commutative multiplicative operators $A^{-1} B^{-1} = (BA)^{-1}$.
%
%

\section{Homology classes and formal variables}\label{sec:homology-conventions}
Different types of homology classes and formal variables are employed in this paper. 
In this appendix we collect the key relations among them.
Let $\Sigma$ be the spectral curve of a class $\CS$ Hitchin system, at a fixed point $u$ in the Hithcin base (or Coulomb branch). Also let $\tSigma$ be the unit tangent bundle over $\Sigma$.
For the most part, we employ charges valued in either the following lattices
\be
	\tGamma : = H_1(\tSigma,\IZ) \,,\qquad \Gamma : = H_1(\Sigma,\IZ) \,.
\ee
These differ slightly from the physical charge lattice of the gauge theory, by a certain projection. 
For simplicity we will ignore projections onto physical sub-lattices in this paper, details on these can be found in \cite{Gaiotto:2009hg, Longhi:2016rjt}.

There is a canonical projection map $\tilde\pi:\tSigma\to\Sigma$, the associated pushforward is denoted $\tilde\pi_* : \tGamma\to\Gamma$. The kernel of $\tilde\pi_*$ is generated by $H\in\tGamma$, a representative for this would be the lift by $\tilde\pi^{-1}$ of a small contractible circle on $\Sigma$, with counter-clockwise orientation with respect to that of $\Sigma$ as a complex curve.
Although not manifest from our notation, the lattices $\tGamma,\Gamma$ are actually local systems over the Hitchin base $\CB$. Locally on $\CB$, there is a distinguished section $\plift:\Gamma\to\tGamma$ introduced in \cite[\S 6.3]{Gaiotto:2012rg}, known as the \emph{preferred lift}.
An important property of this section argued in that reference is
\be
	\left( \plift(\gamma+\gamma') - \plift(\gamma) -\plift(\gamma') \right) \big/ H = \langle\gamma,\gamma'\rangle \ \text{mod}\,2 \,.
\ee
Using the preferred lift we can define the following quadratic refinement\footnote{This should {not} to be confused with the quadratic refinement introduced in \cite{Gaiotto:2009hg}, which was defined on $\Gamma$, not $\tGamma$.}
\be
	\rho(\tilde\gamma) := (-1)^{\left(\plift\circ\pi_*(\tilde\gamma) - \tilde\gamma\right) / H}
\ee
which obeys
\be
	\rho(\tilde\gamma) \rho(\tilde\gamma') = (-1)^{\langle\tilde\gamma,\tilde\gamma'\rangle} \rho(\tilde\gamma+\tilde\gamma')\,.
\ee
Since $H$ belongs to the annihilator of $\langle\,,\,\rangle$, the pairing $\langle\tilde\gamma,\tilde\gamma'\rangle$ is really the same as that of the pushforwards $\langle \gamma,\gamma'\rangle$.

Turning to formal variables, the fundamental ones are the $\hY_{\tilde\gamma}$, associated to elements of $\tGamma$ and subject to the following identities (we follow conventions of \cite{Galakhov:2014xba})\footnote{In that reference $\tilde\Sigma$ was modified by including certain punctures. The punctures were introduced to deal with subtleties tied to properties of relative homotopy classes of open paths. Here we can safely ignore this subtlety because we almost always work with closed homology cycles.}
\be
	\hY_{\tilde\gamma}\hY_{\tilde\gamma'} = y^{\langle\tilde\gamma,\tilde\gamma'\rangle}\hY_{\tilde\gamma+\tilde\gamma'} \,, \qquad \hY_{\tilde\gamma + n H} = (-y)^{n} \hY_{\tilde\gamma}\,.
\ee
We also consider two specializations of these variables.
The first one is for $y\to 1$, denoted by $X_{\tilde\gamma}$ and obeying
\be\label{eq:untwisted-classical-var}
	X_{\tilde\gamma}X_{\tilde\gamma'} =X_{\tilde\gamma+\tilde\gamma'} \,, \qquad X_{\tilde\gamma + n H} = (-1)^{n} X_{\tilde\gamma}\,,
\ee
these are precisely the variables employed in \cite{Gaiotto:2012rg}.
The second specialization is for $y\to -1$, we denote it by $\hX_{\tilde\gamma}$. These satisfy twisted product rules
\be\label{eq:twisted-classical-vars}
	\hX_{\tilde\gamma}\hX_{\tilde\gamma'} =(-1)^{\langle\tilde\gamma,\tilde\gamma'\rangle}  \hX_{\tilde\gamma+\tilde\gamma'} \,, \qquad \hX_{\tilde\gamma + n H} = \hX_{\tilde\gamma}\,.
\ee
It is manifest from (\ref{eq:twisted-classical-vars}) that the $\hX_{\tilde\gamma}$ are really functions of $\gamma=\pi_{*}(\tilde\gamma)$, and therefore will be denoted $\hX_\gamma$. These are the variables employed in \cite{Gaiotto:2009hg}.
The relation between the two specializations is 
\be
	\hX_{\tilde\gamma} = \rho(\tilde\gamma) X_{\tilde\gamma}\,,
\ee
this also clarifies how $\hX$ variables become independent of $H\in\tGamma$: a shift $\tilde\gamma\to \tilde\gamma+H$ is absorbed by the property (\ref{eq:untwisted-classical-var}) together with an extra sign from the quadratic refinement $\rho$.

In the main body of the paper we often switch among different types of variables, their relations are conveniently summarized by the graph (\ref{graph:variables-relations}).

\section{Traffic rules for 2d-4d soliton data}\label{sec:traffic-rules}
For convenience we collect the rules for soliton propagation across the network, both in the American and British resolutions. 
These rules were first derived in \cite[App. A]{Gaiotto:2012rg}, they describe the relations among generating functions of $\CS$-wall ``soliton data'' across nodes of the network: branch points and joints. Generating functions are denoted by $\nu$ or $\tau$ corresponding to solitons flowing into or out of a node, schematically they encode soliton data as $\nu_p = \sum_{a\in\tGamma_{ij}(z)}\mu(a) X_a$, where $z\in p$ is any point on a street $p$ (``streets'' are pieces of $\CS$-walls delimited by nodes of the network).

\subsection{Six-way joints}
\begin{figure}[h!]
\begin{center}
\includegraphics[width=0.65\textwidth]{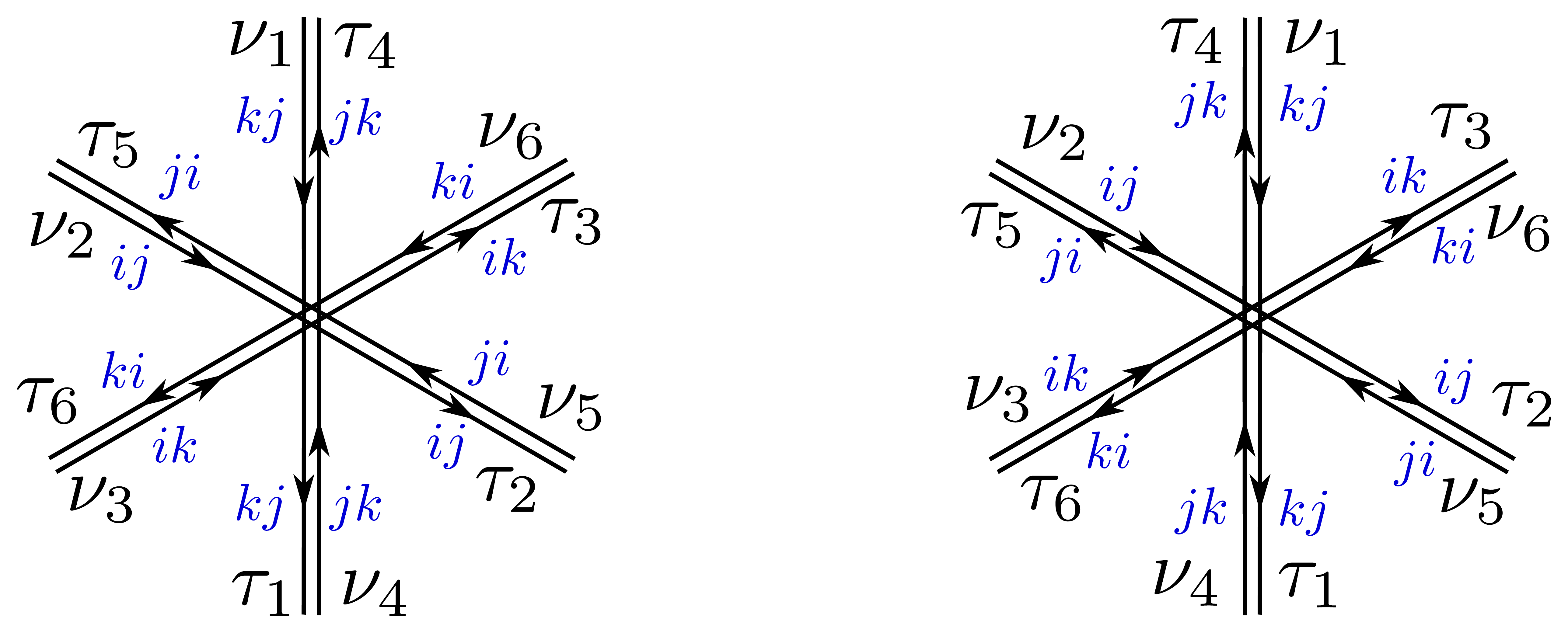}
\caption{Left: the 6-way joint in the American resolution. Right: the British resolution.}
\label{fig:6-way-joints}
\end{center}
\end{figure}
The soliton generating functions of outgoing/ingoing ($\tau_n/\nu_n$) solitons refer to figure \ref{fig:6-way-joints}.
For the American resolution the soliton rules are
\be
\begin{split}
	\tau_1 = \nu_1 + \tau_6 \nu_2 \, , \qquad %
	\tau_2 = \nu_2 + \nu_3 \tau_1 \, , \qquad %
	\tau_3 = \nu_3 + \tau_2 \nu_4 \, , \\
	\tau_4 = \nu_4 + \nu_5 \tau_3 \, , \qquad %
	\tau_5 = \nu_5 + \tau_4 \nu_6 \, , \qquad %
	\tau_6 = \nu_6 + \nu_1 \tau_5 \, .
\end{split}
\ee
For the British resolution the soliton rules are
\be
\begin{split}
	&\tau_1 = \nu_1 + \nu_6 \tau_2 \, , \qquad %
	\tau_2 = \nu_2 + \tau_3 \nu_1 \, , \qquad %
	\tau_3 = \nu_3 + \nu_2 \tau_4 \, , \\
	&\tau_4 = \nu_4 + \tau_5 \nu_3 \, , \qquad %
	\tau_5 = \nu_5 + \nu_4 \tau_6 \, , \qquad %
	\tau_6 = \nu_6 + \tau_1 \nu_5 \, .
\end{split}
\ee

\subsection{Branch Points}
\begin{figure}[h!]
\begin{center}
\includegraphics[width=0.65\textwidth]{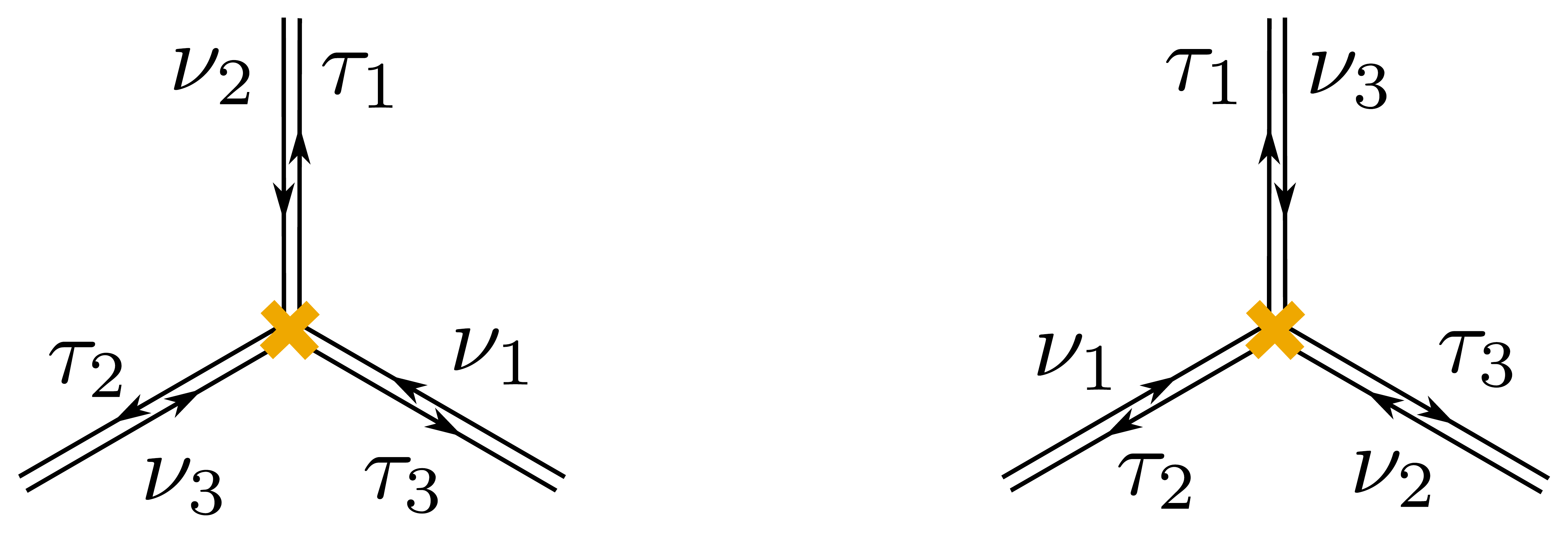}
\caption{Left: the branch point in the American resolution. Right: the British resolution.}
\label{fig:branch-point}
\end{center}
\end{figure}
The soliton generating functions of outgoing/ingoing ($\tau_n/\nu_n$) solitons refer to figure \ref{fig:branch-point}.
In both the American and British resolutions the soliton rules are
\be
\begin{split}
	&\tau_n = X_{a_n} + \nu_n,
\end{split}
\ee
where the different labeling of figure \ref{fig:branch-point} takes into account the change of resolution.

\section{A technique for computing soliton data of networks}
\label{sec:soliton-computations}
In this appendix we present a technique for effectively computing soliton generating functions across a network.
The main difficulties one encounters in computing soliton data is in fact due to the formalism: the equations provided in \cite{Gaiotto:2012rg} do not manifestly keep track of the ``parallel transport" of solitons across streets, which is ubiquitous in most applications of spectral networks.
We will modify the traffic rules, introducing simplified formulae which capture the homological data of soliton paths, but not the writhe. Therefore they will be useful  for computing classical monodromies $\IS$, but not quantum monodromies $\mathbb{U}$.

\begin{figure}[h!]
\begin{center}
\includegraphics[width=0.3\textwidth]{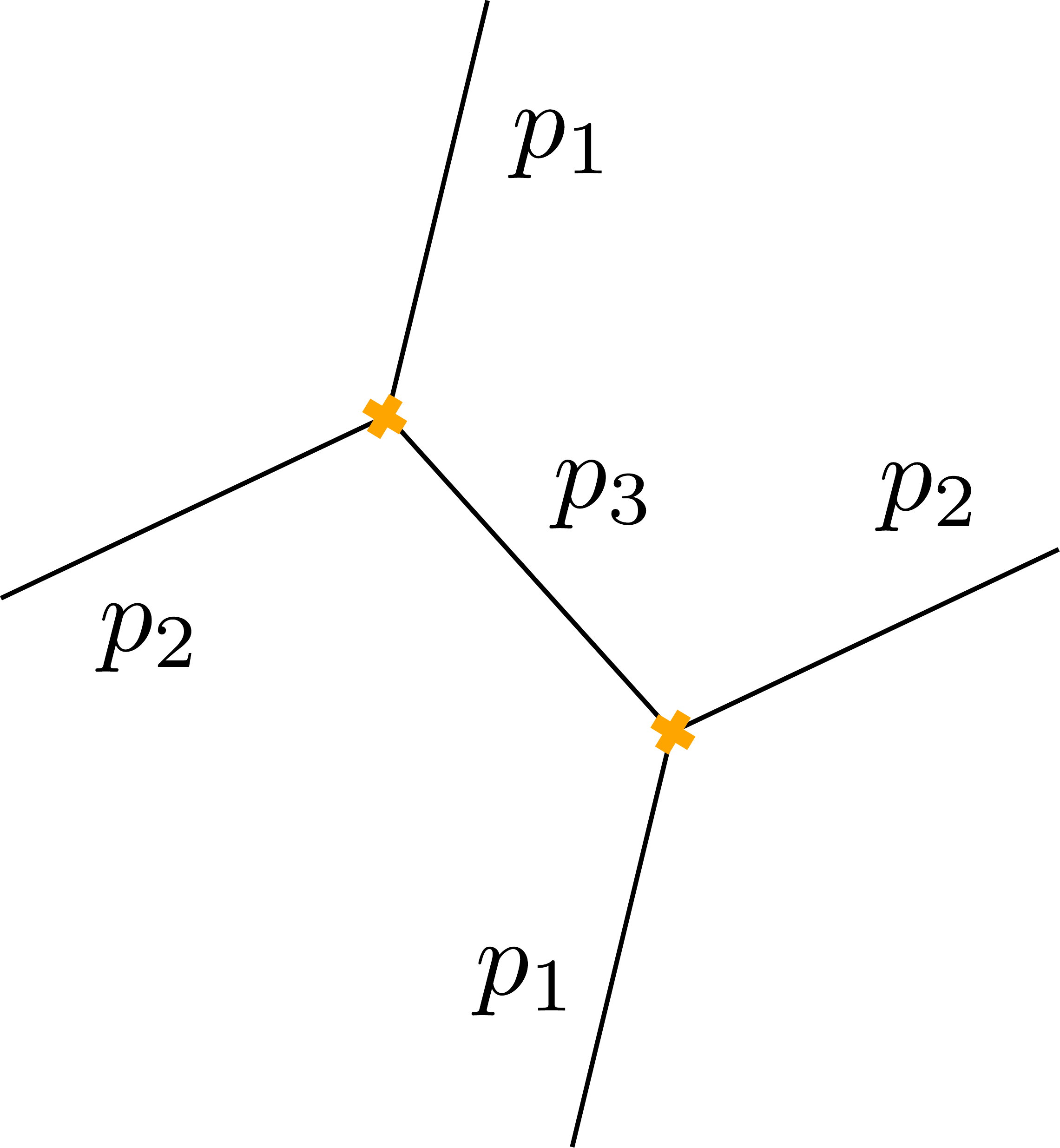}
\caption{Critical graph of $SU(2)$ $\CN=2^*$ theory. The UV curve $C$ is a torus, periodicity of the graph is indicated by street labels.}
\label{fig:N2star_bis}
\end{center}
\end{figure}

To illustrate how this works, we focus on the example of $\CN=2^*$ theory.
Its critical graph is shown in Figure \ref{fig:N2star_bis}, the standard soliton equations for this graph are%
\footnote{To avoid potential confusion, let us comment on the fact that these equations differ slightly from (\ref{eq:N2-star-equations}). 
They would coincide upon switching $\Delta_2 \leftrightarrow \Upsilon_2$ and $a_i\leftrightarrow b_i$. 
This change of variables is simply due to slightly different conventions. 
In fact the graphs in Figures \ref{fig:N2star} and \ref{fig:N2star_bis} have the same topology and framing. 
They are related by a shift by $\pi$ around the imaginary (i.e. vertical) cycle of the torus, this explains the exchange $a_i\leftrightarrow b_i$. 
Moreover the edge $p_2$ is tilted in opposite directions between the two graphs, this causes the switch in conventions for its soliton generating functions. The overall result of the monodromy is independent of these details.
}
\be\label{eq:N2-star-standard-eqs}
\begin{split}
	& \Upsilon_1 = X_{a_1} + \Upsilon_2, \qquad %
	\Upsilon_2 = X_{b_2} + \Delta_3, \qquad %
	\Upsilon_3 = X_{b_3} + \Upsilon_1, \\
	& \Delta_1 = X_{b_1} + \Delta_2, \qquad %
	\Delta_2 = X_{a_2} + \Upsilon_3, \qquad %
	\Delta_3 = X_{a_3} + \Delta_1,  %
\end{split}
\ee
where $\Upsilon_i,\Delta_i$ are 2d soliton generating functions for solitons propagating upwards/downwards on street $p_i$ for $i=1,2,3$ and $a_i, b_i$ are solitons sourced by the upper (resp. lower) branch point on street $p_i$.

Solitons of a street $p$ are relative homology classes valued in $\tGamma_{ij}(z)$ for any $z\in p$, in particular this is a torsor for $\tGamma$, which means that any soliton path can be expressed as $a+\tilde\gamma$ for a fixed $a$, and a suitable $\tilde\gamma$.
In view of this we define the \emph{reduced} soliton generating functions $\tilde\Upsilon,\tilde\Delta$ as
\be
	\Upsilon_1 = X_{a_1} \tilde\Upsilon_1 \,,\qquad \Delta_1 = X_{b_1} \tilde\Delta_1 \,,
\ee
and so on. Note that these have the schematic form $\tilde\Upsilon_1 = 1+ \sum_{\tilde\gamma}c_{\tilde\gamma}X_{\tilde\gamma}$.

Consider now the first equation in (\ref{eq:N2-star-standard-eqs}), if we just substitute $\Upsilon_2$ on the RHS using the second equation we get 
\be
	\Upsilon_1= X_{a_1}+ X_{b_2} +\dots
\ee
which is notationally confusing because $b_2$ is sourced from the lower branch point and is valued in $ \Gamma_{ij}(z)$ with $z\in p_2$,  it does not look like a soliton from the generating function $\Upsilon_1$. 
Implicit in the equations is the fact that one must extend $b_2$ along the lift of $p_2$ and across the upper branch point onto the lift of $p_1$. 
The drawback of this formalism is that it does not keep track of the important information of \emph{how} the transport of a soliton is carried out. 

A trick is to note that only part of such information is needed, in particular the \emph{homology class} of a transported soliton. Other details about the path of transport are unimportant (they would have been important if we wanted to compute writhes of paths, for example).
To keep track of homology, this we modify the network equations by replacements such as
\be
	\Upsilon_2 \to \Upsilon_2^{(p_2)}\,, 
\ee
within the first equation of (\ref{eq:N2-star-standard-eqs}). 
Here  $\Upsilon_2^{(p_2)}$ denotes the generating function of \emph{modified} soliton paths transported along $p_2$, therefore it counts solitons in $\tGamma_{ij}(z)$ with $z\in p_1$. 
Hence it can be expressed as 
\be
	\Upsilon_2^{(p_2)} = X_{a_1} \, \sum_{\tilde\gamma}c_{\tilde\gamma}X_{\tilde\gamma}\,.
\ee
In fact $b_2$ after parallel transport is equivalent to $a_1+\tilde\gamma_2$, therefore 
\be
	\Upsilon_2^{(p_2)} = X_{a_1 + \tilde\gamma_2} \tilde \Upsilon_2 = X_{a_1} \, X_{\tilde\gamma_2} \tilde \Upsilon_2\,.
\ee
Then we can rewrite equations (\ref{eq:N2-star-standard-eqs}) as
\be\label{eq:N2-star-modified-eqs}
\begin{split}
	& \tilde\Upsilon_1 = 1 + X_{\tilde\gamma_2}\tilde\Upsilon_2, \qquad %
	\tilde\Upsilon_2 = 1 + X_{\tilde\gamma_3}\tilde\Delta_3, \qquad %
	\tilde\Upsilon_3 = 1 + X_{\tilde\gamma_1}\tilde\Upsilon_1, \\
	& \tilde\Delta_1 = 1 + X_{\tilde\gamma_2}\tilde\Delta_2, \qquad %
	\tilde\Delta_2 = 1 + X_{\tilde\gamma_3}\tilde\Upsilon_3, \qquad %
	\tilde\Delta_3 = 1 + X_{\tilde\gamma_2}\tilde\Delta_1\,.  %
\end{split}
\ee
These equations manifestly take into account the effect of parallel transporting along the network, moreover they are much easier to manipulate and to implement in computer algebra systems, because all variables are \emph{commutative}.

From the modified traffic rules (\ref{eq:N2-star-modified-eqs}) it's straightforward to derive single-variable equations like
\be
\begin{split}
	\tilde\Upsilon_1 = 1+ X_{\tilde\gamma_2}(1+ X_{\tilde\gamma_3}(1+ X_{\tilde\gamma_1}(1+ X_{\tilde\gamma_2}(1+ X_{\tilde\gamma_3}(1+ X_{\tilde\gamma_1} \, \tilde \Upsilon_1 )))))\,,
\end{split}
\ee
whose solution yields
\be
	\Upsilon_1 = X_{a_1} \frac{1+X_{\tilde\gamma_2}+X_{\tilde\gamma_2+\tilde\gamma_3}+X_{\tilde\gamma_1+\tilde\gamma_2+\tilde\gamma_3}+X_{\tilde\gamma_1+2\tilde\gamma_2+\tilde\gamma_3}+X_{\tilde\gamma_1+2\tilde\gamma_2+2\tilde\gamma_3}}{1-X_{2(\tilde\gamma_1+\tilde\gamma_2+\tilde\gamma_3)}}\,.
\ee
A similar expression holds for $\Delta_1$, so we obtained
\be
\begin{split}
	Q(p_1) & = 1+ \Delta_1 \Upsilon_1  \\
	& = 1+\big(X_{a_1} X_{b_1}\big) \, %
	\left( \frac{1+X_{\tilde\gamma_2}+X_{\tilde\gamma_2+\tilde\gamma_3}+X_{\tilde\gamma_1+\tilde\gamma_2+\tilde\gamma_3}+X_{\tilde\gamma_1+2\tilde\gamma_2+\tilde\gamma_3}+X_{\tilde\gamma_1+2\tilde\gamma_2+2\tilde\gamma_3}}{1-X_{2(\tilde\gamma_1+\tilde\gamma_2+\tilde\gamma_3)}}\right)^2\,,
\end{split}
\ee
noting that $X_{\tilde\gamma_1}=X_{a_1} X_{b_1}$, this is the result claimed in Section \ref{sec:N2-star}.

It is straightforward to apply the same idea to joint equations as well, an application of this idea yields the $T_3$ equations (\ref{eq:T3-eqs-123}).

%
%

\section{Algorithmic factorization of BPS monodromies into quantum dilogarithms}\label{sec:algorithmic-factorization}
In this appendix we give the details behind the procedure presented in Section \ref{sec:solving-monodromy-eqs} for solving equations (\ref{eq:U-eqn}) with the ansatz (\ref{eq:factorized-U}).
In the following we will drop the explicit dependence on a street $p$ of the network, but it is understood that each of the following equations has a counterpart for each $p$.

Since all the BPS states appear at the same phase, there is a canonical choice of half-plane in the complex plane of central charges (up to the choice of particles vs anti-particles).
In turn this determines a half-lattice $\Gamma^+$, inside the charge lattice $\Gamma$ of 4d gauge and flavor charges. All BPS particles have charges contained in $\Gamma^+$.
There is a unique positive integral basis for $\Gamma^+$, let $\gamma_k$ with $k=1,\dots, d$ with $d=\text{rank}(\Gamma)$ be its generators.
Any charge $\gamma\in\Gamma^+$ admits by definition the unique expansion $\gamma = \sum_k a_k\gamma_k$, with $a_k\in\IZ_{\geq0}$.
We use this to define a filtration on $\Gamma^+$: we will say that $\gamma\in\Gamma^+$  is a charge of level $|\gamma|=n$ if $\sum_k a_k = n$.

Given an arbitrary function $F$ of the $\hY_{\tilde\gamma}$ (with $\gamma \equiv \tilde\pi_*(\tilde\gamma)\in\Gamma^+$), we define its series expansion subordinate to $\Gamma^+$ as
\be
	F = \sum_{n\geq 0} \sum_{|\gamma|=n} f_{\tilde\gamma}\, \hY_{\tilde\gamma} = F_0 + F_1 + F_2 + \dots
\ee
where $F_n$ is a finite sum of monomials $f_{\tilde\gamma} \hY_{\tilde\gamma}$ containing only charges of level $|\gamma|=n$.
For example, a quantum dilogarithm has the expansion
\be\label{eq:dilog-expansion}
	\Phi\left((-y)^m \hY_{\tilde\gamma}\right)^{a_m(\gamma)} = 1 -a_m(\gamma) \frac{y}{1-y^2} \, (-y)^m \, \hY_{\tilde\gamma} +O\left(\hY_{2\tilde\gamma}\right)\,.
\ee
Using the filtration, equation (\ref{eq:U-eqn}) can be split into 
\be\label{eq:U-eqn-filtered}
	\sum_{j=0}^n \left(  U_j\, Q_{n-j}^{(-)}  - Q_{n-j}^{(+)} \, U_j  \right) = 0 \,\qquad n=0,1,2,\dots
\ee
where $U_i, Q_i^{(\pm)} $ are coefficients of 
\be
	\mathbb{U} = U_0 +U_1 +U_2 + \dots \qquad Q^{(\pm)} = Q^{(\pm)}_0 + Q^{(\pm)}_1 + Q^{(\pm)}_2 +\dots
\ee
It follows from the ansatz (\ref{eq:factorized-U}) and from (\ref{eq:dilog-expansion}) that $U_0 = Q^{(\pm)}_0 = 1$ , and therefore from (\ref{eq:U-eqn-filtered}) 
\be
	Q_1^{(+)} = Q_1^{(-)} \equiv Q_1
\ee
which defines $Q_1(p)$ for each $p$. Furthermore the generic form of $Q_1$ is determined by $d$ coefficients $q_k$ as
\be
	Q_1 = \sum_{k=1}^d q_k \, \hY_{\tilde\gamma_k}\,.
\ee
To fix the overall power of $y$ in the $q_k$, it's important to stress that $\tilde\gamma$ in  $\hY_{\tilde\gamma}$ is the \emph{preferred lift} of $\gamma$, see Appendix \ref{sec:homology-conventions}.

We will solve for $U_i$ recursively, starting from $U_{j<i}$. It is convenient to reorganize (\ref{eq:U-eqn-filtered}) as (we fix $n=i+1$ here)
\be\label{eq:U-eqn-filtered-step-1}
	U_i \, Q^{(-)}_1  - Q^{(+)}_1  \, U_i + \hat R_{i+1} = 0
\ee
with $\hat R_{i+1} = \sum_{j=0}^{i-1} \left(  U_j\, Q_{i-j+1}^{(-)}  - Q_{i-j+1}^{(+)} \, U_j  \right) $.
Next note that we can split (the $\nwarrow$ denote the phase ordering by central charges, as usual)
\be
	U_i = \tilde U_i + \hat U_i
\ee
with 
\be
\begin{split}
	\tilde U_i & = \left[  \, \prod^{\nwarrow}_{|\gamma|=i}\prod_{m\in\IZ}   \Phi\left(  (-y)^m \hY_{\tilde\gamma}\right)^{a_m(\gamma)}  \,   \right]_{{\rm lvl} = i} \\
	\hat U_i & = \left[  \, \prod^{\nwarrow}_{|\gamma|<i}\prod_{m\in\IZ}   \Phi\left(  (-y)^m \hY_{\tilde\gamma}\right)^{a_m(\gamma)}  \, \right]_{{\rm lvl} = i}
\end{split}
\ee
Note that $\tilde U_i$ has the following explicit dependence on the $a_m(\gamma)$
\be\label{eq:tilde-U-i-expansion}
	\tilde U_i = \frac{y}{1-y^2} \, \sum_{|\gamma|=i, \, m\in\IZ} (-a_m(\gamma)) \, (-y)^m  \, \hY_{\tilde\gamma} \,.
\ee
Then we can further rewrite (\ref{eq:U-eqn-filtered-step-1}) as
\be\label{eq:U-eqn-filtered-step-2}
	\left[\tilde U_i , Q_1\right] + \tilde R_{i+1} = 0
\ee
with 
\be
	\tilde R_{i+1} := \hat R_{i+1}+ \left[\hat U_i,  Q_1\right] = \sum_{|\gamma|=i+1} \tilde r_{\tilde\gamma} \hY_{\tilde\gamma}
\ee
Since $U_{j<i}$ and $\hat U_i$ only depend on $a_m(\gamma)$ for $|\gamma|<i$, the same is true of $\tilde R_{i+1}$.

Plugging (\ref{eq:tilde-U-i-expansion}) into (\ref{eq:U-eqn-filtered-step-2}) and considering each monomial $\hY_{\tilde\gamma}$ separately finally yields (\ref{eq:a-m-eqn})
\be \label{eq:U-eqn-filtered-step-3}
	\tilde r_{\tilde\gamma}  + \sum_{k=1}^{d} q_k \, \frac{ y^{\langle \gamma,\gamma_k\rangle}- y^{-\langle \gamma,\gamma_k\rangle}  }{y-y^{-1}} \sum_{m\in\IZ} (-y)^m\, a_m(\gamma-\gamma_k)  =0 \,.
\ee
Equation (\ref{eq:U-eqn-filtered-step-3}) is linear in the variables $a_m(\gamma)$ with $|\gamma| = i$, and it has an inhomogeneous term $\tilde r_{\tilde\gamma}$ which depends entirely on the  $a_m(\gamma)$ for $|\gamma|<i$. 
Therefore this equation can be used to compute the $a_m(\gamma)$ recursively.

\section{$T_3$ generating functions}\label{sec:T3-Qs}
Here we collect some details on the analysis of soliton generating functions for the $T_3$ theory. We will focus entirely on the \emph{classical} functions. The quantum generating functions can be obtained to arbitrary precision using the software provided with this paper \cite{python-code}.

Thanks to the high degree of symmetry of the graph, it is sufficient to write the equations for 5 of the 30 generating functions. We will work in British resolution, with labels referring to Figure \ref{fig:T3}. The notation $\Upsilon_n$ denotes the generating function of upwards-flowing solitons on street $p_n$. By choice of convention, ``upwards'' means that a soliton flows towards the north pole of the sphere, which coincides with joint $J_1$. For example, $\Upsilon_1 = X_{a_1}+\dots$ where $a_1$ is sourced at branch point $B_1$ and stretches upwards along $p_1$. Similarly, $\Delta_n$ is the generating function of downward-flowing solitons, extending towards the south pole, which coincides with joint $J_2$.

The generating set of equations consists of the following two obtained from joint $J_1$:
\be
	\Delta_1 = \Upsilon_{13}  + \Upsilon_2 \Delta_{15} \,,\qquad \Delta_{15} = \Upsilon_{3} + \Delta_2 \Upsilon_{13}\,,
\ee
together with the following three obtained from branch point $B_1$
\be\label{eq:aux-eq-1}
	\Upsilon_1 = X_{a_1} +\Upsilon_7\,, \qquad%
	\Delta_7 = X_{a_7} +\Upsilon_8 \,,\qquad%
	\Delta_8 = X_{a_8} +\Delta_1 \,,
\ee
where $a_n$ is the 2d-4d soliton charge sourced at $B_1$ and stretching along $p_n$.
The remaining 25 equations are related to these by the $S_3$ symmetry of the graph.

Starting from these equations, we can immediately eliminate $\Upsilon_7, \Upsilon_{10}, \Upsilon_{11}, \Delta_8, \Delta_9,\Delta_{12}$ using the following relations:
\be\label{eq:aux-eq-2}
\begin{split}
	& \Upsilon_7 = X_{b_7} + \Upsilon_4\,,\qquad %
	\Upsilon_{10} =  X_{b_{10}} + \Upsilon_6\,,\qquad %
	\Upsilon_{11} =  X_{b_{11}} + \Upsilon_5\,,\\ %
	& \Delta_8 = X_{a_8} +\Delta_1 \,,\qquad %
	\Delta_9 = X_{a_9} +\Delta_3 \,,\qquad %
	\Delta_{12} = X_{a_{12}} +\Delta_2 \,, %
\end{split}
\ee
where $a_n$ denotes a soliton sourced at one of the upper branch points $B_1, B_2, B_3$ and stretching along street $p_n$, while $b_n$ denotes a soliton sourced at one of the lower branch points.
We already argued in Section \ref{sec:T3} that $\Delta_7, \Delta_8, \Delta_{10}$ and $\Upsilon_{8},\Upsilon_{9},\Upsilon_{12}$ can be obtained in closed form.
A bit more work shows that we can get rid of $\Upsilon_{4,5,6}, \Delta_{4,5,6}$ as follows. Notice that $\Upsilon_7$ appears both in (\ref{eq:aux-eq-1}) and in (\ref{eq:aux-eq-2}) and can be eliminated to obtain the following relation: 
\be
\begin{split}
	& \Upsilon_4 = X_{-\gamma_2-\gamma_3}( \Upsilon_1 -X_{a_1}) - X_{-\gamma_2} X_{b_7}\,,
\end{split}
\ee
where we inserted explicit factors of $X_{-\gamma_i}$ to account for the parallel transport of solitons which is always understood in the traffic rules of spectral networks.
By exploiting the symmetries of the graph it is straightforward to obtain similar equations for the other five generating functions.
Furthermore we can also express $\Upsilon_{1,2,3} , \Delta_{1,2,3}$ in terms of the generating functions of $p_{13},p_{14},p_{15}$ using 
\be
	\Delta_1 = \frac{\Upsilon_{13}+\Delta_{14}\Delta_{15}}{1+\Upsilon_{15}\Delta_{15}}
\ee
and other equations obtained from this by the $S_3$ symmetry.
To summarize, the generating functions of streets $p_1\dots p_{12}$ can be expressed in terms of those of the last three streets $p_{13},p_{14},p_{15}$. Therefore the 30 coupled equations corresponding to the network traffic rules can be reduced to six equations in the six variables $\Delta_{13,14,15}, \Upsilon_{13,14,15}$. 
The catch is that these final six equations are of a high degree, and are hard to solve.
They also admit more than one solution, the correct one can be singled out by requiring that all generating functions have an expansion for small $X_{\tilde\gamma_i}$ which agrees with the perturbative solution  obtained by repeated application of the traffic rules.\footnote{For concreteness, every generating function must have an expansion of the form $1+O(X_{\tilde\gamma_i})$.}
We haven't found a nice compact expression for these generating functions, but it is straightforward to obtain a perturbative solution to arbitrary accuracy (as an expansion in small $X_{\tilde\gamma_i}$) by iterated substitutions.

\nocite{*}
\bibliography{biblio}

\end{document}